\documentclass[aps,pre,twocolumn,titlepage,superscriptaddress,nofootinbib]{revtex4-2}
\usepackage{graphicx}
\usepackage{color}
\usepackage{dcolumn}
\usepackage{latexsym}
\usepackage{cancel}
\usepackage[normalem]{ulem}
\usepackage{hyperref,amssymb}
\usepackage{url}
\usepackage{color,soul}
\usepackage{graphicx}
\usepackage{anyfontsize}
\usepackage{verbatim}
\usepackage{multirow}
\usepackage{amsmath}
\newcommand{\beq}{\begin{eqnarray}}
\newcommand{\eeq}{\end{eqnarray}}
\usepackage{mathrsfs}
\usepackage{float,soul}
\usepackage[dvipsnames]{xcolor}
\usepackage{mathtools}
\usepackage{slashed}
\usepackage{graphicx}   
\usepackage{epstopdf}
\usepackage{subfigure}  
\usepackage{hyperref}   
\usepackage{bbold}
\usepackage{wasysym}
\usepackage{feynmp}

\begin{document}


\title{Dissipation induced elastic-mode instability with topological excitation in holographic non-equilibrium steady cnoidal wave supersolid}

\author{Peng Yang}\email{pengyang23@sjtu.edu.cn}
\affiliation{Wilczek Quantum Center, School of Physics and Astronomy, Shanghai Jiao Tong University, Shanghai 200240, China}
\affiliation{Shanghai Research Center for Quantum Sciences, Shanghai 201315, China}
\author{Yu Tian}\email{ytian@ucas.ac.cn}
\affiliation{School of Physical Sciences, University of Chinese Academy of Sciences, Beijing 100049, China}
\author{Matteo Baggioli}\email{b.matteo@sjtu.edu.cn}
\affiliation{Wilczek Quantum Center, School of Physics and Astronomy, Shanghai Jiao Tong University, Shanghai 200240, China}
\affiliation{Shanghai Research Center for Quantum Sciences, Shanghai 201315, China}

\begin{abstract}
The possible existence of an exotic phase of matter rigid like a solid but able to sustain persistent and dissipation-less flow like a superfluid, a ``supersolid'', has been the subject of intense theoretical and experimental efforts since the discovery of superfluidity in Helium-4. Recently, it has been proposed that nonlinear periodic modulations known as \textit{cnoidal waves}, that naturally emerge in Bose-Einstein condensates, provide a promising platform to find and study supersolidity in non-equilibrium phases of matter. Nevertheless, so far the analysis has been limited to a one-dimensional zero-temperature system. By combining the dissipative Gross-Pitaevskii equation with a finite temperature holographic model, we show that the proposed cnoidal wave supersolid phases of matter are dynamically unstable at finite temperature. We ascribe this instability to the dynamics of the “elastic” Goldstone mode, which arises as a direct consequence of translational order in the presence of dissipation, and establish a direct connection between the elastic-mode instability of the supersolid state and the nucleation of topological excitations during the relaxation towards a homogeneous equilibrium state, which resembles the Landau instability in superfluids. Finally, we numerically confirm the dominant role of the elastic-mode instability in the collision between cnoidal waves in the strong dissipation limit.\\

\noindent\textbf{gauge/gravity duality, holographic supersolid, dissipative Gross-Pitaevskii equation, topological excitation}

\end{abstract}

\maketitle

\section{Introduction}

Symmetries play an indispensable and fundamental role in the language of condensed matter physics. Within the Landau paradigm, phases of matter are indeed classified using symmetries and phase transitions between them are often accompanied by the phenomenon of spontaneous symmetry breaking (SSB) \cite{10.21468/SciPostPhysLectNotes.11}. For instance, the formation of a crystalline solid is related to the spontaneous breaking continuous spatial translational symmetry, the onset of superfluidity in low-temperature quantum systems is rooted in the SSB of a global $U(1)$ symmetry, and so on. Phases of matter are commonly ascribed to equilibrium ground states. Nevertheless, driving a system out of equilibrium allows for novel forms of symmetry breaking (hence, to novel non-equilibrium phases of matter). Time crystals, \textit{non-equilibrium} states that spontaneously break time translational symmetry, are the most representative example of this sort \cite{RevModPhys.95.031001}.

The existence of ``supersolids'', states of matter that simultaneously manifest superfluidity and long-range spatial crystalline order, has been subject of a long controversy \cite{RevModPhys.84.759} due to the seeming contradiction between the dissipationless flowing nature of a superfluid and the rigidity of a solid. However, in the recent years more and more experimental works \cite{2017Natur.543...91L,2017Natur.543...87L,2019PhRvL.122m0405T,2019Natur.574..386G,2019Natur.574..382T,2021Natur.596..357N} have claimed the discovery of supersolidity in Bose-Einstein condensate (BEC) systems. 

In Ref.~\cite{2017Natur.543...91L}, Bragg reflection has been used to detect a stripe phase with supersolid properties in a BEC system with spin-orbit coupling. On the other hand, in Ref.~\cite{2017Natur.543...87L} a supersolid was realized by introducing cavity-mediated long-range interactions into a BEC system. More generally, in dipolar quantum gas systems, supersolid phases have been reported in several instances \cite{2019PhRvL.122m0405T,2019Natur.574..386G,2019Natur.574..382T,2021Natur.596..357N} and their nature confirmed by the number and dispersion of low energy excitations following from Goldstone theorem \cite{2019Natur.574..386G,2019Natur.574..382T}. Two dimensional supersolids have also been experimentally realized \cite{2021Natur.596..357N}, enlarging the spectrum of systems exhibiting this exotic phase of matter. Importantly, whether supersolids are really stable phases of matter or only metastable (and possibly long-lived) ones remains an important question, in particular in non-equilibrium systems.

\begin{figure}[htb]
    \centering
\includegraphics[width=0.7\linewidth]{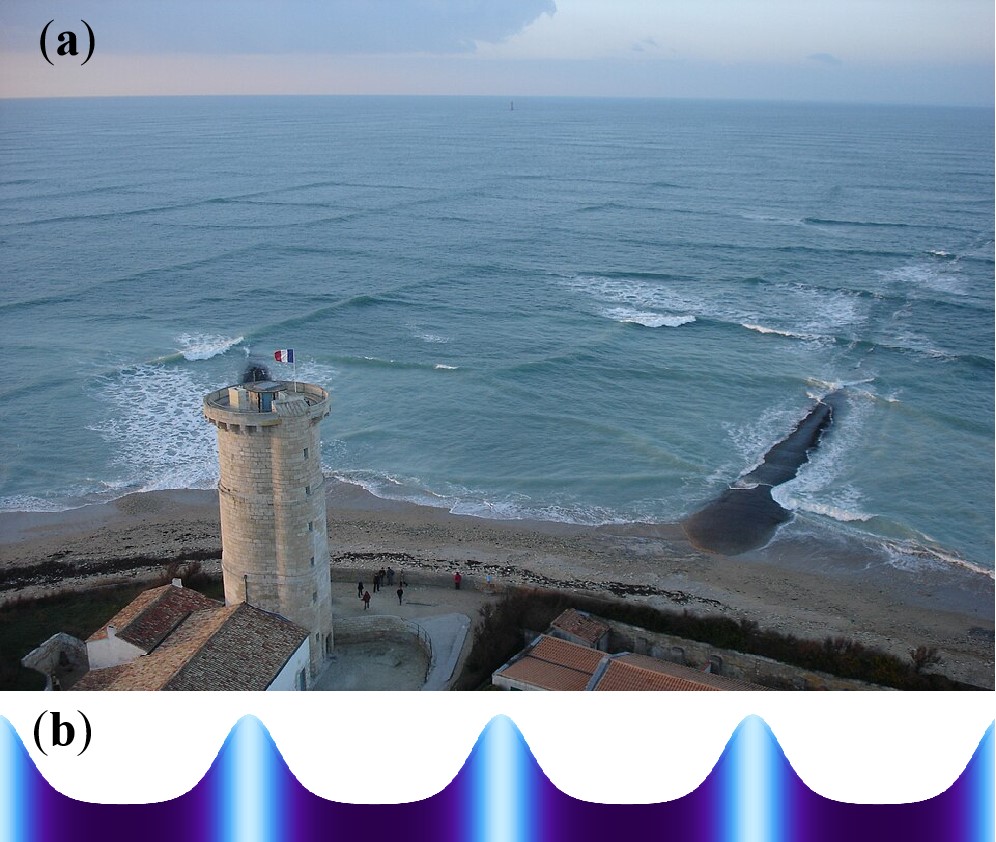}
    \caption{\textbf{Cnoidal waves.} \textbf{(a)} Near-cnoidal wave trains in the Atlantic ocean photographed from the Whale lighthouse (France), Images are taken from \href{https://en.wikipedia.org/wiki/Cnoidal_wave}{Wikipedia}. \textbf{(b)} A cnoidal wave, characterised by sharper crests and flatter troughs than in a sine wave.}
    \label{fig:0}
\end{figure}

Stable supersolids are predicted theoretically in zero-temperature BEC systems \cite{2002PhRvL..88m0402G,2005PhRvL..95l7205W,2010PhRvL.104s5302H,2010PhRvL.105p0403W,2016PhRvX...6b1026O} where their stability is protected by interactions. For instance, dipolar-dipolar interactions \cite{2002PhRvL..88m0402G}, nearest-neighbor repulsion of hard-core bosons on a lattice \cite{2005PhRvL..95l7205W}, Rydberg interactions in Rydberg states \cite{2010PhRvL.104s5302H}, spin-orbit interaction in spin-$1/2$ and spin-1 BEC \cite{2010PhRvL.105p0403W}, nonlinear atom-light interactions (superradiant Rayleigh scattering) \cite{2016PhRvX...6b1026O}, and so on.

Even in absence of these stabilizing interactions, supersolidity has also been proposed and investigated in excited states \cite{2021PhRvR...3a3143M} and even non-equilibrium states \cite{YangP2023}. Among the various proposals, it was recently argued \cite{2021PhRvR...3a3143M} that \textit{cnoidal waves} in a one-dimensional Bose-Einstein condensate are a promising example of supersolidity in non-equilibrium many-body systems. Cnoidal waves are periodic wave solutions of several nonlinear differential equations that can be written in terms of Jacobi elliptic functions and they appear in several physical systems, including the surface of our oceans (see Fig.~\ref{fig:0}). A stability analysis shows that cnoidal waves are energetically unstable \cite{{1974JAM...367,2011JPhA...44B5201B,2015JDE...258.3607G,2017AMR...431}} implying that they can represent only excited (metastable) states. The supersolidity of cnoidal waves has been verified \cite{2021PhRvR...3a3143M} by applying linear perturbations and looking at the nature of the low-energy excitations. In addition to the common superfluid sound mode corresponding to an infinitesimal $U(1)$ transformation of the phase of the condensate wave function, another gapless and linearly dispersing sound mode has been observed and explained from the SSB of translational symmetry. This ``elastic mode'' indeed corresponds to a rigid translation of
the wave function by an infinitesimal displacement.

The proposal of \cite{2021PhRvR...3a3143M} is certainly interesting and captivating. Nevertheless, the analysis therein has been conducted only on a one-dimensional and, most importantly, zero temperature system. Therefore, it remains unclear whether this supersolid state is stable upon finite temperature corrections and, more in general, which are the effects of dissipative mechanisms that are inevitably present at $T\neq 0$. Does the supersolid remain stable at finite temperature? Does it melt at a critical temperature and how? If instabilities occur, how do they develop and which is their ultimate origin? Answering these questions is the main objective of this work.

\section{Gross-Pitaevskii equation and cnoidal wave supersolidity}\label{GPsec}
We start our analysis by considering a one-dimensional zero-temperature BEC system, described by the Gross-Pitaevskii (GP) equation  
\begin{equation}\label{GP}
    i\partial_t\Psi(t,x)=-\frac{1}{2}\partial^2_x\Psi(t,x)-\mu\Psi(t,x)+g|\Psi(t,x)|^2\Psi(t,x),
\end{equation}
in which $\Psi(t,x)$ is the single-particle dimensionless wavefunction, $\mu$ chemical potential and $g$ the coupling constant controlling nonlinear interactions. Using these notations, the condensate density is given by $\rho=|\Psi|^2$. Dissipative coefficients are neglected, but can in principle be introduced using standard phenomenological procedures (see below). Since we are interested in inhomogeneous states with periodicity along the $x$ direction, we choose a Bloch-type ansatz $\Psi= \psi(t,x) e^{i k x}$ for the wavefunction where $k$ represent the Bloch wave-vector and $\psi$ is periodic in $x$ and can be time dependent. The static Bloch solutions are obtained numerically from the GP equation, Eq.~\eqref{GP}, by considering a time independent wave function $\Psi(x)$. For these static 1D Bloch solutions, the continuity equation  following from the global U(1) symmetry implies
\begin{equation}\label{eqeq1}
\cancel{\partial_t\rho}+\partial_x J_x=0,
\end{equation}
where the steady constant current is given by $J_x=(\partial_x \theta +k) \rho$, with $\theta$ the phase of the complex scalar field, $\psi(x)=\sqrt{\rho(x)}e^{i \theta(x)}$. The spatial dependence of the phase $\theta(x)$ reveals the non-zero superflow nature of the inhomogeneous Bloch states. We notice that, despite the density $\rho$ and the phase $\theta$ are in general inhomogeneous functions of the spatial coordinate $x$, in the limit of static density, the current $J$ is forced to be homogeneous by the continuity equation.

The GP equation, Eq. \eqref{GP}, admits an interesting class of nonlinear wave solutions known as \textit{cnoidal waves}, that can be written in terms of Jacobi elliptic function \textit{cn} (from which their name comes from). These solutions have been extensively studied in the literature, see \textit{e.g.} \cite{Tsuzuki1971}. As discussed in detail in Ref.~\cite{2021PhRvR...3a3143M}, the solutions of this equation can be classified according to the spatial properties of the profile of $\rho$ associated to a non-trivial and space-dependent phase $\theta$. The simplest Bloch solutions of Eq.~\eqref{GP} are uniform solutions, where the charge density is constant in space, 
\begin{equation}
    \textit{uniform solution: } \quad\rho(x)=\text{const},\quad \theta(x)=-a x.
\end{equation} 
For uniform solutions, the superfluid velocity $v_x=a$ is constant in space as well.

\begin{figure}[htb]
\centering
\includegraphics[width=0.85\linewidth]{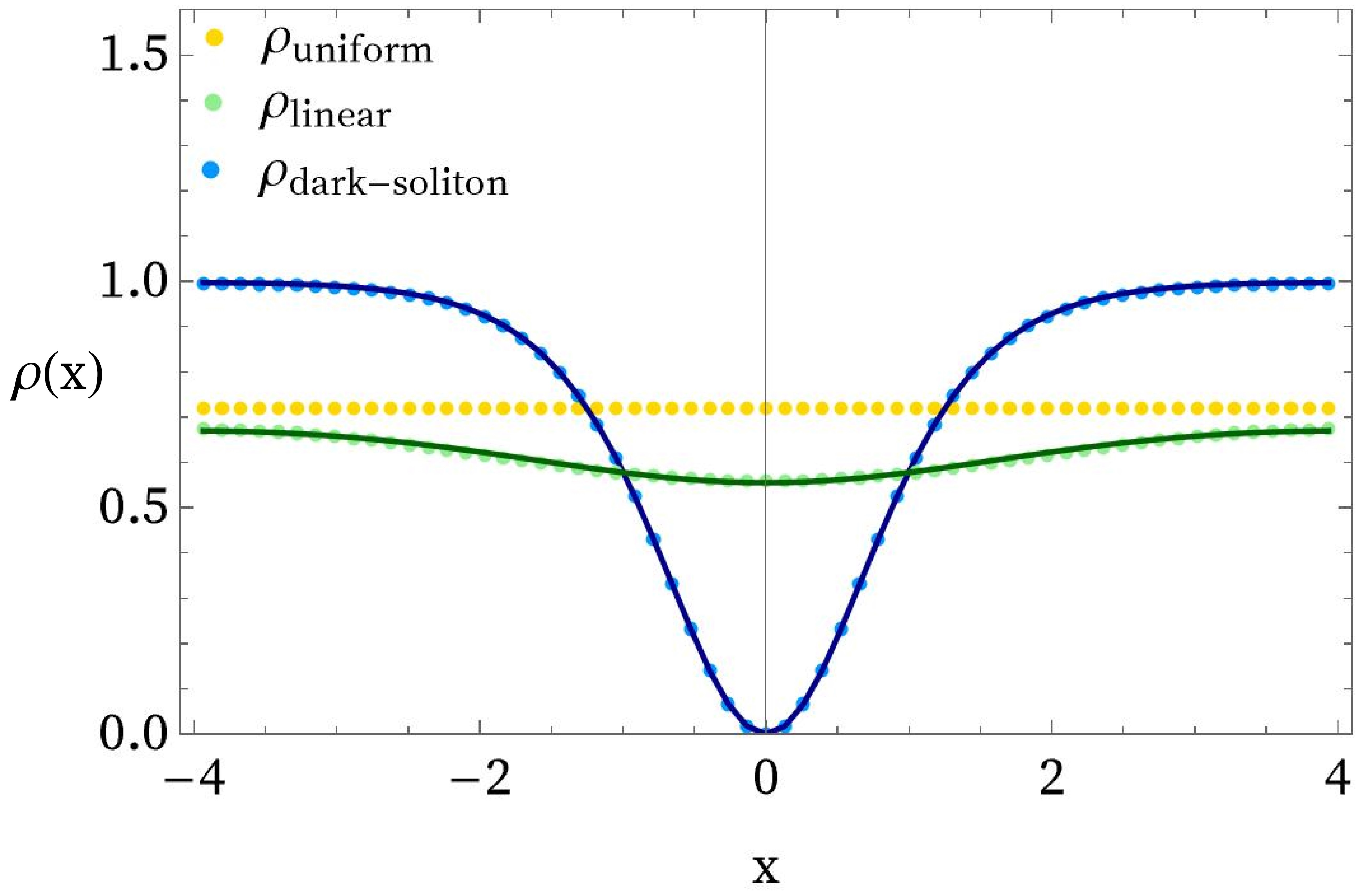}
\caption{\textbf{Inhomogeneous superflow states (cnoidal waves) of the GP equation.} Profile of condensate density $\rho(x)$ with Bloch-wave vector \textcolor{black}{$k=0.5k_0$} for dark-soliton, \textcolor{black}{$k=1.11k_0$} for linear wave and \textcolor{black}{$k=1.25k_0$} for uniform solutions. The green line is a fitting to linear wave solution Eq.~\eqref{linfit} with form $\rho=0.616-0.0573\cos(0.785x)$ while the blue line is the dark-soliton fitting function of Eq.~\eqref{darkfit} with form $\rho=0.997\tanh(1.0021x)^2$.}
\label{GP_cnoidal}
\end {figure}

Moreover, two simplified limiting cases of cnoidal wave solutions exist: linear-wave solutions and dark-soliton ones. For linear-wave solutions, the profile is periodic in space with wavelength $\lambda=2\pi/k_0$, \textit{i.e.}
\begin{equation}
   \textit{linear wave: } \quad\rho(x)-\bar{\rho}\approx \text{cos}(k_0 x),\label{linfit}
\end{equation}
where $k_0$ is a cumbersome function of the GP equation parameters. 
On the other hand, dark-soliton solutions are characterized by the following profile
\begin{equation}
    \textit{dark solition: } \quad\rho(x)\approx \text{tanh}(x/\xi_0)^2,\label{darkfit}
\end{equation}
where $\xi_0$ characterizes the correlation length of the wave function and depends also on the parameters in Eq.~\eqref{GP}. For both the linear wave and dark soliton solutions, the phase $\theta(x)$ is a non-trivial function of $x$, indicating a space-dependent superfluid velocity. In Fig.~\ref{GP_cnoidal}, we show an example of these solutions. For completeness, we also show the uniform solution.

Previous work \cite{2021PhRvR...3a3143M} showed that these cnoidal wave solutions of the 1D GP equation, Eq.~\eqref{GP}, display supersolid characteristics and they represent good candidates for the observation of supersolid phases of matter in nonequilibrium systems.

\begin{figure}[htb]
\centering
\includegraphics[width =0.98\linewidth]{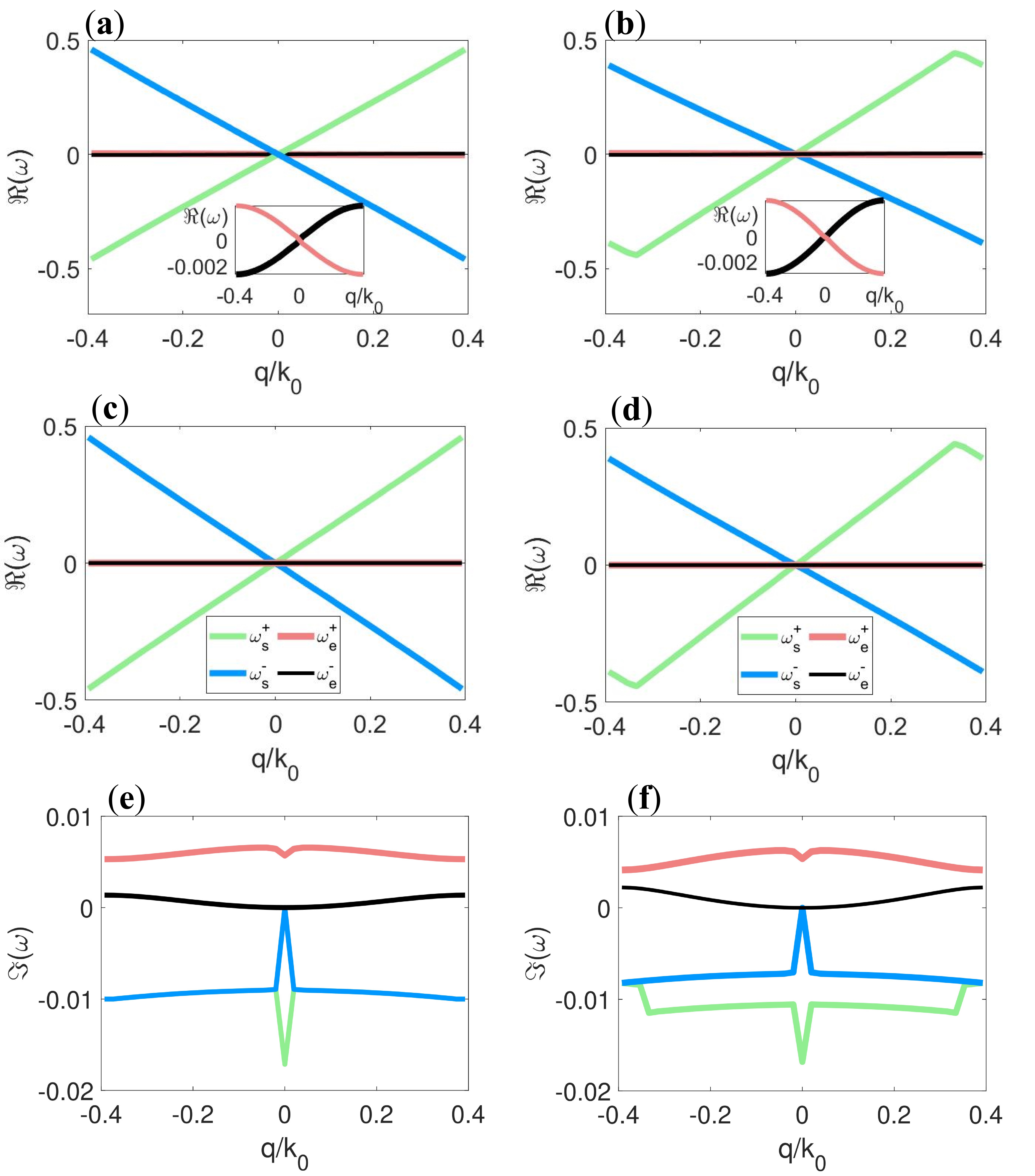}
\caption{\textbf{Linear stability of cnoidal wave supersolidity from GP equation.} Dispersion relation of sound modes $\omega_s^{\pm}$ and elastic modes $\omega_e^{\pm}$ on top of the cnoidal wave solutions of the 1D GP equation, Eq.\eqref{GP}. Panels \textbf{(a)} and \textbf{(b)} are for \textcolor{black}{$k=1/2k_0,2/3k_0$} and $\eta=0$. Insets display zoomed-in views of modes $\omega_e^\pm$. \textcolor{black}{Panels \textbf{(c)} and \textbf{(e)} are the real and imaginary parts of the dispersion relations for  $k=1/2k_0$ with $\eta=0.01$ and panels \textbf{(d)} and \textbf{(f)} are the real and imaginary parts of the dispersion relations for  $k=2/3k_0$ with $\eta=0.01$}.}
\label{GP_QNM}
\end {figure}

To investigate the stability of these inhomogeneous superflow states, we use the Bogoliubov method and perturb the background states as $\Psi(t,x)=\Psi(x)+\delta \Psi(t,x)$. Due to the symmetry of the background solutions, $\delta \Psi(t,x)$ can be expanded as
\begin{equation}
    \delta \Psi(t,x)=\delta u(x) e^{-i\omega t+i q x}+\delta v^*(x) e^{i\omega^* t-i q x}
\end{equation}
where $\delta u, \delta v$ and $\omega$ are complex in general and $^*$ indicates complex conjugation. By doing so, we obtain an eigenvalue problem in terms of an eigenstate vector $(\delta u, \delta v)^T$ obeying the following Bogoliubov equation,
\begin{equation}\label{Bogo}
\begin{pmatrix}
 \mathcal{L}(k+q) & \Psi^2 \\
 \Psi^{*2} & -\mathcal{L}(k-q)
\end{pmatrix}
\begin{pmatrix}
    \delta u \\
    \delta v
\end{pmatrix}
=\omega
\begin{pmatrix}
    \delta u \\
    \delta v
\end{pmatrix},
\end{equation}
where, $\mathcal{L}(p)=-\frac{1}{2}(\partial_x+i p)^2+2g\rho-\mu$ and $\rho$ is the condensate density defined above. By solving Eq.~\eqref{Bogo}, one obtains the dispersion of the low-energy excitations around the background inhomogeneous superflow states, which are shown in Fig.~\ref{GP_QNM}. These excitations involve two types of gapless modes. First, a pair of sound waves $\omega_s^\pm$ appears because of the spontaneous breaking of the U(1) symmetry. Additionally, and as a further proof of supersolidity, another pair of gapless modes emerges due to the spontaneous breaking of translational symmetry (\textit{akin} to acoustic phonons in crystalline solids). Following the analogy with crystals, we label these excitations \textit{elastic modes} and indicate their frequency as $\omega_e^\pm$.

In Fig.~\ref{GP_QNM}(a) and Fig.~\ref{GP_QNM}(b), we show the dispersion relation of the lowest energy excitations on top of the inhomogeneous superflow states at zero temperature for $k=1/2k_0$ and $k=2/3k_0$ respectively. We observe that near $q=0$, all these modes exhibit a linear dispersion relation of the type $\omega=\pm v q$. Nevertheless, the propagation speed of the sound modes is much larger than that of the elastic modes. The imaginary part of the dispersions are zero because of the absence of dissipation.

So far, all the computations in this Section have been performed by neglecting the effects of thermal dissipation and temperature fluctuations. In order to go beyond this approximation, we phenomenologically modify the GP equation \eqref{GP} by using the following substitution
\begin{equation}
    \partial_t \rightarrow \partial_t +i \eta,\label{inin}
\end{equation}
where $\eta$ represents now a damping term (friction) induced by thermal dissipation. We then repeat the same linear stability analysis in presence of $\eta \neq 0$.

When we consider dissipative effects by fixing $\eta=0.01$, we find that the inhomogeneous superflow states are energetic stable with $\Re(\omega_e^+)=\Re(\omega_e^-)=0$ (Fig.~\ref{GP_QNM}(c) for \textcolor{black}{$k=1/2k_0$} and Fig.~\ref{GP_QNM}(d) for \textcolor{black}{$k=2/3k_0$}) but become dynamical unstable due to the positive value of $\Im(\omega_e^{\pm})$ (Fig.~\ref{GP_QNM}(e) for \textcolor{black}{$k=1/2k_0$} and Fig.~\ref{GP_QNM}(f) for \textcolor{black}{$k=0.524$}). In other words, the Goldstone modes arising from the spontaneous symmetry breaking of translational symmetry become unstable and drive the system towards a dynamical instability.

This result suggests that the cnoidal wave supersolid phases proposed in \cite{2021PhRvR...3a3143M} are dynamically unstable as soon as temperature or dissipation are introduced. We remark that the way of introducing temperature used here, Eq.~\eqref{inin}, is rather phenomenological. Therefore, in the next Section we will introduce a more robust formalism to confirm our findings and study the stability of cnoidal wave supersolidity at finite temperature. 

\section{Dissipative supersolid states in a finite-temperature holographic model}
\subsection{Setup}\label{Holographic Model}

In order to investigate inhomogeneous superflow states at finite temperature, we resort to the holographic correspondence, or gauge-gravity duality \cite{ammon2015gauge,baggioli2019applied} -- a framework based on a conjectured duality between gravitational systems and strongly-coupled dissipative field theories. In our case, the advantage of this method is the possibility of introducing without any ad-hoc phenomenological ``trick'' (as done for the GP equation in the previous Section) the effects of dissipation and finite temperature.

The holographic superfluid model was originally introduced in \cite{2008PhRvL.101c1601H} based on the following action,
\begin{equation}\label{L}
S(\Psi,A_{\mu})=\frac{1}{16 \pi G_{4}} \int d^{4} x \sqrt{-g}\left(R+\frac{6}{L^{2}}+\frac{1}{e^{2}} \mathcal{L}_{M}\right),
\end{equation}
in which, a complex scalar field $\Psi$ couples to a $U(1)$ gauge field $A_{\mu}$ in a $(3+1)$-dimensional curved spacetime with negative cosmological constant $\Lambda=-3/{L^2}$. The parameter $G_{4}$ is the gravitational constant in four dimensions and $e$ is the bulk electromagnetic coupling. Moreover, $R$ is the Ricci scalar $R$ and the matter part of the Lagrangian is defined as
\begin{equation}
\mathcal{L}_{M}(\Psi,A_{\mu})=-\frac{1}{4} F_{\mu \nu}F^{\mu \nu}-\left|D_{\mu} \Psi\right|^{2}+2|\Psi|^{2},
\end{equation}
where $D_{\mu}=\partial_{\mu} \Psi-i A_{\mu} \Psi$ and $F=dA$ is the electromagnetic field strength. 

In the limit in which the superfluid couples to a parametrically large thermal bath with constant temperature, the probe limit approximation -- equivalent to ignoring the backreaction of the superfluid on the thermal bath -- is valid (see \cite{Xu:2019msl,Jiang:2023yyn} for examples on how to consider back-reaction in holography). In the gravity side, this approximation corresponds to considering a fix background spacetime metric that is solution of the Einstein equations in the vacuum, originating from the action in Eq.~\eqref{L} after setting $\mathcal{L}_M \rightarrow 0$. The simplest black-brane solution is given by the line element,
\begin{equation}\label{li}
ds^{2}=\frac{L^{2}}{z^{2}}\left(-f(z)dt^{2}+\frac{1}{f(z)}dz^{2}+dx^{2}+dy^{2}\right),
\end{equation}
with $f(z)=1-{z^{3}}/{z_{H}^{3}}$. Here, $z_H$ is the radius of the black-brane horizon at which $f(z_H)=0$ and the radial boundary along $z$ locates at $z=0$, the so-called AdS boundary. The associated Hawking temperature is $T={3}/{4\pi z_{H}}$. Without loss of generality, in the rest of the study, we set $L=z_{H}=1$.

The matter Langrangian $\mathcal{L}_M$ is then considered on top of the fixed background spacetime defined by Eq.~\eqref{li}. The equations of motion for the bulk fields $A_{\mu}$ and $\Psi$ are
\begin{equation}\label{psi}
\frac{1}{\sqrt{-g}} D_{\mu}\left(\sqrt{-g} D^{\mu} \Psi\right)+2 \Psi=0,
\end{equation}
\begin{equation}\label{A_mu}
\frac{1}{\sqrt{-g}} \partial_{\mu}\left(\sqrt{-g} F^{\mu \nu}\right)=i\left(\Psi^{*} D^{\nu} \Psi-\Psi\left(D^{\nu} \Psi\right)^{*}\right).
\end{equation}
A first solution to these equations is given by,
\begin{equation}
    \Psi=0,\qquad A_t(z)\neq 0\,.
\end{equation}
The bulk profile of the complex scalar $\Psi$ is trivial, indicating that the U(1) global symmetry is not broken -- this is the normal phase. As we decrease the temperature, there will be a critical $T_{c}$ below which a solution with nonzero $\Psi$ appears, without introducing any source for the scalar operator ``dual'' to $\Psi$. This solution displays the spontaneous symmetry breaking of the $U(1)$ symmetry and the formation of a superfluid condensate. Hereafter, we will focus on the case with $T<T_{c}$ -- the superfluid phase. 

In general, the asymptotic behavior of the complex bulk field close to the AdS boundary, $z \rightarrow \infty$, is given by 
\begin{equation}\label{didi}
    \Psi(z,x)=\Psi_0(x) z +\Psi_1(x) z^2 +\dots
\end{equation}
where the $\dots$ indicate terms that are subleading in this expansion. Following the holographic dictionary, we will always set $\Psi_0=0$ so that the global U(1) symmetry of the boundary field theory is never broken explicitly (see \cite{Ammon:2021pyz} for the possibility of adding a small source). Under these assumptions, we define the spatially dependent order parameter -- superfluid condensate -- $\langle \mathcal{O}(x) \rangle$ as
\begin{equation}
    \langle \mathcal{O}(x) \rangle \equiv \Psi_1(x),
\end{equation}
which is finite in the superfluid phase. Finally, in the bulk gauge field follows an analogous asymptotic behavior,
\begin{equation}
    A_t(x,z)= \mu (x)-\rho(x) z+\dots\,,
\end{equation}
where $\mu(x)$ is the chemical potential and $\rho(x)$ the associated charge density.

We are interested in supersolid phases that spontaneously break at the same time the global U(1) symmetry at the boundary together with spatial translations (see \cite{RevModPhys.95.011001,2021SCPMA..6470001B} for a review on how to break translations spontaneously in holography). This will force us to consider states with explicit spatial dependence in the superfluid condensate. \textcolor{black}{Previously, different realizations of supersolids in holography can be found in \cite{Baggioli:2022aft,YangP2023}, where the superfluidity emerging on top of a solid is investigated in \cite{Baggioli:2022aft}, and a time-dependent supersolid driven by an external optical lattice is studied in \cite{YangP2023}. For a non-equilibrium steady cnoidal wave supersolid, it hasn't been explored yet holographically. In \cite{Liu_2020,Chen:2024pyy,Yang:2024vga}, a similar momentum-driven, spatially non-uniform non-equilibrium steady state are reported.}

Finally, we emphasize that in the rest of this work we will work at fixed particle number $N$ (see Supplementary Materials for details). We also notice that, despite the condensate density $\langle O(x) \rangle$ will be space dependent, the chemical potential $\mu$ and the current $j_x$ will not be. Also, we emphasize that only phases with space dependent condensate (inhomogeneous states) will correspond to supersolid phases of matter that break also spatial translations spontaneously. The homogeneous solutions, on the contrary, will correspond to superfluid states.

\subsection{Holographic supersolid states with superflow}\label{holo_inhomo}
Within the holographic superfluid model, we can construct inhomogeneous states with finite superflow by using a similar Bloch-type ansatz as done in the previous Section. Due to the $U(1)$ gauge symmetry in the Lagrangian Eq.~\eqref{L}, this is equivalent to switch on a nonzero spatial component for the bulk gauge field $A_x=-k$, where the charge current has been taken along the $x$ direction. In such a condition, the asymptotic behavior of $A_x$ reads $A_x(x,z)= -k+j_x z+\dots\,$ with $j_x$ stands for the charge current in the dual field theory. In general, $j_x$ can be spatially inhomogeneous but for static configurations with constant density that is not the case.

\begin{figure}[htb]
\centering
\includegraphics[width =0.8 \linewidth]{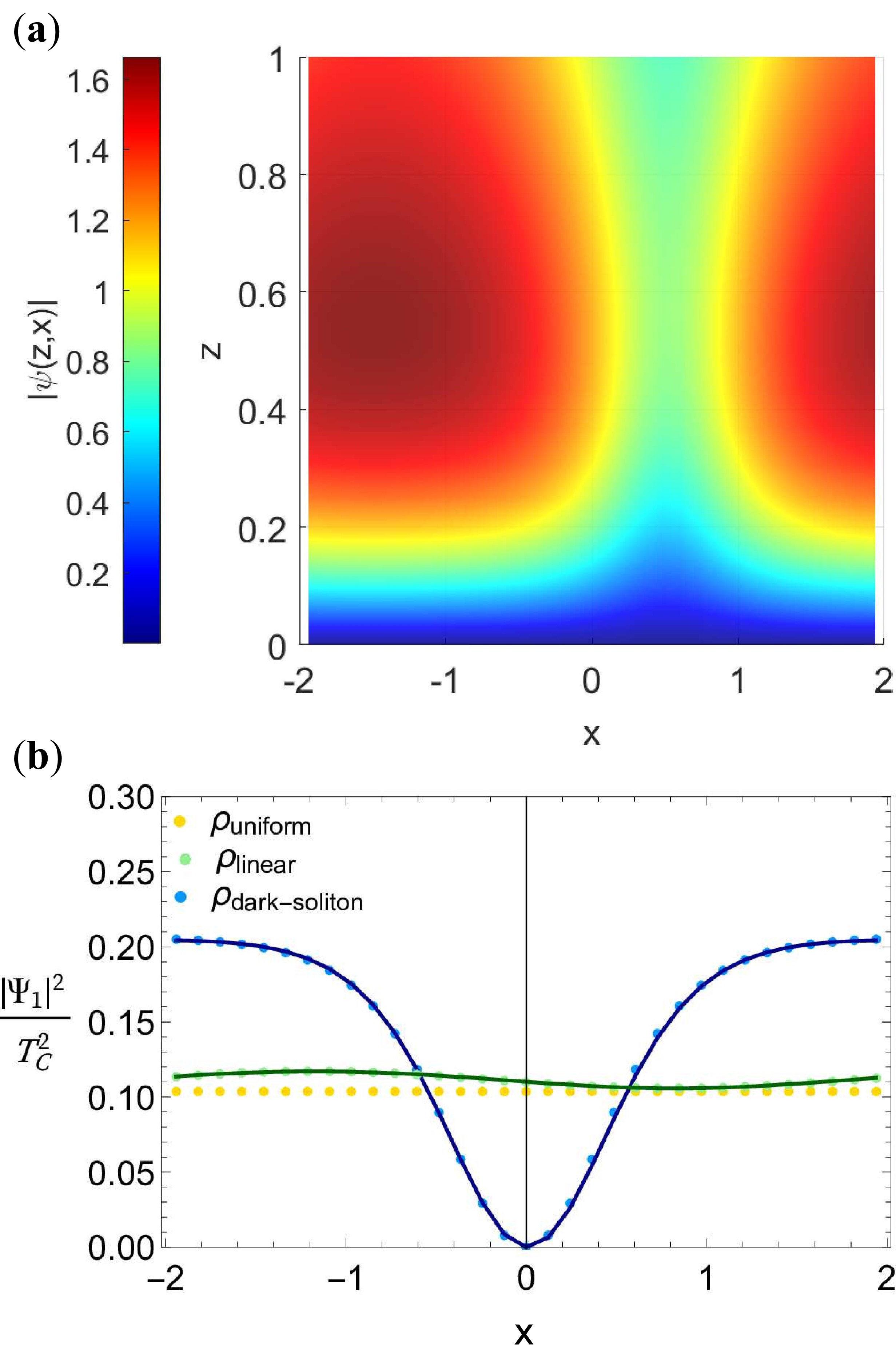}
\caption{\textbf{Cnoidal wave solutions in a finite temperature holographic model.} \textbf{(a)} Amplitude of complex bulk scalar field $\psi(z,x)$ with \textcolor{black}{$k=1.5k_0$}. \textbf{(b)} Profile of the condensate density for a dark-soliton solution (\textcolor{black}{$k=0.5k_0$}), linear-wave solution (\textcolor{black}{$k=1.8k_0$}) and a uniform solution (\textcolor{black}{$k=2k_0$}), respectively. The blue line is fitting function of dark-soliton solution with $\frac{|\Psi_1|^2}{T_c^2}=0.206\tanh(1.633x)^2$ and green lines is fitting function of linear-wave function with $\frac{|\Psi_1|^2}{T_c^2}=0.111-0.056\cos(1.5708x-0.821)$.}
\label{states}
\end {figure}

The characteristics of the inhomogeneous solutions are shown in Fig.~\ref{states}. In panel (a), we show the profile of the bulk complex scalar field $|\psi(z,x)|$ for a typical inhomogeneous solution. On the other hand, in Fig.~\ref{states}(b), we display the profile of the spatial dependent normalized condensate density for three benchmark inhomogeneous solutions, that are obtained for different values of $k$. In addition to the inhomogeneous states, there exists also another branch of solutions that corresponds to homogeneous ground states (yellow symbols in Fig.~\ref{states}(b)).

\begin{figure}[htb]
\centering
\includegraphics[width=0.95 \linewidth]{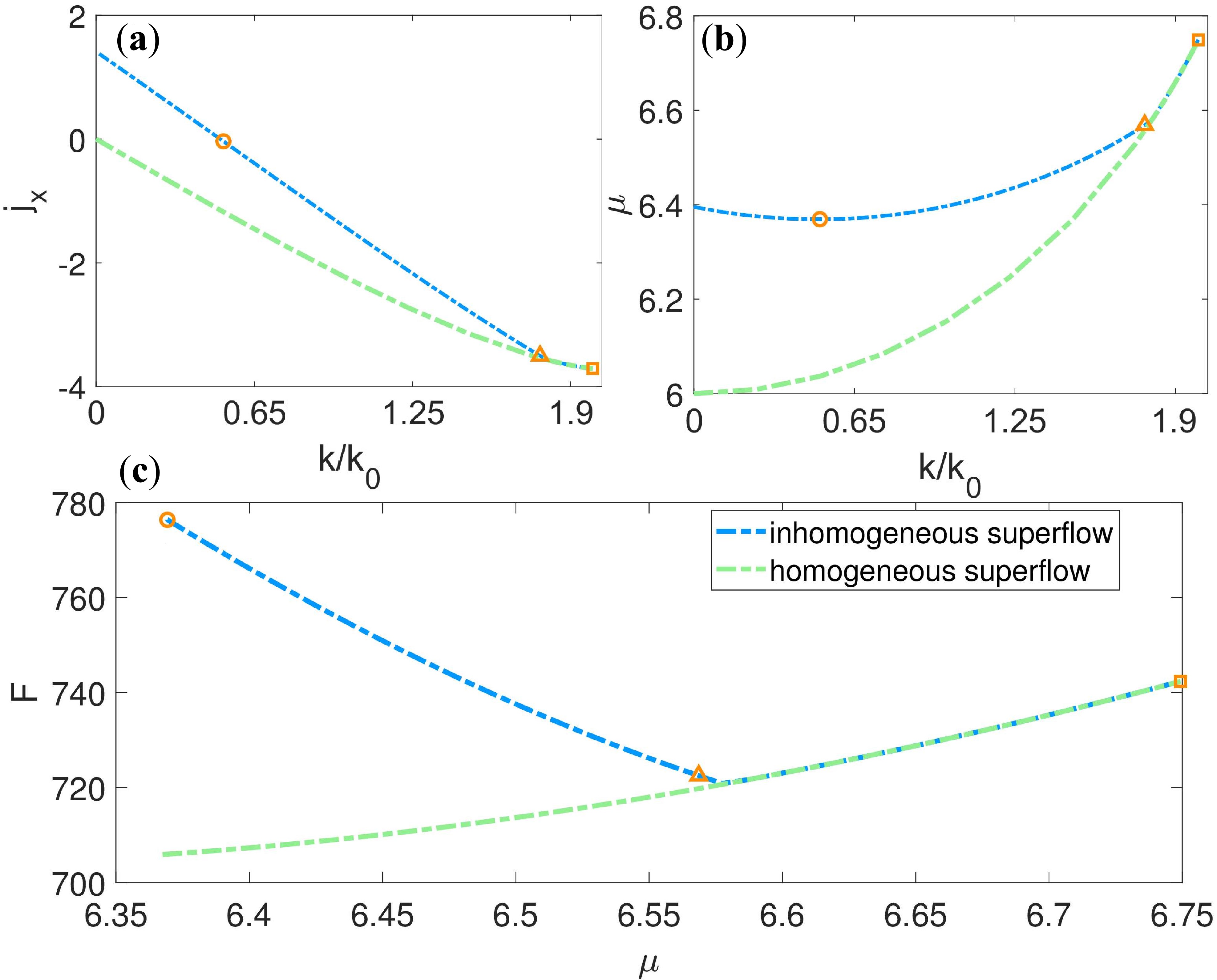}
\caption{\textbf{Supersolid excited states in the holographic model.} Total current density $j_x$ \textbf{(a)} and chemical potential $\mu$ \textbf{(b)} as a function of the superflow parameter $k$. \textbf{(c)} Free energy as a function of $\mu$ for inhomogeneous and homogeneous solutions with superflow. The orange symbols indicate the location of the solutions explicitly shown in Fig.~\ref{states}, where circle is the dark-soliton solution, triangle is the linear-wave solution and square is a uniform solution.}
\label{free}
\end {figure}

In Fig.~\ref{free}, we compare the properties of these different solutions. Fig.~\ref{free}(a) and Fig.~\ref{free}(b) show respectively the total charge current $j_x$ and the chemical potential $\mu$ as a function of the superflow parameter $k$. We notice that the chemical potential of the inhomogeneous solution is always larger than that of the homogeneous ones. The two approach each other only for large values of $k \approx 3$, above which no inhomogeneous solution can be found anymore. This corresponds to a critical value $\mu_c \approx 6.58$ above which no inhomogeneous solution exists anymore. At the same time, the current for the homogeneous state is always larger in absolute value than that of the inhomogeneous solution. Moreover, while the current in the homogeneous state is always negative, that in the inhomogeneous state changes sign at approximately \textcolor{black}{$k=0.5k_0$}.

Following Refs.\cite{Tian__2023,Li_2020}, we then define a generalized free energy functional,
\begin{align}\label{F}
{F}&=\int d\mathbf{x} \left(\frac{A_t^2}{f}|\psi|^2-A_x^2|\psi|^2+A_x \Im(\psi^*\partial_x\psi)\right)\nonumber\\ 
&+\int d\mathbf{r} \frac{1}{2}(\rho\mu-k j_x). 
\end{align}
The possibility of defining a generalized free energy for non-equilibrium states in holographic probe models has been suggested in \cite{Tian__2023}. It has been shown that such a quantity monotonically decreases during a generic dynamical process. Here, we follow their proposal. 

 We show the behavior of $F$ as a function of $\mu$ for both the inhomogeneous and homogeneous solutions in Fig.\ref{free}(c). We observe that below $\mu_c$, where both solutions exist, the homogeneous solution has always lower free energy. This indicates that the inhomogeneous states (corresponding to the supersolid phases) are not thermodynamically favored and they can exist only as excited states, as expected.

\subsection{Stability of nonequilibrium holographic supersolid states}\label{Quasi-normal mode}
The linear stability of the superflow states can be investigated by studying the dispersion relation of the low-energy excitations that, because of their (in general) complex frequencies, are referred to as \textit{quasinormal modes} (QNMs).  Following Refs.~\cite{{2016JHEP...01..016D,2020PhRvL.124c1601G,2020JHEP...02..104L,2021JHEP...11..190Y,2020arXiv201006232L,2024arXiv241013584A}}, we define the fluctuations of the bulk fields as
\begin{equation}
A_\mu=A_{\mu,0}+\delta A_\mu,\qquad 
\psi=\psi_0+\delta \psi.
\end{equation}
The symmetry of the background states suggests a further decomposition into
\begin{equation}
\begin{matrix}
\delta A_\mu=\delta a_\mu e^{-i\omega t +i q x}+\delta a_\mu^* e^{i\omega^* t -i q x},\\
~\\
\delta \psi=\delta u e^{-i\omega t +i q x}+\delta v^* e^{i\omega^* t -i q x}.
\end{matrix}
\end{equation}
Here, $A_{\mu,0}$ and $\psi_0$ are the background solutions discussed in the previous Section, while $\delta\Psi=(\delta a_\mu,\delta u, \delta v)^T$ are perturbation fields on top of these solutions. We work within the linear approximation and neglect nonlinear terms in $\delta \Psi$. In this limit, the equations for the perturbations can be written as an eigenvalue problem,
\begin{equation}\label{L_qnm}
    \mathcal{L}(A_{\mu,0},\psi_0,\omega,k)\delta\Psi=0,
\end{equation}
where the explicit form of the matrix $\mathcal{L}$ is shown in Supplementary Materials.

\begin{figure}[htb]
\centering
\includegraphics[width =0.95 \linewidth]{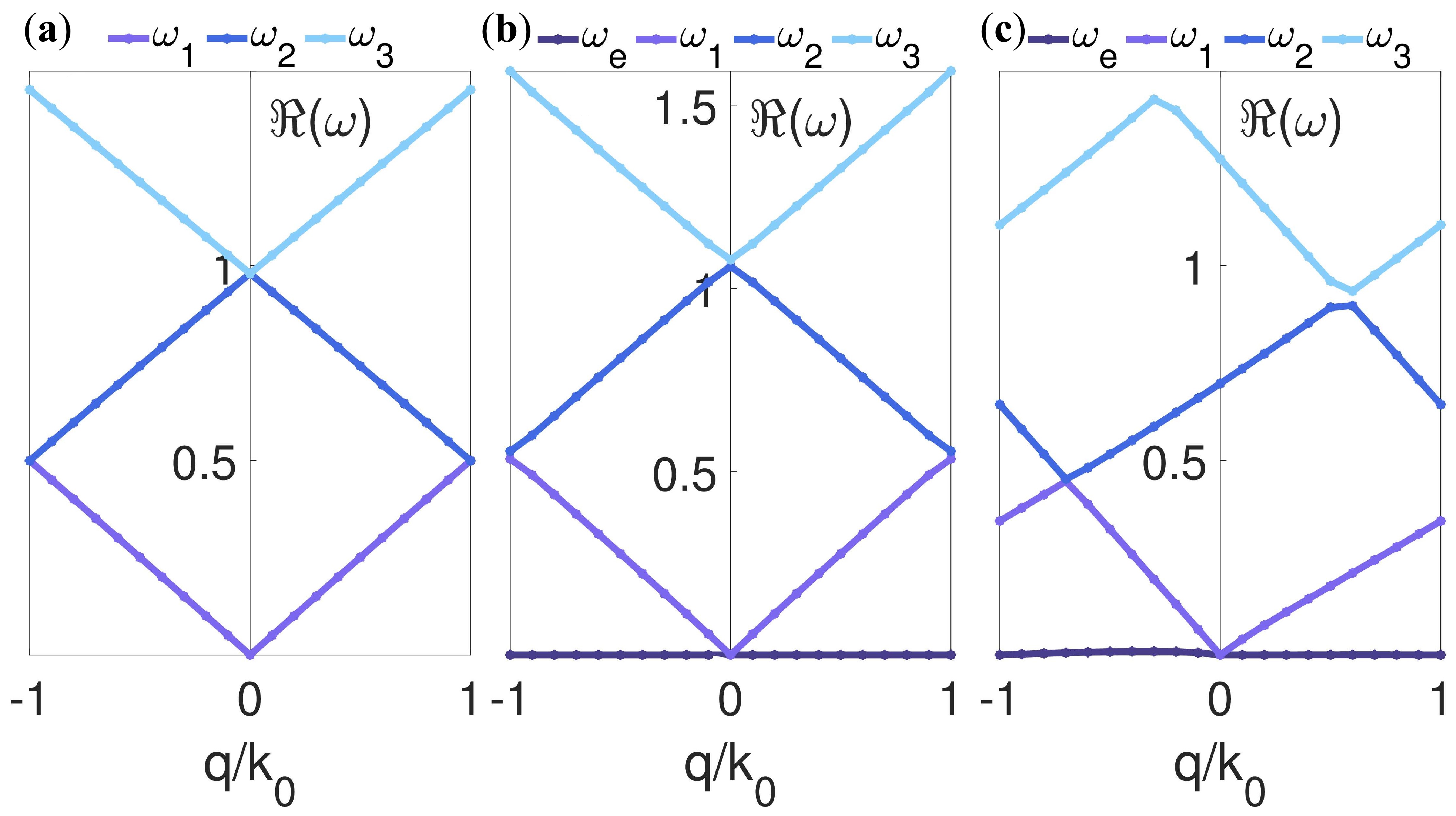}
\caption{\textbf{Low-energy excitations around nonequilibrium superfluid and supersolid states.} \textbf{(a)} Real part of the dispersion relations $\omega_i(q)$ of low-energy excitations around a superfluid state with homogeneous condensate. Panels \textbf{(b)-(c)} illustrate respectively the low-energy modes around nonequilibrium supersolid states with  zero-current (\textcolor{black}{$k=1.5k_0$}) and finite-current (\textcolor{black}{$k=0.5k_0$}). In all plots, $k_0=\pi/L_x$ with $L_x$ the length of $x$ direction (see Supplementary Materials for more details).}
\label{omega}
\end {figure}

We then calculate the dispersion relation of collective modes of both inhomogeneous and homogeneous solutions with superflow. We remind the Reader that the background states, defined by the functions $A_\mu$ and $\psi$, are periodic in the $x$ direction with characteristic wavevector $k_0$. From Eq.\eqref{L_qnm}, $\mathcal{L}$ also inherits the same periodicity properties. Because of this symmetry, any perturbation is still of the Bloch-type and the corresponding Bloch vector $k$ can be chosen within the first Brillouin zone, \textit{i.e.}, $q\in[-k_0,k_0]$. Exploiting this fact, we display the dispersion relations of the low-energy modes using a folded representation in the reduced Brillouin zone. In this way, the low-energy excitations can be classified into different energy bands denoted as $\omega_i$, $i=1,2,3,\dots$, where $\omega_1$ is the lowest energy band and $\omega_{i>1}$ the higher energy bands. For simplicity, we limit ourselves to show the first three energy bands.

In Fig.~\ref{omega}(a), we show the low-energy excitations around the homogeneous solution, corresponding to a superfluid phase. We observe a single gapless excitation with linear dispersion, $\mathrm{Re}(\omega_1)=\pm v q+\dots$. This mode is the superfluid sound arising because of the spontaneous breaking of U(1) symmetry. Its dispersion can be rationalized using superfluid hydrodynamics, that nicely match the holographic data as shown in \cite{Amado:2009ts,Arean:2021tks}. 

In Fig.~\ref{omega}(b), we show the low-energy excitations around the supersolid state with zero-current, corresponding to the exceptional point \textcolor{black}{$k=0.5k_0$}, see panel (a) in Fig.\ref{free}. In addition to a gapless sound mode, similar to the one emerging on top of the superfluid phase shown in panel (a), the spectrum displays an additional gapless mode, indicated as $\omega_e$. This mode is the additional Goldstone mode appearing because of the spontaneous breaking of translational symmetry, and confirming the supersolid nature of these solutions. The characteristics of the spectrum are analogous to those observed in the 1D BEC system in Ref. \cite{2021PhRvR...3a3143M}, where they were also used to confirm the supersolid nature of the cnoidal wave solutions. Finally, we notice that for this specific solution, we have $\mathrm{Re}(\omega_e)=0$. 

In Fig.~\ref{omega}(c), we explore the dispersion of the low-energy excitations around a nonequilibrium supersolid state with finite superflow. Several phenomena have to be noticed. First, the dispersion relation of the superfluid sound mode, $\omega_1(q)$, is not anymore symmetric upon $q \rightarrow - q$ (and the other modes as well). This is the effect of the breaking of parity symmetry ($x \rightarrow -x$) induced by the finite background current $j_x$. The same is true for the elastic mode $\omega_e(q)$, for which $\mathrm{Re}$($\omega)<0$ for $q>0$, and $\mathrm{Re}$($\omega)>0$ for $q<0$. Finally, a level repulsion feature between higher order modes ($\omega_2$ and $\omega_3$) is visible at around half of the Brillouin zone, $q\approx 0.5 k_0$.

\begin{figure}[htb]
\centering
\includegraphics[width =0.9 \linewidth]{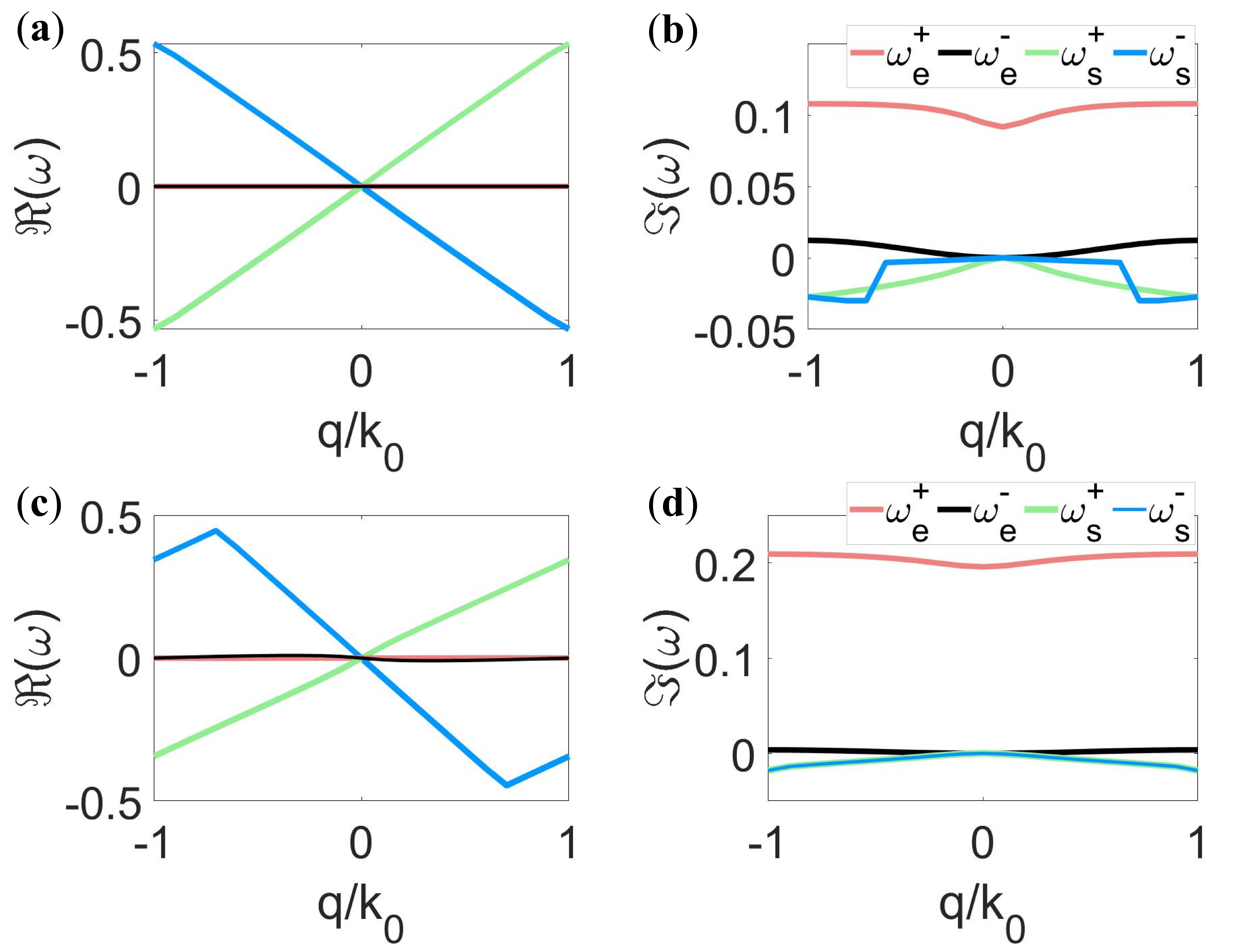}
\caption{\textbf{Stability analysis of nonequilibrium supersolid states.} Real part $\mathrm{Re}(\omega)$ and imaginary part $\mathrm{Im}(\omega)$ of the dispersion relation of the lowest energy excitations in nonequilibrium supersolid states. Panels $\textbf{(a)-(b)}$ refers to the supersolid solution with zero current, while panels $\textbf{(c)-(d)}$ to the one with finite current, $j_x\neq 0$.}
\label{qnm_holo}
\end {figure}

In general, the spectrum in Fig.\ref{omega} confirms the superfluid and supersolid nature of the solutions with respectively homogeneous and inhomogeneous condensates. Nevertheless, in order to study their stability in more detail, the analysis of the QNMs should be expanded to the imaginary part of the dispersion relations. In order to do so, in Fig.~\ref{qnm_holo}, we provide a zoomed view of the real and imaginary parts of the dispersion relation of the lowest energy excitations around the nonequilibrium supersolid states.

From Fig.~\ref{qnm_holo}, the dynamical instability of the supersolid states (with and without superflow) is clear since $\mathrm{Im}(\omega_e)>0$ leading to a perturbation that diverges exponentially in time. This is in agreement with the analysis using the dissipative GP equation performed in Section \ref{GPsec} (see Fig.~\ref{GP_QNM}).

\subsection{Elastic-mode instability of the supersolid states with topological excitations}\label{evoluion}

In the previous Section, we have proved the elastic-mode instability of the cnoidal wave supersolid states at finite temperature. This was done using a linear perturbation analysis around those solutions. In order to capture the full time evolution related to these unstable states, and how this instability develops in time, one need to make a step further and consider the nonlinear dynamical evolution of the system.

We initiate our simulation starting from inhomogeneous states perturbed by unstable elastic modes with different wave vectors $q$. Following Refs. \cite{2020arXiv201006232L,2021JHEP...11..190Y,2023JHEP...05..223L,2023PhRvL.131v1602L,2023PhRvD.107l1901Y,2021JHEP...03..136Z,2019PhRvD.100f1901X}, the evolution scheme is performed using fourth order Runge-Kutta method to solve the following dynamical equations, 
\begin{align}\label{Infalling Jx}
&2\partial_{t} \partial_{z}A_{x}+i(\psi^{*} \partial_{x} \psi-\psi \partial_{x} \psi^{*}-2 i A_{x} \psi^{*} \psi)\nonumber\\
&-\partial_{z} \partial_{x}A_{t}-\partial_{z}(f \partial_{z}A_{x})=0,\\\nonumber~\\
&\partial_{t} \partial_{z}A_{t}+\partial_{t} \partial_{x} A_{x}-f \partial_{x} \partial_{z}A_{x}+i(\psi^{*} \partial_{t} \psi-\psi \partial_{t} \psi^{*})\nonumber\\
&=-\partial_{x}^{2} A_{t}+if(\psi^{*} \partial_{z} \psi+\psi \partial_{z} \psi^{*})-2 \psi^{*}A_{t} \psi,\\\nonumber~\\
&(-\partial_{z}^{2}A_{t}+\partial_{x} \partial_{z} A_{x})+i(\psi^{*} \partial_{z} \psi-\psi \partial_{z} \psi^{*})=0,\\\nonumber~\\
&2\partial_{t}\partial_{z}\psi-f \partial_{z}^{2}\psi-(f^{\prime}+2iA_{t})\partial_{z}\psi+2 i A_{x} \partial_{x}\psi\nonumber\\
&+\partial_{x}^{2}\psi+(i\partial_{x}A_{x}-i\partial_{z}A_{t}+z+A_{x}^{2}) \psi=0,\label{Infalling psi}
\end{align}
where all bulk fields are now functions of $(t,x,z)$.

\begin{figure}[hb]
\centering
\includegraphics[width =0.85\linewidth]{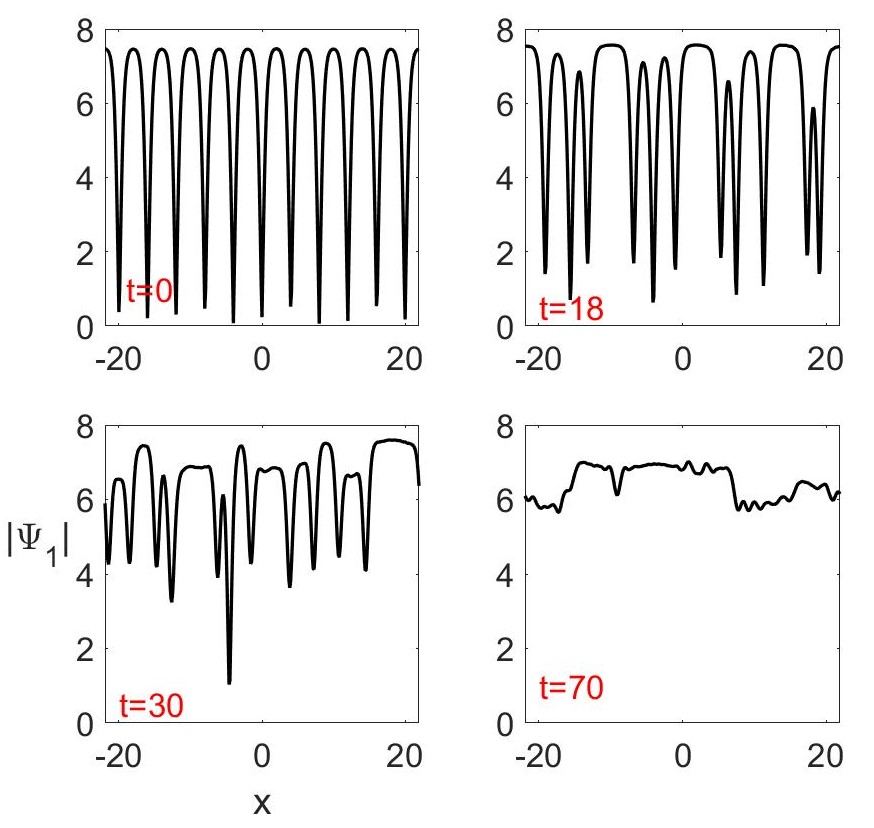}
\caption{\textbf{Elastic-mode instability of the supersolid initial state towards a final homogeneous superfluid state.} The amplitude of the wave-function $\Psi_1(t,x)$ at different instants of time. The initial ($t=0$) solution is perturbed by an elastic mode with $q=0.7k_0$. In the last panel, the system is approaching a final equilibrium state with homogeneous amplitude.}
\label{condensate_0}
\end {figure}

In Fig.~\ref{condensate_0}, we display a series of snapshots of the time evolution of the wave-function amplitude $|\Psi_1|(x,t)$ starting from an initial unstable solution perturbed via an elastic mode with $q=0.7k_0$. The initial solution at $t=0$ displays a strongly inhomogeneous profile that is close to a train of dark solitons, at the core of which the condensate vanishes. After perturbing this initial supersolid state, the wave-function evolves in a complex way and eventually at late time tends towards a uniform phase in which $|\Psi_1(x)|=\text{const}$. The bottom right panel in Fig.~\ref{condensate_0} shows a snapshot before approaching this final equilibrium state. This final equilibrium state is a superfluid with homogeneous condensate and it is stable. In other words, the instability induced by the elastic mode drives the supersolid excited states towards superfluid homogeneous states. 

In order to reveal in more detail the microscopic origin of this instability and its evolution, it is important to discuss in more detail the physical role of the phase of the complex order parameter. For the cnoidal wave inhomogeneous solutions, the phase $\theta(x)$ is a nontrivial function of the spatial coordinate $x$. In particular, it can exhibit local jumps and topological features, such as the presence of  {quantized circulation} in one dimension (see Supplementary Materials for detail), as well. 

In full generality, solitons and anti-solitons are respectively solutions of nonlinear differential equations accompanied by $\pi$ or $-\pi$ local phase jumps. As an explicit example, the GP equation, Eq.~\eqref{GP}, admits this type of solutions. In fact, one can find solitary travelling wave solutions that propagate at velocity $v$ without changing in form,
\begin{equation}
    \Psi(z)= \gamma \,\tanh\left(\frac{\gamma z}{\sqrt{2}}\right)+i \,\frac{v}{c},
\end{equation}
where, $\gamma=\sqrt{1-v^2/c^2}$ is the Lorentz factor, $c$ the sound speed and $z=\sqrt{2}(x-v t)$. In the limit of zero velocity, $v \to 0$, the phase of the complex field $\Psi$ is given by
\begin{equation}
    \theta(x)=\cos^{-1}(-i \,\text{tanh(x)}/\sqrt{-\text{tanh(x)}^2}),
\end{equation}
and it displays a sudden jump $\Delta \theta(0)=\pi$ localized at the soliton core, $x=0$, in the limit of $v=0$.

Cnoidal wave solutions of the Korteweg de Vries equation, written in terms of the Jacobi elliptic function, are also associated to solitons with $\pm\pi$ phase jump in a certain approximate limit (see \textit{e.g.} \cite{KORPEL1981113}). 

In the holographic model, we still decompose the complex order parameter in its amplitude and phase
\begin{equation}
    \Psi_1(x)=|\Psi_1(x)|e^{i \theta(x)},
\end{equation}
where $\Psi_1(x)$ is extracted from the asymptotic behavior of the scalar bulk field according to Eq.~\eqref{didi}.  We then define a gauge-invariant local phase difference,
\begin{equation}
    \Delta \Theta(x)\equiv \int_x^{x+dx} \left[\partial_x\theta(x)-a_x(x)\right]dx,
\end{equation}
where $a_x(x)$ is the leading term in the asymptotic expansion of the bulk gauge field at the boundary and plays the role of the external gauge field in the dual field theory. $\Delta \Theta(x)$ quantifies the local variation of the order parameter phase. Using this quantity, we can define solitons and anti-solitons by using the following criteria
\begin{align}
    &\Delta \Theta(x)=+\pi \, :\,\text{soliton}; \nonumber\\
    &\Delta \Theta(x)=-\pi \, :\,\text{anti-soliton}.\nonumber
\end{align}
\begin{figure}[htb]
\centering
\includegraphics[width =0.9 \linewidth]{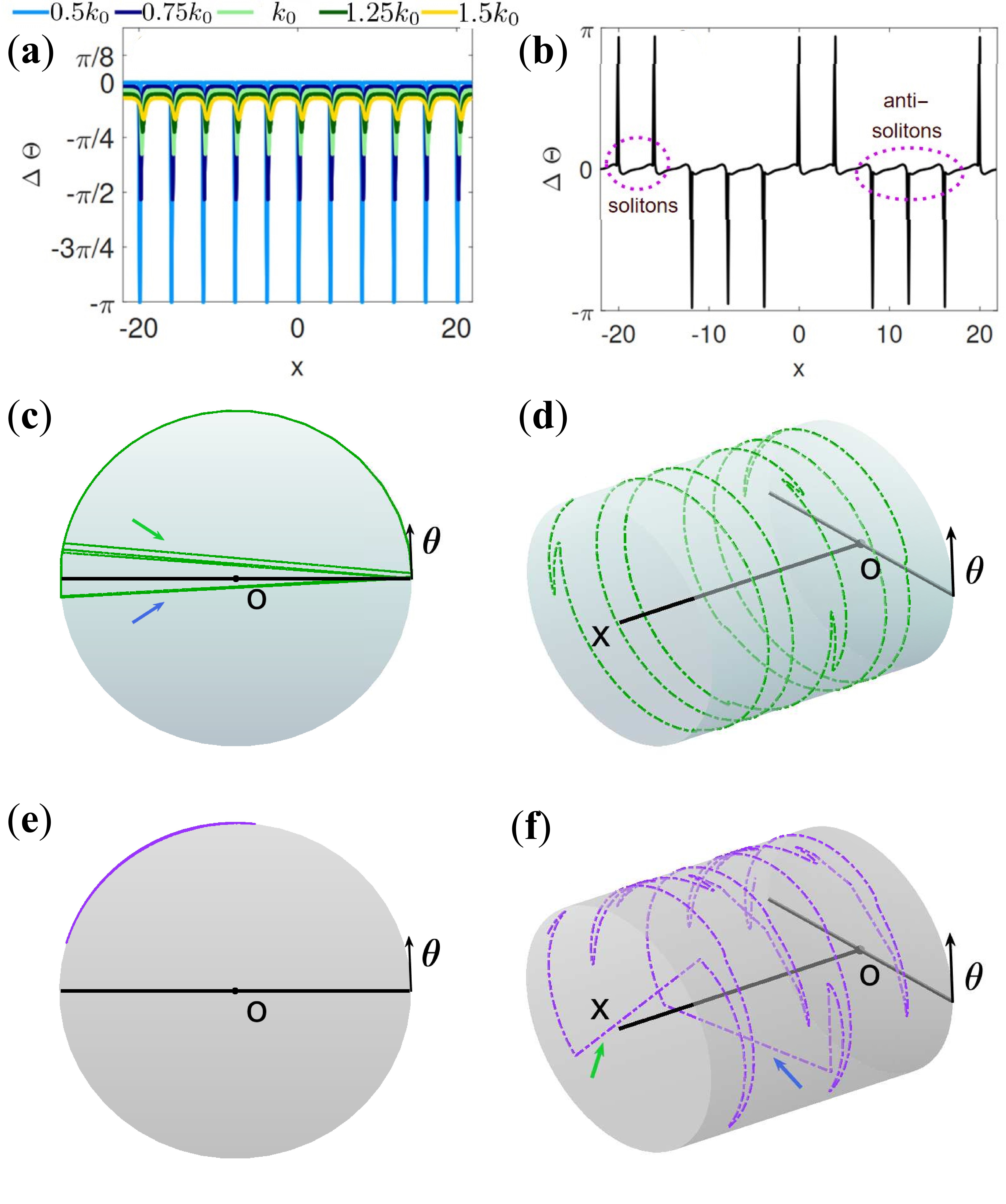}
\caption{\textbf{Spatial structure of the order parameter phase and solitons.} \textbf{(a)} Spatial distribution of the local phase difference $\Delta \Theta$ for the different inhomogeneous initial ($t=0$) solutions with \textcolor{black}{$k=0.5k_0, 0.75k_0, k_0, 1.25k_0, 1.5k_0$}. 
\textbf{(b)} Phase structure after perturbing the inhomogeneous solution \textcolor{black}{$k=0.5k_0$} (dark-soliton) with an elastic mode with $q=0.8k_0$. Here, $t=1$. The purple circles indicate the areas with soliton and anti-soliton phase jumps. \textbf{(c)-(d)} Phase of the order parameter for the inhomogeneous solution with \textcolor{black}{$k=0.5k_0$} (dark-soliton) that has been perturbed by an elastic mode with $q=0.8k_0$ at $t=1$ (top view) and $t=70$ (side view), respectively. (\textbf{c}) displays that the anti-soliton (green arrow) corresponds to a $-\pi$ phase jump before entering into the lower half plane, while the soliton (blue arrow) to a $\pi$ phase jump in lower half plane. (\textbf{d}) displays the phase of order parameter with winding number $\mathcal{W}=5$ at late time when the system is approaching to homogenous stable superflow state. \textbf{(e)-(f)} Phase of the order parameter for the inhomogeneous solution with \textcolor{black}{$k=1.5k_0$} perturbed by an elastic mode with $q=0.8k_0$ at $t=1$  (top view) and $t=32$ (side view), respectively. Green arrow and blue arrow stand for anti-soliton and soliton, respectively.}
\label{Fig_dtheta}
\end{figure}

In panel (a) of Fig.~\ref{Fig_dtheta}, we display the spatial profile of the local phase difference $\Delta \Theta(x)$ for several initial cnoidal wave states with different values of $k$. From there, it is evident that the local phase difference displays already a non-trivial structure. In particular, for small values of $k$ (\textit{e.g.}, the light blue curve with $k=0.5k_0$), the phase difference exhibits a series of sharp jumps from $0$ to $-\pi$. Those correspond to anti-soliton structures that are arranged periodically in space. The position of these jumps corresponds to the location at which the amplitude of the wave-function $\Psi_1(x)$ vanishes (see $t=0$ panel in Fig.~\ref{condensate_0}). By increasing $k$, these jumps remain but their amplitude decreases and becomes smaller than $\pi$. We notice that the initial solutions do not involve positive local phase difference. Hence, no soliton structures exist at $t=0$.

We then perturb this initial solution with an elastic mode and track the spatial structure of $\Delta \theta(x)$ while time evolves. Panel (b) in Fig.~\ref{Fig_dtheta} shows the local phase difference right after perturbing an inhomogeneous supersolid state. It is evident that in extended spatial regions, the initial $-\pi$ phase jumps are flipped into local $+\pi$ jumps. This indicates that transition between anti-soliton to soliton dynamically happens at the onset of the instability. Each of these flips is characterized by a total jump in the phase of $2\pi$, from $-\pi$ to $+\pi$. These soliton/anti-soliton transitions are accompanied by a change in the winding number of the system {defined from {Onsager-Feynman quantization condition} }\cite{PhysRevResearch.2.043065,PhysRevResearch.4.043171,PhysRevLett.47.1840,PhysRevX.8.021021}
\begin{equation}\label{wind}
    \mathcal{W}\equiv \frac{1}{2 \pi} \oint_\mathcal{L} d\theta,
\end{equation}
where the integral is performed over the whole one-dimensional system of length $L_x$ {with periodic boundary condition. For such kind of topological quantities, various experiments have explored them \cite{PhysRevLett.99.260401,PhysRevX.8.021021,2024NatCo..15.4831P}}. A transition from an anti-soliton to a soliton corresponds to a positive change in the winding number $\delta \mathcal{W}=+1$ that can be interpreted therefore as the {increasing of quantized circulation}. On the other hand, a transition from soliton to anti-soliton would correspond to the inverse process with $\delta \mathcal{W}=-1$: {decreasing of quantized circulation}.

In this language, the total winding number accumulated from panel (a) to panel (b) in Fig.~\ref{Fig_dtheta} is $\Delta \mathcal{W}=+5$, since five anti-solitons have transformed into solitons. This indicates that the instability from an inhomogeneous supersolid state to a homogeneous superfluid states depicted in Fig.~\ref{condensate_0} is accompanied by the  {formation of quantized circulation} corresponding to anti-soliton/soliton flips. These dynamics accompanying the generation of topological defects are very similar to those related to the Landau instability of superfluids with superflow current.

To better visualize the dynamics of the phase, we map the phase $\theta(t,x)$ at different times into a cylindrical coordinate system (Panels (c)-(f) of Fig.\ref{Fig_dtheta}) where the axial direction in the cylinder represents the $x$ direction while the angular direction is the phase $\theta$. A positive winding number $\mathcal{W}=+1$ corresponds to the phase wrapping once the surface of cylinder in a clockwise direction. On the other hand, $\mathcal{W}=-1$ corresponds to a winding in anti-clockwise direction. Moreover, phase jumps appear as straight lines across the axis of the cylinder, see for example green and blue arrows in panel (f) of Fig.~\ref{Fig_dtheta}. Fig.~\ref{Fig_dtheta}(c) shows a specific example of phase structure with  anti-solitons and solitons. This transition from anti-soliton to soliton changes the path of the phase along the cylindrical surface and leaves a $2\pi$ phase winding.

Given these facts, the total winding number in the final equilibrium state at time $t_f$ can be directly calculated by the number of flips between anti-solitons and solitons. Then, the final winding number is just given by
\begin{equation}
    \mathcal{W}_f= \mathcal{W}_0+N_{as \rightarrow s}-N_{s \rightarrow as},\label{gigi}
\end{equation}
where $N_{as \rightarrow s}$ is the number of flips between anti-solitons and solitons happened during the time evolution towards the final equilibrium state. $N_{s \rightarrow as}$ represents the number of inverse processes. \textcolor{black}{Due to the resonance between elastic mode $\omega_e^{+}(q)$ and anti-elastic mode $\omega_e^{-}(q)$,
\begin{equation}
\mathcal{E}_{e}^{+}(q)=\mathcal{E}_{e}^{-}(q),
\end{equation}
the spatial distribution of soliton/anti-soliton flips relies only on wave vector $q$ (see Supplementary Materials for details).} Finally, $\mathcal{W}_0$ is the winding number of the initial state, that in our case will always be zero. In the case in which the initial solution displays only anti-solitons (as for the solutions displayed in Fig.~\ref{Fig_dtheta}(a)), then $\mathcal{W}_f$ is simply the number of solitons created during the time evolution. In order to confirm these expectations, in Fig.~\ref{Fig_dtheta}(d) we display the time evolution of the phase from a supersolid state perturbed by an elastic mode ($q=0.8k_0$).

Another example is provided in panels (e)-(f) of Fig.~\ref{Fig_dtheta}.
At initial time ($t=1)$, the phase of the order parameter oscillates in a localized region, as shown in panel \ref{Fig_dtheta}(e). This configuration has clearly zero winding number since the phase does not wrap around the surface of the cylinder. Along the elastic-mode instability, anti-solitons
emerge first and then several anti-soliton/soliton transitions take place subsequently. As an example, in panel \ref{Fig_dtheta}(f), one can see a soliton (blue arrow) appearing at intermediate time. This soliton structure is then followed by a transition to an anti-soliton (green arrow) producing a non-zero winding number $\Delta \mathcal{W}$ that will remain imprinted in the final equilibrium state.

 \begin{figure}[h!]
\centering
\includegraphics[width =0.95\linewidth]{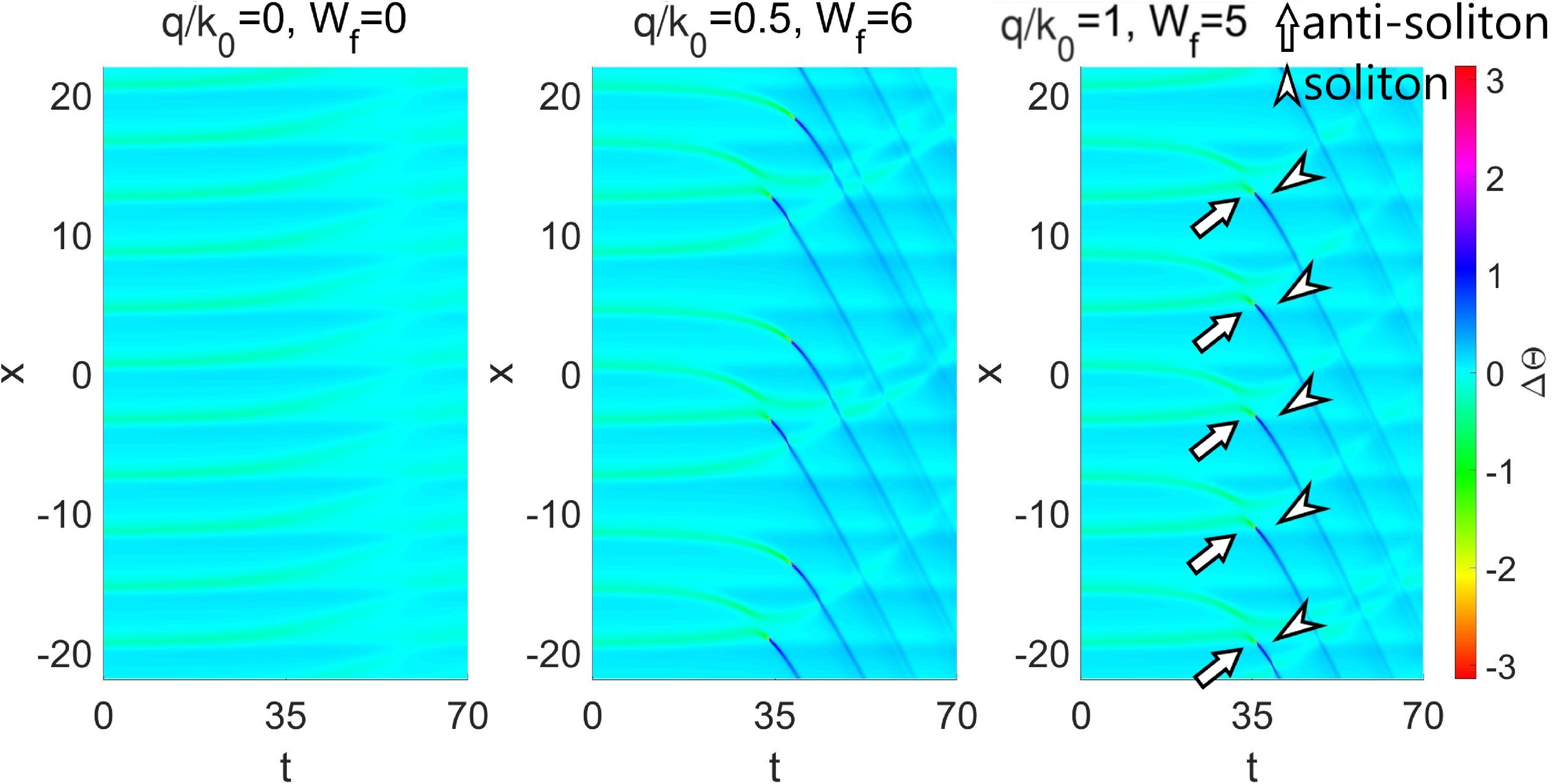}
\caption{\textbf{Spatio-temporal patterns of the phase difference with finite superflow.} Phase difference $\Delta \Theta (t,x)$ for inhomogeneous solutions with finite superflow perturbed by different elastic modes $\omega_e(q)$. Three representative cases have been selected. The arrows indicate the transitions between anti-solitons (green color) to solitons (blue color) that ultimately determine the winding number of the final equilibrium states $\mathcal{W}_f$.}
\label{finite_e}
\end {figure}

In Fig.~\ref{finite_e}, we show the full spatio-temporal profile of the phase $\theta(t,x)$ for an initial supersolid state with finite superflow perturbed by different elastic modes $\omega_e(q)$ (with different wave-vectors). The left panel in Fig.~\ref{finite_e} shows the time evolution with $q=0$. We see that the initial state is characterized by a train of linear-wave solutions with local phase jump $\Delta \Theta<- \pi$. These linear-wave solutions evolve with time and tend to become uniform. Nevertheless, no jumps with positive $\Delta \Theta$ (solitons) appear during the time evolution. Therefore, no quantum circulations transformation and the final state has zero winding number $\mathcal{W}_f=0$.

The situation is different for the center and right panels where the initial state is perturbed with a finite wave-vector elastic mode driving the instability towards a final equilibrium state with no supersolidity. From there, we can clearly observe that during the time evolution some of the anti-solitons (corresponding to green color structures) convert into solitons (blue color) producing a net change in the winding number of $\Delta \mathcal{W}=+1$ each. This {winding number} can be changed only by transitions between solitons and anti-solitons. It is in fact immediate to verify that the final winding number (computed on the final equilibrium solution using Eq.~\eqref{wind}) is given by Eq.~\eqref{gigi}. We notice that the number of anti-soliton/soliton transitions depends on the characteristics of the elastic mode perturbing the initial supersolid phase and it is not universal. In particular, we notice that an elastic mode with larger wave-vector produces a smaller number of these events and therefore a smaller winding number in the final equilibrium state.

In general, by analyzing the spatio-temporal evolution of the order parameter phase during the instability from a supersolid excited state to a equilibrium superfluid state, we have revealed that this instability is accompanied by local anti-soliton/soliton flips and the corresponding production of nonzero winding number. In other words, one could state that this elastic-mode instability that comes from unstable elastic mode can be associated to the {formation of topological excitation} related to these flips, very similarly to the famous Landau instability with unstable sound mode. We also notice a difference between these two scenarion in that the former is characterized by the change the topological structure of the phase by these local flips, while the latter is not.

\subsection{\textcolor{black}{Elastic-mode instability in the strong dissipation limit}}\label{dissipation}
\textcolor{black}{The dynamical instability of the elastic mode is directly related to dissipation, which is confirmed by the linear stability analysis from both the GP equation and the holographic model. In this section, we investigate this elastic-mode instability over a wide range of dissipation strength by tuning the value of $\eta$ in the dissipative GP equation, as defined in Eq.\eqref{Bogo}-\eqref{inin}.}

\begin{figure}[h!]
\centering
\includegraphics[width=0.75\linewidth]{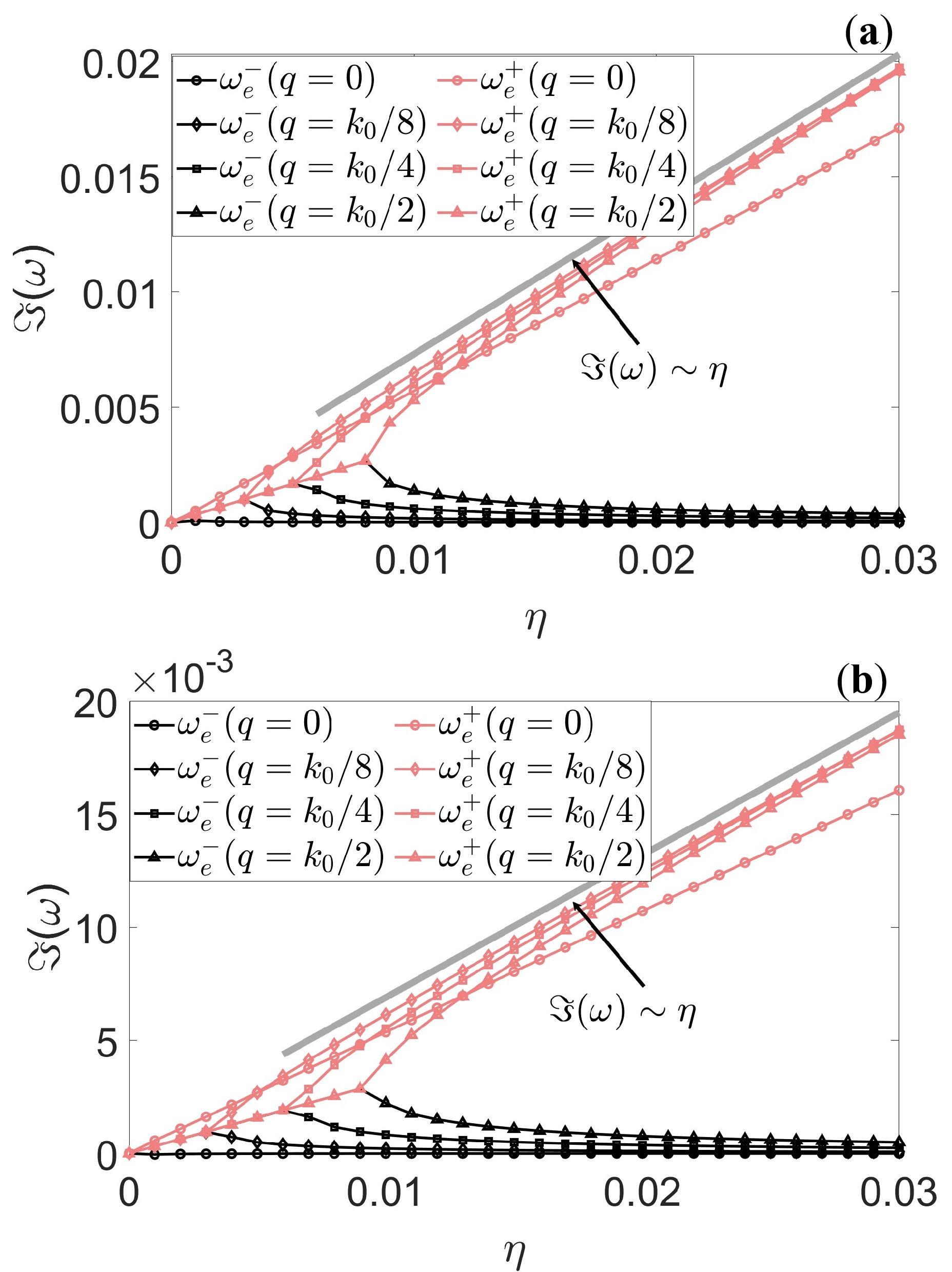}
\caption{\textcolor{black}{\textbf{The relaxation rate of elastic mode $\mathrm{Im}(\omega_e)$ as function of dissipation strength $\eta$}. \textbf{(a)}  Cnoidal wave with $k=1/2k_0$. \textbf{(b)}  Cnoidal wave with $k=4/3k_0$. The dominaint unstable elastic mode transitions from $q=0$ to a  finite $q$. The corresponding relaxation rate increases and scales as $\mathrm{Im}(\omega)\sim\eta$.}}
\label{relaxation_rate}
\end{figure}

\textcolor{black}{In Fig.~\ref{relaxation_rate}, the relaxation rate of the elastic mode $\mathrm{Im}(\omega_e)$ is calculated from Eq.~\eqref{Bogo}-~\eqref{inin}, with the dissipation strength $\eta$ increasing from $0$ to $0.03$. The dominant unstable elastic mode is $\omega_e(q=0)$ at weak dissipation and shifts to a finite wave vector $q$ close to $1/8k_0$ at strong dissipation. The relaxation rate scales linearly with $\eta$ at strong dissipation, $\mathrm{Im}(\omega_e)\sim\eta$, which implies that the elastic-mode instability becomes more significant at strong dissipation.}

\begin{figure}[htb]
\centering
\includegraphics[width =0.8\linewidth]{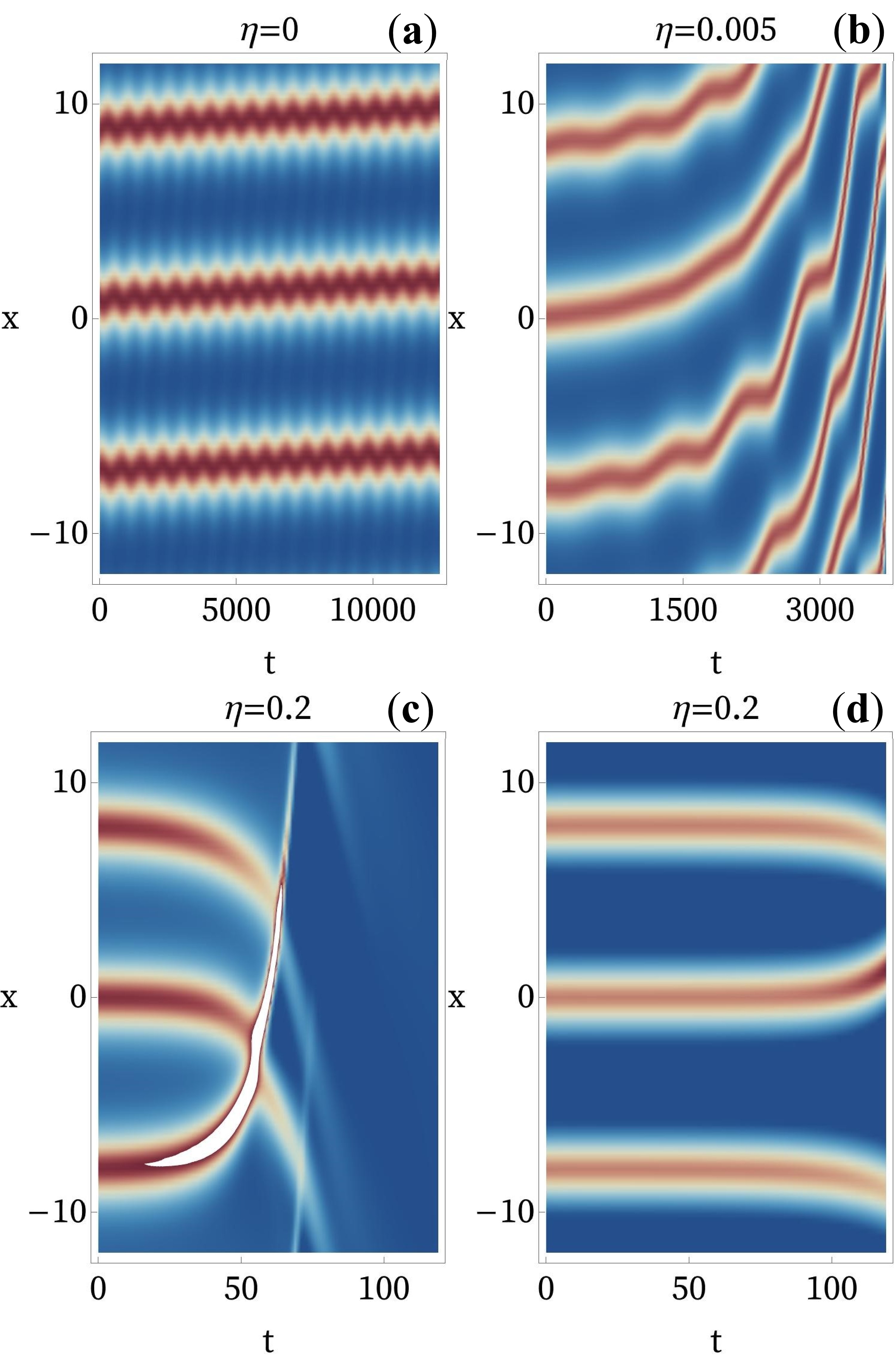}
\caption{\textcolor{black}{\textbf{Spatio-temporal patterns of the condensate}. (\textbf{a}) Self-acceleration of cnoidal wave without dissipation. (\textbf{b}) Self-acceleration of cnoidal wave with weak dissipation $\eta=0.005$. (\textbf{c}) Elastic-mode instability of cnoidal wave under strong dissipation $\eta=0.2$. (\textbf{d}) Thermodynamic relaxation of stable-mode perturbed cnoidal wave under strong dissipation $\eta=0.2$.}}
\label{thermal}
\end {figure}

\textcolor{black}{In section \ref{GPsec}, we show that the cnoidal wave is thermodynamically (energetically) unstable due to the negative value $\mathrm{Re}({\omega_e^{+}})$, which agrees with the results in \cite{2021PhRvR...3a3143M}. As a numerical test, in Fig.\ref{thermal}, we display the time evolution of the cnoidal wave with $k=0.7 k_0$ under the perturbation of elastic mode with $q=0.3k_0$ based on GP equation.  The response of the cnoidal wave to the energetically unstable elastic mode is an acceleration in the absence of dissipation as shown in Fig.\ref{thermal}(a), which resembles the self-acceleration instability of soliton in \cite{PhysRevLett.84.2298}. However, this instability usually won't destroy the soliton on a laboratory-measurable time scale as {\cite{PhysRevLett.84.2298}} mentions}.

\textcolor{black}{To further investigate the competition between thermodynamic instability and the elastic-mode instability, we first simulate the evolution of the same cnoidal wave under weak dissipation in Fig.\ref{thermal}(b), from which we can see that the weak dissipation significantly enhances the self-acceleration before the soliton/anti-soliton flip. However, this enhancement is replaced by elastic-mode instability under strong dissipation as shown in Fig.\ref{thermal}(c)-(d), where the time scale of elastic-mode instability induced soliton/anti-soliton flip (Fig.\ref{thermal}(c)) occurs earlier than thermodynamic relaxation (Fig.\ref{thermal}(d)). The spatio-temporal pattern of elastic-mode instability induced soliton/anti-soliton flip is similar to the results from the holographic model, which  reflects the nature of the strong dissipation in the holographic superfluid system, as investigated in \cite{2023PhRvD.107l1901Y}.}

\section{Outlook}\label{sum}
In this work, we have considered the recently proposed cnoidal wave supersolid phase \cite{2021PhRvR...3a3143M} and investigated in detail the effects of dissipation and finite temperature on its dynamics and stability. By combining a phenomenological GP equation with a full-fledged holographic model we have demonstrated that cnoidal wave supersolid states are both thermodynamically and dynamically unstable. More in detail, we have verified that this instability is caused by the elastic Goldstone mode appearing because of the spontaneous breaking of translational symmetry. \textcolor{black}{Furthermore, from a dynamical point of view, we have confirmed that the elastic-mode instability is accompanied by the production of topological excitations in the form of a persistent current with an integer winding number, which corresponds to local transitions between anti-soliton and soliton structures in real space. Finally, we numerically test that the elastic-mode instability dominates the dynamics of the cnoidal wave, competing with thermodynamic instability under strong dissipation in the dissipative GP equation, which is consistent with the results from the strongly dissipative holographic superfluid model.}

{It is worth noting that, a zero-temperature quantum system, such as Bose-Einstein condensate, whose dynamics
is governed by GP equation, is equal to thermodynamically unstable if it’s in a excited state.
However, this equivalence does not hold for dynamical instability, no matter it is at zero-temperature or touching with a thermal bath \cite{PhysRevA.64.061603,PhysRevA.84.043615,PhysRevA.87.063610,PhysRevResearch.6.023048,PhysRevA.110.033307}. This phenomenon can be explained within the framework of energy landscape \cite{PhysRevA.66.063603} and requires extension to the free energy landscape \cite{2024JHEP...02..184Z} when thermal effect are considered. An excited state resides at local minimum of the (free) energy is dynamical stable and small perturbation can not push it over energy barrier to reach the ground state. Only when the perturbation is
large enough can it lead to a dynamical transition to the ground state, as confirmed by ultracold-gas experiments \cite{PhysRevA.95.021602}. Although our heuristic finite-temperature holographic model and phenomenological GP
model show consistent results regarding the dynamical stability of supersolid cnoidal wave states, various experiments \cite{2014NaPho...8..145H,PhysRevResearch.2.033528,trypogeorgos2025emerging} have realized such kind of non-equilibrium supersolids in nonlinear optics platform, explained within the framework of non-linear Schr$\ddot o$dinger equation. In this sense, further investigation into finite-temperature
non-equilibrium supersolidity, particularly with respect to thermodynamics, is essential. The finite-temperature holographic model, with its well-defined Black hole thermodynamics \cite{Carlip:2014pma}, provides a valuable framework for exploring this question in the future.}


\section*{Acknowledgement}
We thank M.~Boninsegni for illumating lectures and discussions about supersolidity. We thank Biao Wu and Li-Chen Zhao for helpful discussion. M.B. acknowledges the support of the Shanghai Municipal Science and Technology Major Project (Grant No.2019SHZDZX01) and the sponsorship from the Yangyang Development Fund. Y.T. acknowledges the support of the National Natural Science Foundation of China (Grant No. 12361141825,
12035016 and 12375058). P.Y. acknowledges the support of the National Natural Science Foundation of China (Grant No. 12405021), the China Postdoctoral Science Foundation (Grant Number 2024T170545) and the Shanghai Post-doctoral Excellence Program (No. 2024380).

 \noindent \textbf{Conflict of interest}\quad {The authors declare that they have no conflict of interest.}

\bibliography{biblio}

\newpage
\onecolumngrid
\appendix 
\clearpage

\begin{center}
    {\Large \bf Supplementary Materials for ``Dissipation induced elastic-mode instability with topological excitation in holographic non-equilibrium steady cnoidal wave supersolid''}
\end{center}
\begin{center}
    {\large Peng Yang, Yu Tian, Matteo Baggioli}\\
    
    Corresponding Authors: \color{blue}pengyang23@sjtu.edu.cn
\end{center}
In this Supplementary Materials, we provide further details about the Brillouin zone representation, the holographic model, the soliton/anti-soliton pairs under elastic-model perturbation and the topological defects in superfluid system.\\

\twocolumngrid

\renewcommand\thefigure{S\arabic{figure}}    
\setcounter{figure}{0} 
\renewcommand{\theequation}{S\arabic{equation}}
\setcounter{equation}{0}
\renewcommand{\thesubsection}{SM\arabic{subsection}}

\section*{The first Brillouin zone representation and the extended Brillouin representation}\label{appd_0}
\begin{figure}[h!]
\centering
\includegraphics[width=0.7\linewidth]{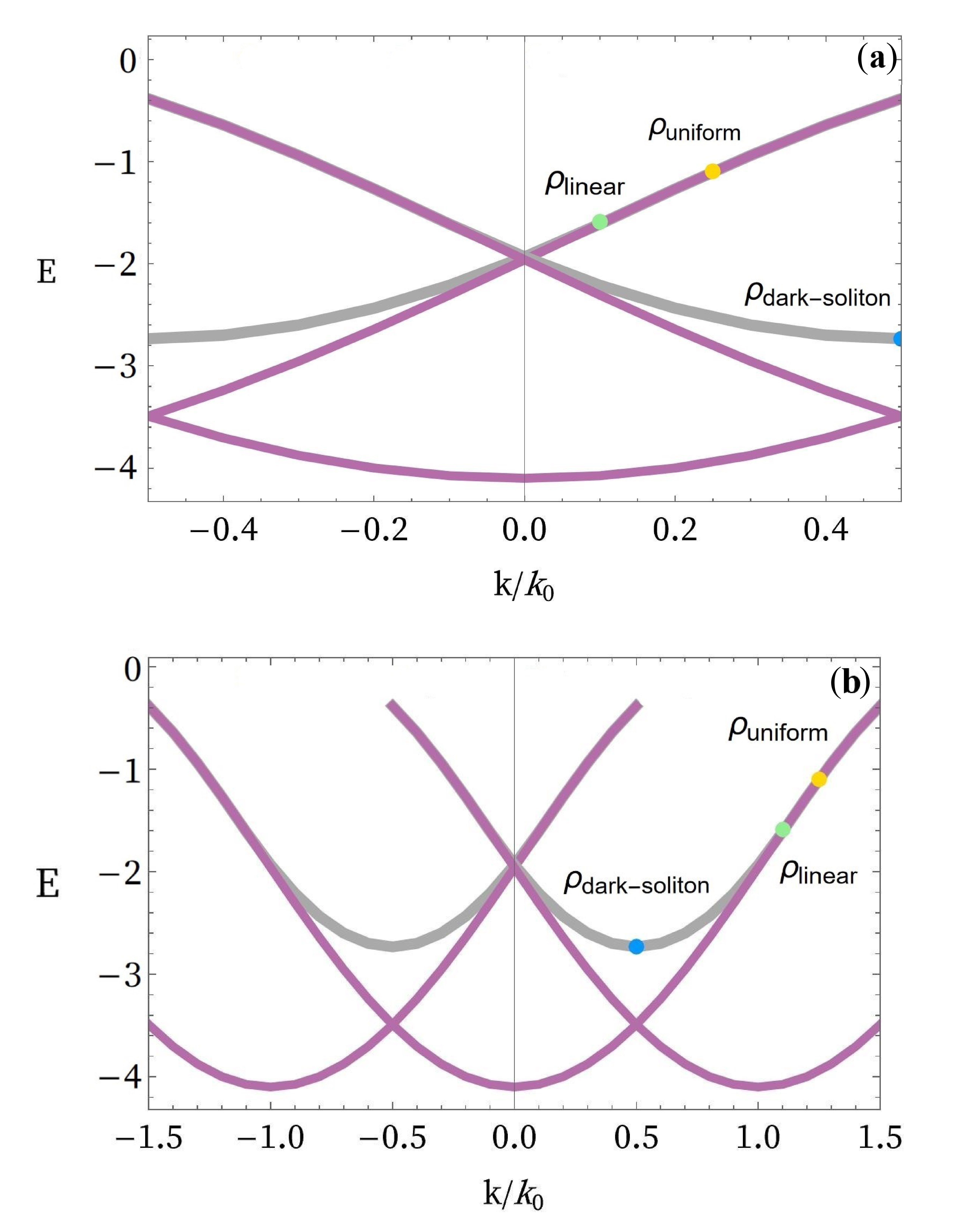}
\caption{\textcolor{black}{\textbf{The energy bands for homogeneous superflow states and cnoidal wave states.} The purple curves represent homogeneous superflow states, and the gray curves represent cnoidal wave states. (\textbf{a}) In the first Brillouin zone representation, $\rho_\text{linear}$ is located at $k = 0.11k_0$, $\rho_\text{uniform}$ is located at $k = 0.25k_0$, and $\rho_\text{dark-soliton}$ is located at $k = 0.5k_0$. (\textbf{b}) In the extended Brillouin zone representation, $\rho_\text{linear}$ is located at $k = 1.11k_0$, $\rho_\text{uniform}$ is located at $k = 1.25k_0$, and $\rho_\text{dark-soliton}$ is located at $k = 0.5k_0$.}}
\label{brill}
\end{figure}

\textcolor{black}{In the main text, we consider Bloch-type solutions with the ansatz $\Psi = \psi e^{i k x}$, in which the Bloch-wave vector is well-defined in the first Brillouin zone, i.e., $k \in [-\frac{1}{2}k_0, \frac{1}{2}k_0]$, due to the periodicity $\psi(x + \frac{2\pi}{k_0}) = \psi(x)$. In Fig.~\ref{brill}(a), we show the energy band $E(k)$, in which the two limiting cases of the cnoidal wave — the linear wave state, the dark-soliton state, and the uniform superflow state — can be identified by fitting the density profile $\rho(x) = |\Psi|^2$. Furthermore, by checking the local phase jump around the soliton core of the dark soliton state, we find that at $k = 0.5k_0$, the soliton has $\delta\theta = -\pi$ (anti-soliton), and at $k = -0.5k_0$, the soliton has $\delta\theta = \pi$ (soliton). Considering that the cnoidal wave states shown in Fig.~2 of the main text belong to the same branch of anti-solitons, the Bloch band in the extended Brillouin zone is therefore necessary and is shown in Fig.~\ref{brill}(b).}

\begin{figure}[h!]
\centering
\includegraphics[width=0.85\linewidth]{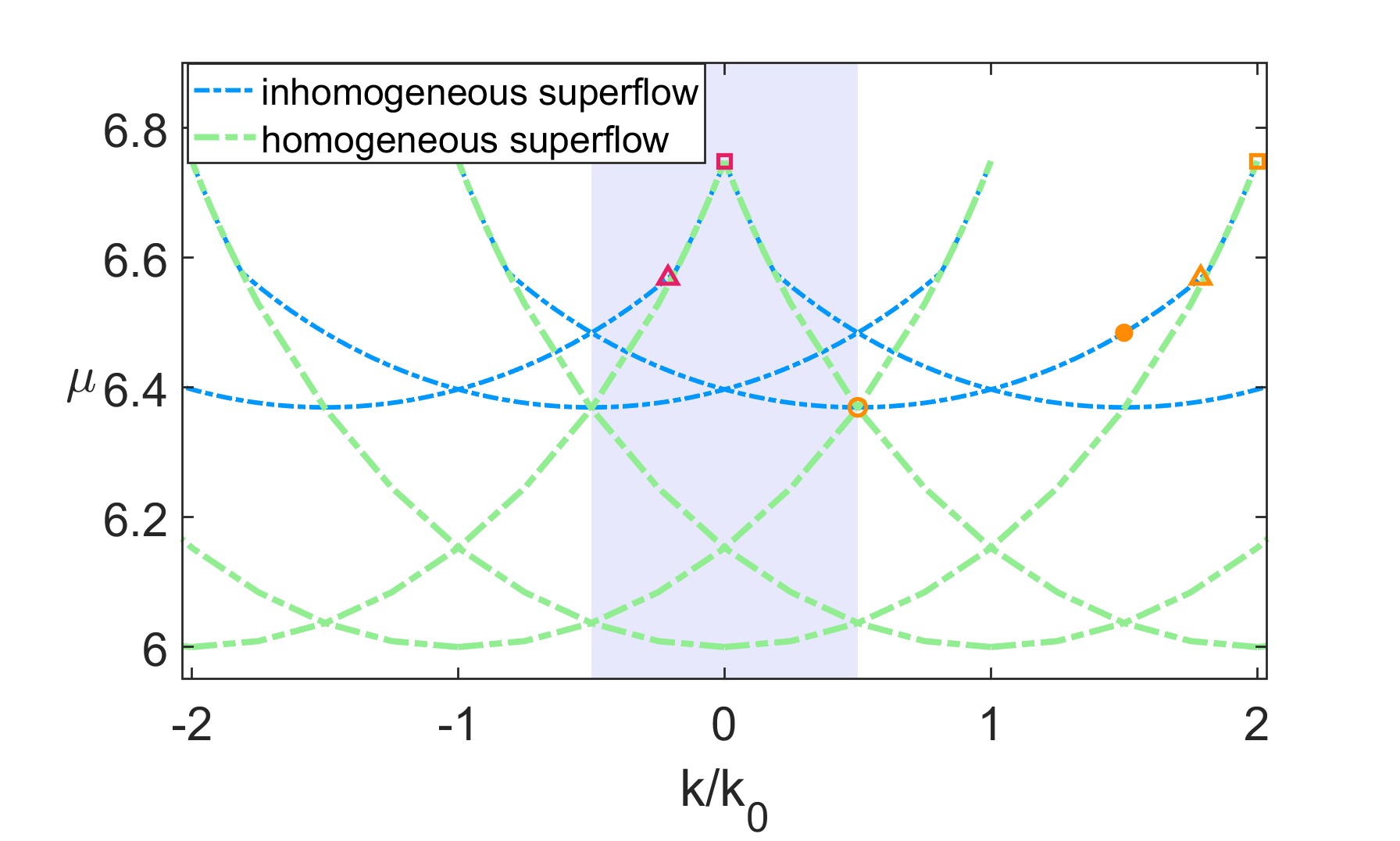}
\caption{\textcolor{black}{\textbf{The chemical potential of homogenous and inhomogenous superflow states.} The green curves represent homogeneous superflow states, and the blue curves represent inhomogeneous superflow states. The first Brillouin zone is shown with a purple background. In the first Brillouin zone representation, $\rho_\text{linear}$ is located at $k = -0.188 k_0$ (purple triangle), $\rho_\text{uniform}$ is located at $k = 0$ (purple square), and $\rho_\text{dark-soliton}$ (orange circle) is located at $k = 0.5k_0$. In the extended Brillouin zone representation, $\rho_\text{linear}$ is located at $k = 1.812 k_0$ (orange triangle), $\rho_\text{uniform}$ is located at $k = 2k_0$ (orange square), and $\rho_\text{dark-soliton}$ is located at $k = 0.5k_0$ (orange circle). The finite-current background supersolid state in Fig.~2 of main text is located at $k = 1.5 k_0$ (orange solid circle).}}
\label{brill_holo}
\end{figure}

\textcolor{black}{Based on the same representation and Bloch-type ansatz in the holographic model, the chemical potential $\mu(k)$ is periodic, similar to the energy band. In Fig.~\ref{brill_holo}, we show the chemical potential as a function of the superflow parameter $k$ in both the first Brillouin zone and the extended Brillouin zone representations.}

\section*{Details of the holographic model and computations}\label{appd}
The equations of motion for the static solutions with spatial dependence only in the $x$ direction are given by
\begin{align}
&z^{2}(\partial_{z} f \partial_{z} A_{x}+f \partial_{z}^{2} A_{x})-i(\Psi^{*} \partial_{x} \Psi-\Psi \partial_{x} \Psi^{*})\nonumber\label{static_Appd}\\
&+2 i \Psi A_{x} \Psi^{*}=0,\\~\nonumber\\
&z^{2} \partial_{x} \partial_{z} A_{x}-i\left(\Psi^{*} \partial_{z} \Psi-\Psi\partial_{z} \Psi^{*}\right)=0,\\~\nonumber\\
&z^{2}\left(f \partial_{z}^{2} A_{t}+\partial_{x}^{2} A_{t}\right)-2 A_{t} \Psi^{*} \Psi=0,\\~\nonumber\\
&\frac{z^{2}}{f} A_{t}^{2} \Psi+z^{4} \partial_{z}\left(z^{-2}\right) \partial_{z} \Psi\nonumber\\
&-z^{2} f \partial_{z}^{2} \Psi+z^{2}\left(\partial_{x}-i A_{x}\right)^{2} \Psi+2 \Psi=0\label{static_Appd4}.
\end{align}

For convenience, we redefine $\Psi=z\psi$ and separate it with real and imaginary parts $\psi=(\psi_R+i \psi_I) e^{i k x}$. Substituting the above definition into the static equations of motion, we get the following five differential equations,
\begin{align}
&\partial_{z} f \partial_{z} A_x + f \partial_{z}^{2} A_{x}+2(\psi_{R} \partial_{x} \psi_{I}-\psi_{I} A_{x} \psi_{R})\nonumber\\
&+2(k-A_{x})(\psi_{R}^{2}+\psi_{I}^{2})=0,
\end{align}

\begin{align}
&\partial_{x} \partial_{z} A_{x}-2(\psi_{R} \partial_{z} \psi_{I}-\psi_{I} \partial_{z} \psi_{R})=0,
\end{align}

\begin{align}
&f \partial_{z}^{2}A_{t}+\partial_{x}^{2}A_{t}-2 A_{t}(\psi_{R}^{2}+\psi_{I}^{2})=0,
\end{align}

\begin{align}
&({f^{-1}}A_{t}^{2}-z-(A_{x}-k)^{2}) \psi_{R}+\partial_{z}(f \partial_{z} \psi_{R})\nonumber\\
&+(\partial_{x}^{2} \psi_{R}-2(k-A_{x}) \partial_{x}\psi_{I}+\psi_{I} \partial_{x} A_{x})=0,
\end{align}

\begin{align}
&({f^{-1}} A_{x}^2-z-(A_{x}-k)^{2}) \psi_{I}+\partial_{z}(f \partial_{z} \psi_{I})\nonumber\\
&+(\partial_{x}^{2} \psi_{I}+2(k-A_{x}) \partial_{x} \psi_{R}- \partial_{x}\psi_{R} A_{t})=0.
\end{align}
We will solve these equation with the condition that the source of the bulk gauge field $A_x$ at the boundary is finite and given by $-k$, which indicates that the superflow states we will obtain are supported by external constant driving.

To obtain the superflow states under constant driving, we impose the source free boundary conditions for $\Psi$ at the conformal boundary, regular conditions for $\Psi$ and $A_{t}=0$ at the horizon and periodic boundary conditions for $A_{t,x}$ and $\Psi$ in the $x$ direction. We work at fixed particle number, where the total particle number $N$ is defined as
\begin{equation}\label{N0}
N=\int^{\frac{L_x}{2}}_{-\frac{L_x}{2}} dx \rho(x).
\end{equation}
We set spatial length $L_x=4$ and fix $N=N_0$, which corresponds to the total particle at zero-current homogeneous state with chemical potential equals to $\mu=6$. We solve the static equations of motion numerically by the Newton-Raphson method.

\textcolor{black}{Following \cite{Li_2020}, the generalized free energy functional can be defined from the potential density,
\begin{align}
V&=-\mathcal{L}_{M}(\partial_t=0),\nonumber\\
&=g^{ii}|D_i\psi|^2+(m^2+g^{tt} A_t^2)|\psi|^2+\frac{1}{2}g^{tt}g^{ii}(\partial_tA_t)^2\nonumber\\
&+\frac{1}{4}g^{ii}g^{jj}F_{ij}F_{ij},
\end{align}
where, the indexes $\{i,j\}$ stands for the spatial coordinate in the dual field theory. The static free energy obtained directly by integration of $V$ in full space
\begin{align}
F&=\int d\mathbf{x} V,\nonumber\\
&=\int d\mathbf{x} \left(\frac{A_t^2}{f}|\psi|^2-A_x^2|\psi|^2+A_x \Im(\psi^*\partial_x\psi)\right)\nonumber\\ 
&+\int d\mathbf{r} \frac{1}{2}(\rho\mu-k j_x). 
\end{align}
To obtain the final  expression, the static equations of motion Eq.\eqref{static_Appd}-\eqref{static_Appd4} are used.}

The equations for the linearized perturbations defined in the main text are given by
\begin{align}
&(i v \partial_{z} \psi+i \psi^{*} \partial_{z} u)-(i u \partial_{z} \psi^{*}+i \psi \partial_{z} v)-\partial_{z}^{2} a\nonumber\\
&\partial_{z}(\partial_{x} b+i q b)=0,
\\~\nonumber\\
&(2 b \psi \psi^{*}+2(A_{x}-\frac{1}{2} q) \psi^{*}u+2\left(A_{x}+\frac{1}{2} q\right) \psi v)\nonumber\\
&+i(v \partial_{x} \psi+\psi^{*} \partial_{x} u)-i(u \partial_{x} \psi^{*}+\psi \partial_{x}v)\nonumber\\
&-\partial_{z}(f \partial_{z} b)-\partial_{z}(\partial_{x}a+i q a)=0,\\~\nonumber\\
&(z+i(\partial_{x} A_{x}-\partial_{z} A_{t})) u+(i(\partial_{x} b+i q b-\partial_{z} a)) \psi\nonumber\\
&-(2 i \omega+f^{\prime}+2 i A_{t}) \partial_{z}u+2i(A_{x}-q) \partial_{x} u\nonumber\\
&-\partial_{x}^{2} u-2 i a \partial_{z} \psi+2 ib\partial_{x} \psi+(A_x-q)^2u=0,\\~\nonumber\\
&((A_{x}+q)^{2}-i(\partial_{x}A_{x}-\partial_{z} A_{t})+z) v-2i b \partial_{x} \psi^{*}\nonumber\\
&+(2 A_{x} b-i(\partial_{x}b+iqb-\partial_{z} a)) \psi^{*}+2 i a \partial_{z} \psi^{*}\nonumber\\
&-(2 i \omega+f^{\prime}-2i A_{t}) \partial_{z} v-2 i(A_{x}+q) \partial_{x}v\nonumber\\
&-f \partial_{z}^{2} v-\partial_{x}^{2} v=0,
\end{align}
where the following coordinate system has been used,
\begin{equation}
    ds^{2}=\frac{1}{z^{2}}(-f(z) d t^{2}-2dtdz+d\Omega^{2}_2)
\end{equation}
and $\omega$ is the complex frequency.

\section*{Wave vector $p$ of elastic-mode induced soliton/anti-soliton transition}\label{appd_c}

\begin{figure}[htb]
\centering
\includegraphics[width =0.9\linewidth]{fig_S3.jpg}
\caption{\textcolor{black}{\textbf{Spatial distribution and wave vector $p$ of perturbed $\Delta \Theta$.} (\textbf{a}) Phase difference $\Delta \Theta (x)$ of the anti-soliton train (blue line) and soliton/anti-soliton pairs (yellow line). (\textbf{b}) Fourier spectrum of $\Delta \Theta$ with the peak located at $p = \pm 0.182 k_0$. (\textbf{c}) The asymptotic value of the wave vector $p$ for the elastic-node induced $\Delta \Theta$ in the thermodynamic limit, with the number of anti-solitons $N_{as}$ increasing in an extended anti-soliton train. (\textbf{d}) The corresponding relation between $p$ and $q$ within one period of $q$.}}
\label{resona}
\end {figure}

\noindent \textcolor{black}{In the main text, the cnoidal wave becomes an anti-soliton train in the zero-current limit, as shown in Fig.~9(a). The anti-soliton train transforms into soliton/anti-soliton pairs due to the perturbation of unstable elastic modes, as shown in Fig.~9(b). In this section, we investigate the relation between soliton/anti-soliton pairs and the wave vector $q$ of elastic modes, and the results are shown in Fig.~\ref{resona}. As an example, we display the local phase difference $\Delta \Theta$ of the anti-soliton train (blue lines in Fig.~\ref{resona}(a)) and the perturbed anti-soliton train due to the unstable elastic mode $\omega_e^{\pm}(q = 0.8k_0)$ (yellow lines in Fig.~\ref{resona}(a)). The effect of the elastic mode is to break the discrete periodicity of $\Delta \Theta$ into another period. The wave vector of the perturbed $\Delta \Theta$ is $p = 0.182 k_0$, as shown in Fig.~\ref{resona}(b). In the thermodynamic limit, this wave vector approaches 0.2, as shown in Fig.~\ref{resona}(c), which is directly induced by the wave vector $q$ of the elastic mode. Since, in the first Brillouin zone representation, the perturbed wave vector is mapped to $q = -0.2k_0$. Furthermore, we numerically check the corresponding relation $p - q$ within one period of $q$ in Fig.~\ref{resona}(d). As expected, the wave vector of the perturbed $\Delta \Theta$ is equal to the wave vector of the elastic mode in the first Brillouin zone representation. In Fig.~7(a) in the main text, the real part of the eigenvalue $\mathcal{E}_{e} = \mathrm{Re}(\omega_e)$ of the dynamically unstable elastic modes is zero for the cnoidal wave solutions in the zero-current limit. The resonance between the elastic and anti-elastic modes satisfies a similar resonance condition as in \cite{PhysRevA.64.061603}.
\begin{equation}
\mathcal{E}_{e}^{+}(q)=\mathcal{E}_{e}^{-}(q).
\end{equation}}

\section*{Notes on topological defects in superfluid}\label{appd_b}

Topological excitations are common in condensed matter systems, such as in solid crystals, spin chain, superfluid and superconductor systems. Based on the number of spatial dimension of systems that be considered, one can classify topological defects as winding number in 1D, singular point in 2D and defect line in 3D (see Fig.~\ref{defec}), respectively. To provide a brief note about topological defects in superfluid, We follow the definition in textbook \cite{selingerintroduction}.

\begin{figure}[hb]
\centering
\includegraphics[width =0.8\linewidth]{fig_S4.jpg}
\caption{\textbf{Schematic diagram of topological defects of magnetic moment $\vec\mu$.} (\textbf{a}) shows topological defect as singular point. (\textbf{b}) shows topological defect as winding number and (\textbf{c}) plots the topological defect as defect line.}
\label{defec}
\end {figure}

Let's firstly consider a spin system, in which the magnetic moment $\vec{\mu}$ is free to rotate in a 2D plane. In such a system, there exist two typical phases, one is the disordered phase with $\vec{\mu}=0$ at higher temperature and the other is the ordered phase where there is a nonzero magnetic moment with unified orientation $\vec{\mu}=\mu \cdot \hat {e}_n$ below critical temperature, $\hat {e}_n$ is a unit vector in 2D plane. To simplify but without loss of generality, we assume that $\mu$ is constant and only the degree of freedom comes from the direction $\hat {e}_n$. 

In the absence of external field, for ordered phase at equilibrium, any direction of $\hat {e}_n$ has equal free energy. Due to the rotational symmetry being spontaneously broken, the equilibrium state select a random direction with minimized free energy, i.e., $\hat{e}_n(\mathbf{r})=\hat {e}_n$. Apart from the above configuration, there exist another kind of configuration that minimized the free energy, in which the magnetic moment direction $\hat{e}_n(\mathbf{r})$ is space-dependent just like Fig.~\ref{defec}. Fig.~\ref{defec}(a) is a representative example of space-dependent configuration of $\hat{e}_n(\mathbf{r})$, in which there is a singular point. To characterize this configuration with singular point, we use a close loop $\mathcal{C}$ to define a winding number $W$,

\begin{equation}
    \mathcal{W}=\frac{1}{2\pi}\int_{\mathcal{C}}(\epsilon_{3ij}n_i\partial_kn_j)dl_k,
\end{equation}
where $\epsilon_{ijk}$ is the Levi-Civita symbol. The winding number measure the the number of times that the direction of magnetic moment rotates along $\mathcal{C}$ in a clockwise direction. As we can see, by moving the loop $\mathcal{C}$ from the purple one to the orange one (Fig.~\ref{defec}(a)), $\mathcal{W}$ jump from 0 to $-1$. Since we consider  $\vec\mu$ rotating in 2D angular space, the measurement of winding number $\mathcal{W}$ is straightforwardly generalized to one-dimension ring (Fig.~\ref{defec}(b)) and three-dimension system (Fig.~\ref{defec}(c)). For a further generalization one suggest the readers to refer to second part of \cite{selingerintroduction}, to the case with $\vec\mu$ rotating in full space and some other systems, superfluid system for instance. As \cite{selingerintroduction} points out, the genralization from spin system to superfluid system is given by mapping the direction of magnetic moment in real space to the phase of superfluid in order parameter space,
\begin{equation}
    \hat{e}_n(\mathbf{r})=(\text{cos}(\theta(\mathbf{r})),\text{sin}(\theta(\mathbf{r}))),
\end{equation}
where $\theta$ comes from the phase of order parameter of superfluid $\psi=|\psi|e^{i \theta}$. With this mapping in mind, the definition of  winding number for superfluid system is given by
\begin{align}
    \mathcal{W} &=\frac{1}{2\pi}\int_{{ring}}\frac{d\theta}{dx}dx\\
    &=\frac{1}{2\pi}\int_{loop}\nabla\theta\cdot d\mathbf{l},
\end{align}
where the first equation is for one-dimension ring and the second equation is for two-dimension plane.

\bibliographystyle{apsrev4-2}
\bibliography{biblio}

\begin{thebibliography}{69}%
\makeatletter
\providecommand \@ifxundefined [1]{%
 \@ifx{#1\undefined}
}%
\providecommand \@ifnum [1]{%
 \ifnum #1\expandafter \@firstoftwo
 \else \expandafter \@secondoftwo
 \fi
}%
\providecommand \@ifx [1]{%
 \ifx #1\expandafter \@firstoftwo
 \else \expandafter \@secondoftwo
 \fi
}%
\providecommand \natexlab [1]{#1}%
\providecommand \enquote  [1]{``#1''}%
\providecommand \bibnamefont  [1]{#1}%
\providecommand \bibfnamefont [1]{#1}%
\providecommand \citenamefont [1]{#1}%
\providecommand \href@noop [0]{\@secondoftwo}%
\providecommand \href [0]{\begingroup \@sanitize@url \@href}%
\providecommand \@href[1]{\@@startlink{#1}\@@href}%
\providecommand \@@href[1]{\endgroup#1\@@endlink}%
\providecommand \@sanitize@url [0]{\catcode `\\12\catcode `\$12\catcode
  `\&12\catcode `\#12\catcode `\^12\catcode `\_12\catcode `\%12\relax}%
\providecommand \@@startlink[1]{}%
\providecommand \@@endlink[0]{}%
\providecommand \url  [0]{\begingroup\@sanitize@url \@url }%
\providecommand \@url [1]{\endgroup\@href {#1}{\urlprefix }}%
\providecommand \urlprefix  [0]{URL }%
\providecommand \Eprint [0]{\href }%
\providecommand \doibase [0]{https://doi.org/}%
\providecommand \selectlanguage [0]{\@gobble}%
\providecommand \bibinfo  [0]{\@secondoftwo}%
\providecommand \bibfield  [0]{\@secondoftwo}%
\providecommand \translation [1]{[#1]}%
\providecommand \BibitemOpen [0]{}%
\providecommand \bibitemStop [0]{}%
\providecommand \bibitemNoStop [0]{.\EOS\space}%
\providecommand \EOS [0]{\spacefactor3000\relax}%
\providecommand \BibitemShut  [1]{\csname bibitem#1\endcsname}%
\let\auto@bib@innerbib\@empty
\bibitem [{\citenamefont {Beekman}\ \emph {et~al.}(2019)\citenamefont
  {Beekman}, \citenamefont {Rademaker},\ and\ \citenamefont {van
  Wezel}}]{10.21468/SciPostPhysLectNotes.11}%
  \BibitemOpen
  \bibfield  {author} {\bibinfo {author} {\bibfnamefont {A.~J.}\ \bibnamefont
  {Beekman}}, \bibinfo {author} {\bibfnamefont {L.}~\bibnamefont {Rademaker}},\
  and\ \bibinfo {author} {\bibfnamefont {J.}~\bibnamefont {van Wezel}},\ }\href
  {https://doi.org/10.21468/SciPostPhysLectNotes.11} {\bibfield  {journal}
  {\bibinfo  {journal} {SciPost Phys. Lect. Notes}\ ,\ \bibinfo {pages} {11}}
  (\bibinfo {year} {2019})}\BibitemShut {NoStop}%
\bibitem [{\citenamefont {Zaletel}\ \emph {et~al.}(2023)\citenamefont
  {Zaletel}, \citenamefont {Lukin}, \citenamefont {Monroe}, \citenamefont
  {Nayak}, \citenamefont {Wilczek},\ and\ \citenamefont
  {Yao}}]{RevModPhys.95.031001}%
  \BibitemOpen
  \bibfield  {author} {\bibinfo {author} {\bibfnamefont {M.~P.}\ \bibnamefont
  {Zaletel}}, \bibinfo {author} {\bibfnamefont {M.}~\bibnamefont {Lukin}},
  \bibinfo {author} {\bibfnamefont {C.}~\bibnamefont {Monroe}}, \bibinfo
  {author} {\bibfnamefont {C.}~\bibnamefont {Nayak}}, \bibinfo {author}
  {\bibfnamefont {F.}~\bibnamefont {Wilczek}},\ and\ \bibinfo {author}
  {\bibfnamefont {N.~Y.}\ \bibnamefont {Yao}},\ }\href
  {https://doi.org/10.1103/RevModPhys.95.031001} {\bibfield  {journal}
  {\bibinfo  {journal} {Rev. Mod. Phys.}\ }\textbf {\bibinfo {volume} {95}},\
  \bibinfo {pages} {031001} (\bibinfo {year} {2023})}\BibitemShut {NoStop}%
\bibitem [{\citenamefont {Boninsegni}\ and\ \citenamefont
  {Prokof'ev}(2012)}]{RevModPhys.84.759}%
  \BibitemOpen
  \bibfield  {author} {\bibinfo {author} {\bibfnamefont {M.}~\bibnamefont
  {Boninsegni}}\ and\ \bibinfo {author} {\bibfnamefont {N.~V.}\ \bibnamefont
  {Prokof'ev}},\ }\href {https://doi.org/10.1103/RevModPhys.84.759} {\bibfield
  {journal} {\bibinfo  {journal} {Rev. Mod. Phys.}\ }\textbf {\bibinfo {volume}
  {84}},\ \bibinfo {pages} {759} (\bibinfo {year} {2012})}\BibitemShut
  {NoStop}%
\bibitem [{\citenamefont {{Li}}\ \emph {et~al.}(2017)\citenamefont {{Li}},
  \citenamefont {{Lee}}, \citenamefont {{Huang}}, \citenamefont {{Burchesky}},
  \citenamefont {{Shteynas}}, \citenamefont {{Top}}, \citenamefont
  {{Jamison}},\ and\ \citenamefont {{Ketterle}}}]{2017Natur.543...91L}%
  \BibitemOpen
  \bibfield  {author} {\bibinfo {author} {\bibfnamefont {J.-R.}\ \bibnamefont
  {{Li}}}, \bibinfo {author} {\bibfnamefont {J.}~\bibnamefont {{Lee}}},
  \bibinfo {author} {\bibfnamefont {W.}~\bibnamefont {{Huang}}}, \bibinfo
  {author} {\bibfnamefont {S.}~\bibnamefont {{Burchesky}}}, \bibinfo {author}
  {\bibfnamefont {B.}~\bibnamefont {{Shteynas}}}, \bibinfo {author}
  {\bibfnamefont {F.~{\c{C}}.}\ \bibnamefont {{Top}}}, \bibinfo {author}
  {\bibfnamefont {A.~O.}\ \bibnamefont {{Jamison}}},\ and\ \bibinfo {author}
  {\bibfnamefont {W.}~\bibnamefont {{Ketterle}}},\ }\href
  {https://doi.org/10.1038/nature21431} {\bibfield  {journal} {\bibinfo
  {journal} {\nat}\ }\textbf {\bibinfo {volume} {543}},\ \bibinfo {pages} {91}
  (\bibinfo {year} {2017})},\ \Eprint {https://arxiv.org/abs/1610.08194}
  {arXiv:1610.08194 [cond-mat.quant-gas]} \BibitemShut {NoStop}%
\bibitem [{\citenamefont {{L{\'e}onard}}\ \emph {et~al.}(2017)\citenamefont
  {{L{\'e}onard}}, \citenamefont {{Morales}}, \citenamefont {{Zupancic}},
  \citenamefont {{Esslinger}},\ and\ \citenamefont
  {{Donner}}}]{2017Natur.543...87L}%
  \BibitemOpen
  \bibfield  {author} {\bibinfo {author} {\bibfnamefont {J.}~\bibnamefont
  {{L{\'e}onard}}}, \bibinfo {author} {\bibfnamefont {A.}~\bibnamefont
  {{Morales}}}, \bibinfo {author} {\bibfnamefont {P.}~\bibnamefont
  {{Zupancic}}}, \bibinfo {author} {\bibfnamefont {T.}~\bibnamefont
  {{Esslinger}}},\ and\ \bibinfo {author} {\bibfnamefont {T.}~\bibnamefont
  {{Donner}}},\ }\href {https://doi.org/10.1038/nature21067} {\bibfield
  {journal} {\bibinfo  {journal} {\nat}\ }\textbf {\bibinfo {volume} {543}},\
  \bibinfo {pages} {87} (\bibinfo {year} {2017})},\ \Eprint
  {https://arxiv.org/abs/1609.09053} {arXiv:1609.09053 [cond-mat.quant-gas]}
  \BibitemShut {NoStop}%
\bibitem [{\citenamefont {{Tanzi}}\ \emph
  {et~al.}(2019{\natexlab{a}})\citenamefont {{Tanzi}}, \citenamefont
  {{Lucioni}}, \citenamefont {{Fam{\`a}}}, \citenamefont {{Catani}},
  \citenamefont {{Fioretti}}, \citenamefont {{Gabbanini}}, \citenamefont
  {{Bisset}}, \citenamefont {{Santos}},\ and\ \citenamefont
  {{Modugno}}}]{2019PhRvL.122m0405T}%
  \BibitemOpen
  \bibfield  {author} {\bibinfo {author} {\bibfnamefont {L.}~\bibnamefont
  {{Tanzi}}}, \bibinfo {author} {\bibfnamefont {E.}~\bibnamefont {{Lucioni}}},
  \bibinfo {author} {\bibfnamefont {F.}~\bibnamefont {{Fam{\`a}}}}, \bibinfo
  {author} {\bibfnamefont {J.}~\bibnamefont {{Catani}}}, \bibinfo {author}
  {\bibfnamefont {A.}~\bibnamefont {{Fioretti}}}, \bibinfo {author}
  {\bibfnamefont {C.}~\bibnamefont {{Gabbanini}}}, \bibinfo {author}
  {\bibfnamefont {R.~N.}\ \bibnamefont {{Bisset}}}, \bibinfo {author}
  {\bibfnamefont {L.}~\bibnamefont {{Santos}}},\ and\ \bibinfo {author}
  {\bibfnamefont {G.}~\bibnamefont {{Modugno}}},\ }\href
  {https://doi.org/10.1103/PhysRevLett.122.130405} {\bibfield  {journal}
  {\bibinfo  {journal} {\prl}\ }\textbf {\bibinfo {volume} {122}},\ \bibinfo
  {eid} {130405} (\bibinfo {year} {2019}{\natexlab{a}})},\ \Eprint
  {https://arxiv.org/abs/1811.02613} {arXiv:1811.02613 [cond-mat.quant-gas]}
  \BibitemShut {NoStop}%
\bibitem [{\citenamefont {{Guo}}\ \emph {et~al.}(2019)\citenamefont {{Guo}},
  \citenamefont {{B{\"o}ttcher}}, \citenamefont {{Hertkorn}}, \citenamefont
  {{Schmidt}}, \citenamefont {{Wenzel}}, \citenamefont {{B{\"u}chler}},
  \citenamefont {{Langen}},\ and\ \citenamefont
  {{Pfau}}}]{2019Natur.574..386G}%
  \BibitemOpen
  \bibfield  {author} {\bibinfo {author} {\bibfnamefont {M.}~\bibnamefont
  {{Guo}}}, \bibinfo {author} {\bibfnamefont {F.}~\bibnamefont
  {{B{\"o}ttcher}}}, \bibinfo {author} {\bibfnamefont {J.}~\bibnamefont
  {{Hertkorn}}}, \bibinfo {author} {\bibfnamefont {J.-N.}\ \bibnamefont
  {{Schmidt}}}, \bibinfo {author} {\bibfnamefont {M.}~\bibnamefont {{Wenzel}}},
  \bibinfo {author} {\bibfnamefont {H.~P.}\ \bibnamefont {{B{\"u}chler}}},
  \bibinfo {author} {\bibfnamefont {T.}~\bibnamefont {{Langen}}},\ and\
  \bibinfo {author} {\bibfnamefont {T.}~\bibnamefont {{Pfau}}},\ }\href
  {https://doi.org/10.1038/s41586-019-1569-5} {\bibfield  {journal} {\bibinfo
  {journal} {\nat}\ }\textbf {\bibinfo {volume} {574}},\ \bibinfo {pages} {386}
  (\bibinfo {year} {2019})},\ \Eprint {https://arxiv.org/abs/1906.04633}
  {arXiv:1906.04633 [cond-mat.quant-gas]} \BibitemShut {NoStop}%
\bibitem [{\citenamefont {{Tanzi}}\ \emph
  {et~al.}(2019{\natexlab{b}})\citenamefont {{Tanzi}}, \citenamefont
  {{Roccuzzo}}, \citenamefont {{Lucioni}}, \citenamefont {{Fam{\`a}}},
  \citenamefont {{Fioretti}}, \citenamefont {{Gabbanini}}, \citenamefont
  {{Modugno}}, \citenamefont {{Recati}},\ and\ \citenamefont
  {{Stringari}}}]{2019Natur.574..382T}%
  \BibitemOpen
  \bibfield  {author} {\bibinfo {author} {\bibfnamefont {L.}~\bibnamefont
  {{Tanzi}}}, \bibinfo {author} {\bibfnamefont {S.~M.}\ \bibnamefont
  {{Roccuzzo}}}, \bibinfo {author} {\bibfnamefont {E.}~\bibnamefont
  {{Lucioni}}}, \bibinfo {author} {\bibfnamefont {F.}~\bibnamefont
  {{Fam{\`a}}}}, \bibinfo {author} {\bibfnamefont {A.}~\bibnamefont
  {{Fioretti}}}, \bibinfo {author} {\bibfnamefont {C.}~\bibnamefont
  {{Gabbanini}}}, \bibinfo {author} {\bibfnamefont {G.}~\bibnamefont
  {{Modugno}}}, \bibinfo {author} {\bibfnamefont {A.}~\bibnamefont
  {{Recati}}},\ and\ \bibinfo {author} {\bibfnamefont {S.}~\bibnamefont
  {{Stringari}}},\ }\href {https://doi.org/10.1038/s41586-019-1568-6}
  {\bibfield  {journal} {\bibinfo  {journal} {\nat}\ }\textbf {\bibinfo
  {volume} {574}},\ \bibinfo {pages} {382} (\bibinfo {year}
  {2019}{\natexlab{b}})},\ \Eprint {https://arxiv.org/abs/1906.02791}
  {arXiv:1906.02791 [cond-mat.quant-gas]} \BibitemShut {NoStop}%
\bibitem [{\citenamefont {{Norcia}}\ \emph {et~al.}(2021)\citenamefont
  {{Norcia}}, \citenamefont {{Politi}}, \citenamefont {{Klaus}}, \citenamefont
  {{Poli}}, \citenamefont {{Sohmen}}, \citenamefont {{Mark}}, \citenamefont
  {{Bisset}}, \citenamefont {{Santos}},\ and\ \citenamefont
  {{Ferlaino}}}]{2021Natur.596..357N}%
  \BibitemOpen
  \bibfield  {author} {\bibinfo {author} {\bibfnamefont {M.~A.}\ \bibnamefont
  {{Norcia}}}, \bibinfo {author} {\bibfnamefont {C.}~\bibnamefont {{Politi}}},
  \bibinfo {author} {\bibfnamefont {L.}~\bibnamefont {{Klaus}}}, \bibinfo
  {author} {\bibfnamefont {E.}~\bibnamefont {{Poli}}}, \bibinfo {author}
  {\bibfnamefont {M.}~\bibnamefont {{Sohmen}}}, \bibinfo {author}
  {\bibfnamefont {M.~J.}\ \bibnamefont {{Mark}}}, \bibinfo {author}
  {\bibfnamefont {R.~N.}\ \bibnamefont {{Bisset}}}, \bibinfo {author}
  {\bibfnamefont {L.}~\bibnamefont {{Santos}}},\ and\ \bibinfo {author}
  {\bibfnamefont {F.}~\bibnamefont {{Ferlaino}}},\ }\href
  {https://doi.org/10.1038/s41586-021-03725-7} {\bibfield  {journal} {\bibinfo
  {journal} {\nat}\ }\textbf {\bibinfo {volume} {596}},\ \bibinfo {pages} {357}
  (\bibinfo {year} {2021})},\ \Eprint {https://arxiv.org/abs/2102.05555}
  {arXiv:2102.05555 [cond-mat.quant-gas]} \BibitemShut {NoStop}%
\bibitem [{\citenamefont {{Giovanazzi}}\ \emph {et~al.}(2002)\citenamefont
  {{Giovanazzi}}, \citenamefont {{O'dell}},\ and\ \citenamefont
  {{Kurizki}}}]{2002PhRvL..88m0402G}%
  \BibitemOpen
  \bibfield  {author} {\bibinfo {author} {\bibfnamefont {S.}~\bibnamefont
  {{Giovanazzi}}}, \bibinfo {author} {\bibfnamefont {D.}~\bibnamefont
  {{O'dell}}},\ and\ \bibinfo {author} {\bibfnamefont {G.}~\bibnamefont
  {{Kurizki}}},\ }\href {https://doi.org/10.1103/PhysRevLett.88.130402}
  {\bibfield  {journal} {\bibinfo  {journal} {\prl}\ }\textbf {\bibinfo
  {volume} {88}},\ \bibinfo {eid} {130402} (\bibinfo {year} {2002})},\ \Eprint
  {https://arxiv.org/abs/cond-mat/0108046} {arXiv:cond-mat/0108046 [cond-mat]}
  \BibitemShut {NoStop}%
\bibitem [{\citenamefont {{Wessel}}\ and\ \citenamefont
  {{Troyer}}(2005)}]{2005PhRvL..95l7205W}%
  \BibitemOpen
  \bibfield  {author} {\bibinfo {author} {\bibfnamefont {S.}~\bibnamefont
  {{Wessel}}}\ and\ \bibinfo {author} {\bibfnamefont {M.}~\bibnamefont
  {{Troyer}}},\ }\href {https://doi.org/10.1103/PhysRevLett.95.127205}
  {\bibfield  {journal} {\bibinfo  {journal} {\prl}\ }\textbf {\bibinfo
  {volume} {95}},\ \bibinfo {eid} {127205} (\bibinfo {year} {2005})},\ \Eprint
  {https://arxiv.org/abs/cond-mat/0505298} {arXiv:cond-mat/0505298
  [cond-mat.other]} \BibitemShut {NoStop}%
\bibitem [{\citenamefont {{Henkel}}\ \emph {et~al.}(2010)\citenamefont
  {{Henkel}}, \citenamefont {{Nath}},\ and\ \citenamefont
  {{Pohl}}}]{2010PhRvL.104s5302H}%
  \BibitemOpen
  \bibfield  {author} {\bibinfo {author} {\bibfnamefont {N.}~\bibnamefont
  {{Henkel}}}, \bibinfo {author} {\bibfnamefont {R.}~\bibnamefont {{Nath}}},\
  and\ \bibinfo {author} {\bibfnamefont {T.}~\bibnamefont {{Pohl}}},\ }\href
  {https://doi.org/10.1103/PhysRevLett.104.195302} {\bibfield  {journal}
  {\bibinfo  {journal} {\prl}\ }\textbf {\bibinfo {volume} {104}},\ \bibinfo
  {eid} {195302} (\bibinfo {year} {2010})},\ \Eprint
  {https://arxiv.org/abs/1001.3250} {arXiv:1001.3250 [physics.atom-ph]}
  \BibitemShut {NoStop}%
\bibitem [{\citenamefont {{Wang}}\ \emph {et~al.}(2010)\citenamefont {{Wang}},
  \citenamefont {{Gao}}, \citenamefont {{Jian}},\ and\ \citenamefont
  {{Zhai}}}]{2010PhRvL.105p0403W}%
  \BibitemOpen
  \bibfield  {author} {\bibinfo {author} {\bibfnamefont {C.}~\bibnamefont
  {{Wang}}}, \bibinfo {author} {\bibfnamefont {C.}~\bibnamefont {{Gao}}},
  \bibinfo {author} {\bibfnamefont {C.-M.}\ \bibnamefont {{Jian}}},\ and\
  \bibinfo {author} {\bibfnamefont {H.}~\bibnamefont {{Zhai}}},\ }\href
  {https://doi.org/10.1103/PhysRevLett.105.160403} {\bibfield  {journal}
  {\bibinfo  {journal} {\prl}\ }\textbf {\bibinfo {volume} {105}},\ \bibinfo
  {eid} {160403} (\bibinfo {year} {2010})},\ \Eprint
  {https://arxiv.org/abs/1006.5148} {arXiv:1006.5148 [cond-mat.quant-gas]}
  \BibitemShut {NoStop}%
\bibitem [{\citenamefont {{Ostermann}}\ \emph {et~al.}(2016)\citenamefont
  {{Ostermann}}, \citenamefont {{Piazza}},\ and\ \citenamefont
  {{Ritsch}}}]{2016PhRvX...6b1026O}%
  \BibitemOpen
  \bibfield  {author} {\bibinfo {author} {\bibfnamefont {S.}~\bibnamefont
  {{Ostermann}}}, \bibinfo {author} {\bibfnamefont {F.}~\bibnamefont
  {{Piazza}}},\ and\ \bibinfo {author} {\bibfnamefont {H.}~\bibnamefont
  {{Ritsch}}},\ }\href {https://doi.org/10.1103/PhysRevX.6.021026} {\bibfield
  {journal} {\bibinfo  {journal} {Physical Review X}\ }\textbf {\bibinfo
  {volume} {6}},\ \bibinfo {eid} {021026} (\bibinfo {year} {2016})},\ \Eprint
  {https://arxiv.org/abs/1601.04900} {arXiv:1601.04900 [quant-ph]} \BibitemShut
  {NoStop}%
\bibitem [{\citenamefont {{Martone}}\ \emph {et~al.}(2021)\citenamefont
  {{Martone}}, \citenamefont {{Recati}},\ and\ \citenamefont
  {{Pavloff}}}]{2021PhRvR...3a3143M}%
  \BibitemOpen
  \bibfield  {author} {\bibinfo {author} {\bibfnamefont {G.~I.}\ \bibnamefont
  {{Martone}}}, \bibinfo {author} {\bibfnamefont {A.}~\bibnamefont
  {{Recati}}},\ and\ \bibinfo {author} {\bibfnamefont {N.}~\bibnamefont
  {{Pavloff}}},\ }\href {https://doi.org/10.1103/PhysRevResearch.3.013143}
  {\bibfield  {journal} {\bibinfo  {journal} {Physical Review Research}\
  }\textbf {\bibinfo {volume} {3}},\ \bibinfo {eid} {013143} (\bibinfo {year}
  {2021})},\ \Eprint {https://arxiv.org/abs/2008.00795} {arXiv:2008.00795
  [cond-mat.quant-gas]} \BibitemShut {NoStop}%
\bibitem [{\citenamefont {{Yang}}\ \emph {et~al.}(2023)\citenamefont {{Yang}},
  \citenamefont {{Baggioli}}, \citenamefont {{Cai}}, \citenamefont {{Tian}},\
  and\ \citenamefont {{Zhang}}}]{YangP2023}%
  \BibitemOpen
  \bibfield  {author} {\bibinfo {author} {\bibfnamefont {P.}~\bibnamefont
  {{Yang}}}, \bibinfo {author} {\bibfnamefont {M.}~\bibnamefont {{Baggioli}}},
  \bibinfo {author} {\bibfnamefont {Z.}~\bibnamefont {{Cai}}}, \bibinfo
  {author} {\bibfnamefont {Y.}~\bibnamefont {{Tian}}},\ and\ \bibinfo {author}
  {\bibfnamefont {H.}~\bibnamefont {{Zhang}}},\ }\href
  {https://doi.org/10.1103/PhysRevLett.131.221601} {\bibfield  {journal}
  {\bibinfo  {journal} {\prl}\ }\textbf {\bibinfo {volume} {131}},\ \bibinfo
  {eid} {221601} (\bibinfo {year} {2023})},\ \Eprint
  {https://arxiv.org/abs/2304.02534} {arXiv:2304.02534 [hep-th]} \BibitemShut
  {NoStop}%
\bibitem [{\citenamefont {Rowlands}(1974)}]{1974JAM...367}%
  \BibitemOpen
  \bibfield  {author} {\bibinfo {author} {\bibfnamefont {G.}~\bibnamefont
  {Rowlands}},\ }\href@noop {} {\bibfield  {journal} {\bibinfo  {journal} {IMA
  Journal of Applied Mathematics}\ }\textbf {\bibinfo {volume} {13}},\ \bibinfo
  {pages} {367} (\bibinfo {year} {1974})}\BibitemShut {NoStop}%
\bibitem [{\citenamefont {{Bottman}}\ \emph {et~al.}(2011)\citenamefont
  {{Bottman}}, \citenamefont {{Deconinck}},\ and\ \citenamefont
  {{Nivala}}}]{2011JPhA...44B5201B}%
  \BibitemOpen
  \bibfield  {author} {\bibinfo {author} {\bibfnamefont {N.}~\bibnamefont
  {{Bottman}}}, \bibinfo {author} {\bibfnamefont {B.}~\bibnamefont
  {{Deconinck}}},\ and\ \bibinfo {author} {\bibfnamefont {M.}~\bibnamefont
  {{Nivala}}},\ }\href {https://doi.org/10.1088/1751-8113/44/28/285201}
  {\bibfield  {journal} {\bibinfo  {journal} {Journal of Physics A Mathematical
  General}\ }\textbf {\bibinfo {volume} {44}},\ \bibinfo {eid} {285201}
  (\bibinfo {year} {2011})}\BibitemShut {NoStop}%
\bibitem [{\citenamefont {{Gallay}}\ and\ \citenamefont
  {{Pelinovsky}}(2015)}]{2015JDE...258.3607G}%
  \BibitemOpen
  \bibfield  {author} {\bibinfo {author} {\bibfnamefont {T.}~\bibnamefont
  {{Gallay}}}\ and\ \bibinfo {author} {\bibfnamefont {D.}~\bibnamefont
  {{Pelinovsky}}},\ }\href {https://doi.org/10.1016/j.jde.2015.01.018}
  {\bibfield  {journal} {\bibinfo  {journal} {Journal of Differential
  Equations}\ }\textbf {\bibinfo {volume} {258}},\ \bibinfo {pages} {3607}
  (\bibinfo {year} {2015})},\ \Eprint {https://arxiv.org/abs/1409.6453}
  {arXiv:1409.6453 [math.AP]} \BibitemShut {NoStop}%
\bibitem [{\citenamefont {Gustafson}\ \emph {et~al.}(2017)\citenamefont
  {Gustafson}, \citenamefont {Le~Coz},\ and\ \citenamefont
  {Tsai}}]{2017AMR...431}%
  \BibitemOpen
  \bibfield  {author} {\bibinfo {author} {\bibfnamefont {S.}~\bibnamefont
  {Gustafson}}, \bibinfo {author} {\bibfnamefont {S.}~\bibnamefont {Le~Coz}},\
  and\ \bibinfo {author} {\bibfnamefont {T.-P.}\ \bibnamefont {Tsai}},\
  }\href@noop {} {\bibfield  {journal} {\bibinfo  {journal} {Applied
  Mathematics Research eXpress}\ }\textbf {\bibinfo {volume} {2017}},\ \bibinfo
  {pages} {431} (\bibinfo {year} {2017})}\BibitemShut {NoStop}%
\bibitem [{\citenamefont {Tsuzuki}(1971)}]{Tsuzuki1971}%
  \BibitemOpen
  \bibfield  {author} {\bibinfo {author} {\bibfnamefont {T.}~\bibnamefont
  {Tsuzuki}},\ }\href {https://doi.org/10.1007/BF00628744} {\bibfield
  {journal} {\bibinfo  {journal} {Journal of Low Temperature Physics}\ }\textbf
  {\bibinfo {volume} {4}},\ \bibinfo {pages} {441} (\bibinfo {year}
  {1971})}\BibitemShut {NoStop}%
\bibitem [{\citenamefont {Ammon}\ and\ \citenamefont
  {Erdmenger}(2015)}]{ammon2015gauge}%
  \BibitemOpen
  \bibfield  {author} {\bibinfo {author} {\bibfnamefont {M.}~\bibnamefont
  {Ammon}}\ and\ \bibinfo {author} {\bibfnamefont {J.}~\bibnamefont
  {Erdmenger}},\ }\href@noop {} {\emph {\bibinfo {title} {Gauge/gravity
  duality: Foundations and applications}}}\ (\bibinfo  {publisher} {Cambridge
  University Press},\ \bibinfo {year} {2015})\BibitemShut {NoStop}%
\bibitem [{\citenamefont {Baggioli}()}]{baggioli2019applied}%
  \BibitemOpen
  \bibfield  {author} {\bibinfo {author} {\bibfnamefont {M.}~\bibnamefont
  {Baggioli}},\ }\href@noop {} {\emph {\bibinfo {title} {Applied holography: a
  practical mini-course}}}\ (\bibinfo  {publisher} {Springer})\BibitemShut
  {NoStop}%
\bibitem [{\citenamefont {{Hartnoll}}\ \emph {et~al.}(2008)\citenamefont
  {{Hartnoll}}, \citenamefont {{Herzog}},\ and\ \citenamefont
  {{Horowitz}}}]{2008PhRvL.101c1601H}%
  \BibitemOpen
  \bibfield  {author} {\bibinfo {author} {\bibfnamefont {S.~A.}\ \bibnamefont
  {{Hartnoll}}}, \bibinfo {author} {\bibfnamefont {C.~P.}\ \bibnamefont
  {{Herzog}}},\ and\ \bibinfo {author} {\bibfnamefont {G.~T.}\ \bibnamefont
  {{Horowitz}}},\ }\href {https://doi.org/10.1103/PhysRevLett.101.031601}
  {\bibfield  {journal} {\bibinfo  {journal} {\prl}\ }\textbf {\bibinfo
  {volume} {101}},\ \bibinfo {eid} {031601} (\bibinfo {year} {2008})},\ \Eprint
  {https://arxiv.org/abs/0803.3295} {arXiv:0803.3295 [hep-th]} \BibitemShut
  {NoStop}%
\bibitem [{\citenamefont {Xu}\ \emph {et~al.}(2020{\natexlab{a}})\citenamefont
  {Xu}, \citenamefont {Du}, \citenamefont {Erdmenger}, \citenamefont {Meyer},
  \citenamefont {Tian},\ and\ \citenamefont {Xian}}]{Xu:2019msl}%
  \BibitemOpen
  \bibfield  {author} {\bibinfo {author} {\bibfnamefont {Z.}~\bibnamefont
  {Xu}}, \bibinfo {author} {\bibfnamefont {Y.}~\bibnamefont {Du}}, \bibinfo
  {author} {\bibfnamefont {J.}~\bibnamefont {Erdmenger}}, \bibinfo {author}
  {\bibfnamefont {R.}~\bibnamefont {Meyer}}, \bibinfo {author} {\bibfnamefont
  {Y.}~\bibnamefont {Tian}},\ and\ \bibinfo {author} {\bibfnamefont {Z.-Y.}\
  \bibnamefont {Xian}},\ }\href {https://doi.org/10.1103/PhysRevD.101.086011}
  {\bibfield  {journal} {\bibinfo  {journal} {Phys. Rev. D}\ }\textbf {\bibinfo
  {volume} {101}},\ \bibinfo {pages} {086011} (\bibinfo {year}
  {2020}{\natexlab{a}})},\ \Eprint {https://arxiv.org/abs/1910.09253}
  {arXiv:1910.09253 [hep-th]} \BibitemShut {NoStop}%
\bibitem [{\citenamefont {Jiang}\ \emph {et~al.}(2024)\citenamefont {Jiang},
  \citenamefont {Chen}, \citenamefont {Liu}, \citenamefont {Tian},
  \citenamefont {Xiong}, \citenamefont {Zhang},\ and\ \citenamefont
  {Wang}}]{Jiang:2023yyn}%
  \BibitemOpen
  \bibfield  {author} {\bibinfo {author} {\bibfnamefont {J.-Y.}\ \bibnamefont
  {Jiang}}, \bibinfo {author} {\bibfnamefont {Q.}~\bibnamefont {Chen}},
  \bibinfo {author} {\bibfnamefont {Y.}~\bibnamefont {Liu}}, \bibinfo {author}
  {\bibfnamefont {Y.}~\bibnamefont {Tian}}, \bibinfo {author} {\bibfnamefont
  {W.}~\bibnamefont {Xiong}}, \bibinfo {author} {\bibfnamefont {C.-Y.}\
  \bibnamefont {Zhang}},\ and\ \bibinfo {author} {\bibfnamefont
  {B.}~\bibnamefont {Wang}},\ }\href
  {https://doi.org/10.1007/s11433-023-2231-5} {\bibfield  {journal} {\bibinfo
  {journal} {Sci. China Phys. Mech. Astron.}\ }\textbf {\bibinfo {volume}
  {67}},\ \bibinfo {pages} {220411} (\bibinfo {year} {2024})},\ \Eprint
  {https://arxiv.org/abs/2306.10371} {arXiv:2306.10371 [gr-qc]} \BibitemShut
  {NoStop}%
\bibitem [{\citenamefont {Ammon}\ \emph {et~al.}(2022)\citenamefont {Ammon},
  \citenamefont {Arean}, \citenamefont {Baggioli}, \citenamefont {Gray},\ and\
  \citenamefont {Grieninger}}]{Ammon:2021pyz}%
  \BibitemOpen
  \bibfield  {author} {\bibinfo {author} {\bibfnamefont {M.}~\bibnamefont
  {Ammon}}, \bibinfo {author} {\bibfnamefont {D.}~\bibnamefont {Arean}},
  \bibinfo {author} {\bibfnamefont {M.}~\bibnamefont {Baggioli}}, \bibinfo
  {author} {\bibfnamefont {S.}~\bibnamefont {Gray}},\ and\ \bibinfo {author}
  {\bibfnamefont {S.}~\bibnamefont {Grieninger}},\ }\href
  {https://doi.org/10.1007/JHEP03(2022)015} {\bibfield  {journal} {\bibinfo
  {journal} {JHEP}\ }\textbf {\bibinfo {volume} {03}},\ \bibinfo {pages}
  {015}},\ \Eprint {https://arxiv.org/abs/2111.10305} {arXiv:2111.10305
  [hep-th]} \BibitemShut {NoStop}%
\bibitem [{\citenamefont {Baggioli}\ and\ \citenamefont
  {Gout\'eraux}(2023)}]{RevModPhys.95.011001}%
  \BibitemOpen
  \bibfield  {author} {\bibinfo {author} {\bibfnamefont {M.}~\bibnamefont
  {Baggioli}}\ and\ \bibinfo {author} {\bibfnamefont {B.}~\bibnamefont
  {Gout\'eraux}},\ }\href {https://doi.org/10.1103/RevModPhys.95.011001}
  {\bibfield  {journal} {\bibinfo  {journal} {Rev. Mod. Phys.}\ }\textbf
  {\bibinfo {volume} {95}},\ \bibinfo {pages} {011001} (\bibinfo {year}
  {2023})}\BibitemShut {NoStop}%
\bibitem [{\citenamefont {{Baggioli}}\ \emph {et~al.}(2021)\citenamefont
  {{Baggioli}}, \citenamefont {{Kim}}, \citenamefont {{Li}},\ and\
  \citenamefont {{Li}}}]{2021SCPMA..6470001B}%
  \BibitemOpen
  \bibfield  {author} {\bibinfo {author} {\bibfnamefont {M.}~\bibnamefont
  {{Baggioli}}}, \bibinfo {author} {\bibfnamefont {K.-Y.}\ \bibnamefont
  {{Kim}}}, \bibinfo {author} {\bibfnamefont {L.}~\bibnamefont {{Li}}},\ and\
  \bibinfo {author} {\bibfnamefont {W.-J.}\ \bibnamefont {{Li}}},\ }\href
  {https://doi.org/10.1007/s11433-021-1681-8} {\bibfield  {journal} {\bibinfo
  {journal} {Science China Physics, Mechanics, and Astronomy}\ }\textbf
  {\bibinfo {volume} {64}},\ \bibinfo {eid} {270001} (\bibinfo {year}
  {2021})},\ \Eprint {https://arxiv.org/abs/2101.01892} {arXiv:2101.01892
  [hep-th]} \BibitemShut {NoStop}%
\bibitem [{\citenamefont {Baggioli}\ and\ \citenamefont
  {Frangi}(2022)}]{Baggioli:2022aft}%
  \BibitemOpen
  \bibfield  {author} {\bibinfo {author} {\bibfnamefont {M.}~\bibnamefont
  {Baggioli}}\ and\ \bibinfo {author} {\bibfnamefont {G.}~\bibnamefont
  {Frangi}},\ }\href {https://doi.org/10.1007/JHEP06(2022)152} {\bibfield
  {journal} {\bibinfo  {journal} {JHEP}\ }\textbf {\bibinfo {volume} {06}},\
  \bibinfo {pages} {152}},\ \Eprint {https://arxiv.org/abs/2202.03745}
  {arXiv:2202.03745 [hep-th]} \BibitemShut {NoStop}%
\bibitem [{\citenamefont {Liu}\ and\ \citenamefont {Sonner}(2019)}]{Liu_2020}%
  \BibitemOpen
  \bibfield  {author} {\bibinfo {author} {\bibfnamefont {H.}~\bibnamefont
  {Liu}}\ and\ \bibinfo {author} {\bibfnamefont {J.}~\bibnamefont {Sonner}},\
  }\href {https://doi.org/10.1088/1361-6633/ab4f91} {\bibfield  {journal}
  {\bibinfo  {journal} {Reports on Progress in Physics}\ }\textbf {\bibinfo
  {volume} {83}},\ \bibinfo {pages} {016001} (\bibinfo {year}
  {2019})}\BibitemShut {NoStop}%
\bibitem [{\citenamefont {Chen}\ \emph {et~al.}(2025)\citenamefont {Chen},
  \citenamefont {Liu}, \citenamefont {Tian}, \citenamefont {Wu},\ and\
  \citenamefont {Zhang}}]{Chen:2024pyy}%
  \BibitemOpen
  \bibfield  {author} {\bibinfo {author} {\bibfnamefont {Q.}~\bibnamefont
  {Chen}}, \bibinfo {author} {\bibfnamefont {Y.}~\bibnamefont {Liu}}, \bibinfo
  {author} {\bibfnamefont {Y.}~\bibnamefont {Tian}}, \bibinfo {author}
  {\bibfnamefont {X.}~\bibnamefont {Wu}},\ and\ \bibinfo {author}
  {\bibfnamefont {H.}~\bibnamefont {Zhang}},\ }\href
  {https://doi.org/10.1007/s11433-025-2633-y} {\bibfield  {journal} {\bibinfo
  {journal} {Sci. China Phys. Mech. Astron.}\ }\textbf {\bibinfo {volume}
  {68}},\ \bibinfo {pages} {260414} (\bibinfo {year} {2025})},\ \Eprint
  {https://arxiv.org/abs/2408.09679} {arXiv:2408.09679 [hep-th]} \BibitemShut
  {NoStop}%
\bibitem [{\citenamefont {Yang}\ \emph {et~al.}(2025)\citenamefont {Yang},
  \citenamefont {Lan}, \citenamefont {Tian}, \citenamefont {Yan},\ and\
  \citenamefont {Zhang}}]{Yang:2024vga}%
  \BibitemOpen
  \bibfield  {author} {\bibinfo {author} {\bibfnamefont {P.}~\bibnamefont
  {Yang}}, \bibinfo {author} {\bibfnamefont {S.}~\bibnamefont {Lan}}, \bibinfo
  {author} {\bibfnamefont {Y.}~\bibnamefont {Tian}}, \bibinfo {author}
  {\bibfnamefont {Y.-K.}\ \bibnamefont {Yan}},\ and\ \bibinfo {author}
  {\bibfnamefont {H.}~\bibnamefont {Zhang}},\ }\href
  {https://doi.org/10.1103/lw69-gp12} {\bibfield  {journal} {\bibinfo
  {journal} {Phys. Rev. D}\ }\textbf {\bibinfo {volume} {112}},\ \bibinfo
  {pages} {026032} (\bibinfo {year} {2025})},\ \Eprint
  {https://arxiv.org/abs/2412.18320} {arXiv:2412.18320 [hep-th]} \BibitemShut
  {NoStop}%
\bibitem [{\citenamefont {Tian}\ \emph {et~al.}(2023)\citenamefont {Tian},
  \citenamefont {Wu},\ and\ \citenamefont {Zhang}}]{Tian__2023}%
  \BibitemOpen
  \bibfield  {author} {\bibinfo {author} {\bibfnamefont {Y.}~\bibnamefont
  {Tian}}, \bibinfo {author} {\bibfnamefont {X.-N.}\ \bibnamefont {Wu}},\ and\
  \bibinfo {author} {\bibfnamefont {H.}~\bibnamefont {Zhang}},\ }\href
  {https://doi.org/10.1088/0256-307x/40/10/100402} {\bibfield  {journal}
  {\bibinfo  {journal} {Chinese Physics Letters}\ }\textbf {\bibinfo {volume}
  {40}},\ \bibinfo {pages} {100402} (\bibinfo {year} {2023})}\BibitemShut
  {NoStop}%
\bibitem [{\citenamefont {Li}\ \emph {et~al.}(2020)\citenamefont {Li},
  \citenamefont {Nie},\ and\ \citenamefont {Tian}}]{Li_2020}%
  \BibitemOpen
  \bibfield  {author} {\bibinfo {author} {\bibfnamefont {X.}~\bibnamefont
  {Li}}, \bibinfo {author} {\bibfnamefont {Z.-Y.}\ \bibnamefont {Nie}},\ and\
  \bibinfo {author} {\bibfnamefont {Y.}~\bibnamefont {Tian}},\ }\bibfield
  {journal} {\bibinfo  {journal} {Journal of High Energy Physics}\ }\textbf
  {\bibinfo {volume} {2020}},\ \href {https://doi.org/10.1007/jhep09(2020)063}
  {10.1007/jhep09(2020)063} (\bibinfo {year} {2020})\BibitemShut {NoStop}%
\bibitem [{\citenamefont {{Du}}\ \emph {et~al.}(2016)\citenamefont {{Du}},
  \citenamefont {{Lan}}, \citenamefont {{Tian}},\ and\ \citenamefont
  {{Zhang}}}]{2016JHEP...01..016D}%
  \BibitemOpen
  \bibfield  {author} {\bibinfo {author} {\bibfnamefont {Y.}~\bibnamefont
  {{Du}}}, \bibinfo {author} {\bibfnamefont {S.-Q.}\ \bibnamefont {{Lan}}},
  \bibinfo {author} {\bibfnamefont {Y.}~\bibnamefont {{Tian}}},\ and\ \bibinfo
  {author} {\bibfnamefont {H.}~\bibnamefont {{Zhang}}},\ }\href
  {https://doi.org/10.1007/JHEP01(2016)016} {\bibfield  {journal} {\bibinfo
  {journal} {Journal of High Energy Physics}\ }\textbf {\bibinfo {volume}
  {2016}},\ \bibinfo {eid} {16} (\bibinfo {year} {2016})},\ \Eprint
  {https://arxiv.org/abs/1511.07179} {arXiv:1511.07179 [hep-th]} \BibitemShut
  {NoStop}%
\bibitem [{\citenamefont {{Guo}}\ \emph {et~al.}(2020)\citenamefont {{Guo}},
  \citenamefont {{Keski-Vakkuri}}, \citenamefont {{Liu}}, \citenamefont
  {{Tian}},\ and\ \citenamefont {{Zhang}}}]{2020PhRvL.124c1601G}%
  \BibitemOpen
  \bibfield  {author} {\bibinfo {author} {\bibfnamefont {M.}~\bibnamefont
  {{Guo}}}, \bibinfo {author} {\bibfnamefont {E.}~\bibnamefont
  {{Keski-Vakkuri}}}, \bibinfo {author} {\bibfnamefont {H.}~\bibnamefont
  {{Liu}}}, \bibinfo {author} {\bibfnamefont {Y.}~\bibnamefont {{Tian}}},\ and\
  \bibinfo {author} {\bibfnamefont {H.}~\bibnamefont {{Zhang}}},\ }\href
  {https://doi.org/10.1103/PhysRevLett.124.031601} {\bibfield  {journal}
  {\bibinfo  {journal} {\prl}\ }\textbf {\bibinfo {volume} {124}},\ \bibinfo
  {eid} {031601} (\bibinfo {year} {2020})}\BibitemShut {NoStop}%
\bibitem [{\citenamefont {{Li}}\ \emph {et~al.}(2020)\citenamefont {{Li}},
  \citenamefont {{Tian}},\ and\ \citenamefont {{Zhang}}}]{2020JHEP...02..104L}%
  \BibitemOpen
  \bibfield  {author} {\bibinfo {author} {\bibfnamefont {X.}~\bibnamefont
  {{Li}}}, \bibinfo {author} {\bibfnamefont {Y.}~\bibnamefont {{Tian}}},\ and\
  \bibinfo {author} {\bibfnamefont {H.}~\bibnamefont {{Zhang}}},\ }\href
  {https://doi.org/10.1007/JHEP02(2020)104} {\bibfield  {journal} {\bibinfo
  {journal} {Journal of High Energy Physics}\ }\textbf {\bibinfo {volume}
  {2020}},\ \bibinfo {eid} {104} (\bibinfo {year} {2020})},\ \Eprint
  {https://arxiv.org/abs/1904.05497} {arXiv:1904.05497 [hep-th]} \BibitemShut
  {NoStop}%
\bibitem [{\citenamefont {{Yang}}\ \emph {et~al.}(2021)\citenamefont {{Yang}},
  \citenamefont {{Li}},\ and\ \citenamefont {{Tian}}}]{2021JHEP...11..190Y}%
  \BibitemOpen
  \bibfield  {author} {\bibinfo {author} {\bibfnamefont {P.}~\bibnamefont
  {{Yang}}}, \bibinfo {author} {\bibfnamefont {X.}~\bibnamefont {{Li}}},\ and\
  \bibinfo {author} {\bibfnamefont {Y.}~\bibnamefont {{Tian}}},\ }\href
  {https://doi.org/10.1007/JHEP11(2021)190} {\bibfield  {journal} {\bibinfo
  {journal} {Journal of High Energy Physics}\ }\textbf {\bibinfo {volume}
  {2021}},\ \bibinfo {eid} {190} (\bibinfo {year} {2021})},\ \Eprint
  {https://arxiv.org/abs/2109.09080} {arXiv:2109.09080 [hep-th]} \BibitemShut
  {NoStop}%
\bibitem [{\citenamefont {{Lan}}\ \emph {et~al.}(2020)\citenamefont {{Lan}},
  \citenamefont {{Liu}}, \citenamefont {{Tian}},\ and\ \citenamefont
  {{Zhang}}}]{2020arXiv201006232L}%
  \BibitemOpen
  \bibfield  {author} {\bibinfo {author} {\bibfnamefont {S.}~\bibnamefont
  {{Lan}}}, \bibinfo {author} {\bibfnamefont {H.}~\bibnamefont {{Liu}}},
  \bibinfo {author} {\bibfnamefont {Y.}~\bibnamefont {{Tian}}},\ and\ \bibinfo
  {author} {\bibfnamefont {H.}~\bibnamefont {{Zhang}}},\ }\href
  {https://doi.org/10.48550/arXiv.2010.06232} {\bibfield  {journal} {\bibinfo
  {journal} {arXiv e-prints}\ ,\ \bibinfo {eid} {arXiv:2010.06232}} (\bibinfo
  {year} {2020})},\ \Eprint {https://arxiv.org/abs/2010.06232}
  {arXiv:2010.06232 [hep-th]} \BibitemShut {NoStop}%
\bibitem [{\citenamefont {{Arean}}\ and\ \citenamefont
  {{Garcia-Fari{\~n}a}}(2024)}]{2024arXiv241013584A}%
  \BibitemOpen
  \bibfield  {author} {\bibinfo {author} {\bibfnamefont {D.}~\bibnamefont
  {{Arean}}}\ and\ \bibinfo {author} {\bibfnamefont {D.}~\bibnamefont
  {{Garcia-Fari{\~n}a}}},\ }\href {https://doi.org/10.48550/arXiv.2410.13584}
  {\bibfield  {journal} {\bibinfo  {journal} {arXiv e-prints}\ ,\ \bibinfo
  {eid} {arXiv:2410.13584}} (\bibinfo {year} {2024})},\ \Eprint
  {https://arxiv.org/abs/2410.13584} {arXiv:2410.13584 [hep-th]} \BibitemShut
  {NoStop}%
\bibitem [{\citenamefont {Amado}\ \emph {et~al.}(2009)\citenamefont {Amado},
  \citenamefont {Kaminski},\ and\ \citenamefont {Landsteiner}}]{Amado:2009ts}%
  \BibitemOpen
  \bibfield  {author} {\bibinfo {author} {\bibfnamefont {I.}~\bibnamefont
  {Amado}}, \bibinfo {author} {\bibfnamefont {M.}~\bibnamefont {Kaminski}},\
  and\ \bibinfo {author} {\bibfnamefont {K.}~\bibnamefont {Landsteiner}},\
  }\href {https://doi.org/10.1088/1126-6708/2009/05/021} {\bibfield  {journal}
  {\bibinfo  {journal} {JHEP}\ }\textbf {\bibinfo {volume} {05}},\ \bibinfo
  {pages} {021}},\ \Eprint {https://arxiv.org/abs/0903.2209} {arXiv:0903.2209
  [hep-th]} \BibitemShut {NoStop}%
\bibitem [{\citenamefont {Arean}\ \emph {et~al.}(2021)\citenamefont {Arean},
  \citenamefont {Baggioli}, \citenamefont {Grieninger},\ and\ \citenamefont
  {Landsteiner}}]{Arean:2021tks}%
  \BibitemOpen
  \bibfield  {author} {\bibinfo {author} {\bibfnamefont {D.}~\bibnamefont
  {Arean}}, \bibinfo {author} {\bibfnamefont {M.}~\bibnamefont {Baggioli}},
  \bibinfo {author} {\bibfnamefont {S.}~\bibnamefont {Grieninger}},\ and\
  \bibinfo {author} {\bibfnamefont {K.}~\bibnamefont {Landsteiner}},\ }\href
  {https://doi.org/10.1007/JHEP11(2021)206} {\bibfield  {journal} {\bibinfo
  {journal} {JHEP}\ }\textbf {\bibinfo {volume} {11}},\ \bibinfo {pages}
  {206}},\ \Eprint {https://arxiv.org/abs/2107.08802} {arXiv:2107.08802
  [hep-th]} \BibitemShut {NoStop}%
\bibitem [{\citenamefont {{Lan}}\ \emph
  {et~al.}(2023{\natexlab{a}})\citenamefont {{Lan}}, \citenamefont {{Li}},
  \citenamefont {{Mo}}, \citenamefont {{Tian}}, \citenamefont {{Yan}},
  \citenamefont {{Yang}},\ and\ \citenamefont {{Zhang}}}]{2023JHEP...05..223L}%
  \BibitemOpen
  \bibfield  {author} {\bibinfo {author} {\bibfnamefont {S.}~\bibnamefont
  {{Lan}}}, \bibinfo {author} {\bibfnamefont {X.}~\bibnamefont {{Li}}},
  \bibinfo {author} {\bibfnamefont {J.}~\bibnamefont {{Mo}}}, \bibinfo {author}
  {\bibfnamefont {Y.}~\bibnamefont {{Tian}}}, \bibinfo {author} {\bibfnamefont
  {Y.-K.}\ \bibnamefont {{Yan}}}, \bibinfo {author} {\bibfnamefont
  {P.}~\bibnamefont {{Yang}}},\ and\ \bibinfo {author} {\bibfnamefont
  {H.}~\bibnamefont {{Zhang}}},\ }\href
  {https://doi.org/10.1007/JHEP05(2023)223} {\bibfield  {journal} {\bibinfo
  {journal} {Journal of High Energy Physics}\ }\textbf {\bibinfo {volume}
  {2023}},\ \bibinfo {eid} {223} (\bibinfo {year} {2023}{\natexlab{a}})},\
  \Eprint {https://arxiv.org/abs/2301.03203} {arXiv:2301.03203 [hep-th]}
  \BibitemShut {NoStop}%
\bibitem [{\citenamefont {{Lan}}\ \emph
  {et~al.}(2023{\natexlab{b}})\citenamefont {{Lan}}, \citenamefont {{Li}},
  \citenamefont {{Tian}}, \citenamefont {{Yang}},\ and\ \citenamefont
  {{Zhang}}}]{2023PhRvL.131v1602L}%
  \BibitemOpen
  \bibfield  {author} {\bibinfo {author} {\bibfnamefont {S.}~\bibnamefont
  {{Lan}}}, \bibinfo {author} {\bibfnamefont {X.}~\bibnamefont {{Li}}},
  \bibinfo {author} {\bibfnamefont {Y.}~\bibnamefont {{Tian}}}, \bibinfo
  {author} {\bibfnamefont {P.}~\bibnamefont {{Yang}}},\ and\ \bibinfo {author}
  {\bibfnamefont {H.}~\bibnamefont {{Zhang}}},\ }\href
  {https://doi.org/10.1103/PhysRevLett.131.221602} {\bibfield  {journal}
  {\bibinfo  {journal} {\prl}\ }\textbf {\bibinfo {volume} {131}},\ \bibinfo
  {eid} {221602} (\bibinfo {year} {2023}{\natexlab{b}})},\ \Eprint
  {https://arxiv.org/abs/2311.01316} {arXiv:2311.01316 [cond-mat.quant-gas]}
  \BibitemShut {NoStop}%
\bibitem [{\citenamefont {{Yan}}\ \emph {et~al.}(2023)\citenamefont {{Yan}},
  \citenamefont {{Lan}}, \citenamefont {{Tian}}, \citenamefont {{Yang}},
  \citenamefont {{Yao}},\ and\ \citenamefont {{Zhang}}}]{2023PhRvD.107l1901Y}%
  \BibitemOpen
  \bibfield  {author} {\bibinfo {author} {\bibfnamefont {Y.-K.}\ \bibnamefont
  {{Yan}}}, \bibinfo {author} {\bibfnamefont {S.}~\bibnamefont {{Lan}}},
  \bibinfo {author} {\bibfnamefont {Y.}~\bibnamefont {{Tian}}}, \bibinfo
  {author} {\bibfnamefont {P.}~\bibnamefont {{Yang}}}, \bibinfo {author}
  {\bibfnamefont {S.}~\bibnamefont {{Yao}}},\ and\ \bibinfo {author}
  {\bibfnamefont {H.}~\bibnamefont {{Zhang}}},\ }\href
  {https://doi.org/10.1103/PhysRevD.107.L121901} {\bibfield  {journal}
  {\bibinfo  {journal} {\prd}\ }\textbf {\bibinfo {volume} {107}},\ \bibinfo
  {eid} {L121901} (\bibinfo {year} {2023})},\ \Eprint
  {https://arxiv.org/abs/2207.02814} {arXiv:2207.02814 [hep-th]} \BibitemShut
  {NoStop}%
\bibitem [{\citenamefont {{Zeng}}\ \emph {et~al.}(2021)\citenamefont {{Zeng}},
  \citenamefont {{Xia}},\ and\ \citenamefont {{Zhang}}}]{2021JHEP...03..136Z}%
  \BibitemOpen
  \bibfield  {author} {\bibinfo {author} {\bibfnamefont {H.-B.}\ \bibnamefont
  {{Zeng}}}, \bibinfo {author} {\bibfnamefont {C.-Y.}\ \bibnamefont {{Xia}}},\
  and\ \bibinfo {author} {\bibfnamefont {H.-Q.}\ \bibnamefont {{Zhang}}},\
  }\href {https://doi.org/10.1007/JHEP03(2021)136} {\bibfield  {journal}
  {\bibinfo  {journal} {Journal of High Energy Physics}\ }\textbf {\bibinfo
  {volume} {2021}},\ \bibinfo {eid} {136} (\bibinfo {year} {2021})}\BibitemShut
  {NoStop}%
\bibitem [{\citenamefont {{Xia}}\ \emph {et~al.}(2019)\citenamefont {{Xia}},
  \citenamefont {{Zeng}}, \citenamefont {{Zhang}}, \citenamefont {{Nie}},
  \citenamefont {{Tian}},\ and\ \citenamefont {{Li}}}]{2019PhRvD.100f1901X}%
  \BibitemOpen
  \bibfield  {author} {\bibinfo {author} {\bibfnamefont {C.-Y.}\ \bibnamefont
  {{Xia}}}, \bibinfo {author} {\bibfnamefont {H.-B.}\ \bibnamefont {{Zeng}}},
  \bibinfo {author} {\bibfnamefont {H.-Q.}\ \bibnamefont {{Zhang}}}, \bibinfo
  {author} {\bibfnamefont {Z.-Y.}\ \bibnamefont {{Nie}}}, \bibinfo {author}
  {\bibfnamefont {Y.}~\bibnamefont {{Tian}}},\ and\ \bibinfo {author}
  {\bibfnamefont {X.}~\bibnamefont {{Li}}},\ }\href
  {https://doi.org/10.1103/PhysRevD.100.061901} {\bibfield  {journal} {\bibinfo
   {journal} {\prd}\ }\textbf {\bibinfo {volume} {100}},\ \bibinfo {eid}
  {061901} (\bibinfo {year} {2019})},\ \Eprint
  {https://arxiv.org/abs/1904.10925} {arXiv:1904.10925 [hep-th]} \BibitemShut
  {NoStop}%
\bibitem [{\citenamefont {Korpel}\ and\ \citenamefont
  {Banerjee}(1981)}]{KORPEL1981113}%
  \BibitemOpen
  \bibfield  {author} {\bibinfo {author} {\bibfnamefont {A.}~\bibnamefont
  {Korpel}}\ and\ \bibinfo {author} {\bibfnamefont {P.}~\bibnamefont
  {Banerjee}},\ }\href
  {https://doi.org/https://doi.org/10.1016/0375-9601(81)90925-7} {\bibfield
  {journal} {\bibinfo  {journal} {Physics Letters A}\ }\textbf {\bibinfo
  {volume} {82}},\ \bibinfo {pages} {113} (\bibinfo {year} {1981})}\BibitemShut
  {NoStop}%
\bibitem [{\citenamefont {Yatsuta}\ \emph {et~al.}(2020)\citenamefont
  {Yatsuta}, \citenamefont {Malomed},\ and\ \citenamefont
  {Yakimenko}}]{PhysRevResearch.2.043065}%
  \BibitemOpen
  \bibfield  {author} {\bibinfo {author} {\bibfnamefont {I.}~\bibnamefont
  {Yatsuta}}, \bibinfo {author} {\bibfnamefont {B.}~\bibnamefont {Malomed}},\
  and\ \bibinfo {author} {\bibfnamefont {A.}~\bibnamefont {Yakimenko}},\ }\href
  {https://doi.org/10.1103/PhysRevResearch.2.043065} {\bibfield  {journal}
  {\bibinfo  {journal} {Phys. Rev. Res.}\ }\textbf {\bibinfo {volume} {2}},\
  \bibinfo {pages} {043065} (\bibinfo {year} {2020})}\BibitemShut {NoStop}%
\bibitem [{\citenamefont {Bland}\ \emph {et~al.}(2022)\citenamefont {Bland},
  \citenamefont {Yatsuta}, \citenamefont {Edwards}, \citenamefont {Nikolaieva},
  \citenamefont {Oliinyk}, \citenamefont {Yakimenko},\ and\ \citenamefont
  {Proukakis}}]{PhysRevResearch.4.043171}%
  \BibitemOpen
  \bibfield  {author} {\bibinfo {author} {\bibfnamefont {T.}~\bibnamefont
  {Bland}}, \bibinfo {author} {\bibfnamefont {I.~V.}\ \bibnamefont {Yatsuta}},
  \bibinfo {author} {\bibfnamefont {M.}~\bibnamefont {Edwards}}, \bibinfo
  {author} {\bibfnamefont {Y.~O.}\ \bibnamefont {Nikolaieva}}, \bibinfo
  {author} {\bibfnamefont {A.~O.}\ \bibnamefont {Oliinyk}}, \bibinfo {author}
  {\bibfnamefont {A.~I.}\ \bibnamefont {Yakimenko}},\ and\ \bibinfo {author}
  {\bibfnamefont {N.~P.}\ \bibnamefont {Proukakis}},\ }\href
  {https://doi.org/10.1103/PhysRevResearch.4.043171} {\bibfield  {journal}
  {\bibinfo  {journal} {Phys. Rev. Res.}\ }\textbf {\bibinfo {volume} {4}},\
  \bibinfo {pages} {043171} (\bibinfo {year} {2022})}\BibitemShut {NoStop}%
\bibitem [{\citenamefont {Haldane}(1981)}]{PhysRevLett.47.1840}%
  \BibitemOpen
  \bibfield  {author} {\bibinfo {author} {\bibfnamefont {F.~D.~M.}\
  \bibnamefont {Haldane}},\ }\href
  {https://doi.org/10.1103/PhysRevLett.47.1840} {\bibfield  {journal} {\bibinfo
   {journal} {Phys. Rev. Lett.}\ }\textbf {\bibinfo {volume} {47}},\ \bibinfo
  {pages} {1840} (\bibinfo {year} {1981})}\BibitemShut {NoStop}%
\bibitem [{\citenamefont {Eckel}\ \emph {et~al.}(2018)\citenamefont {Eckel},
  \citenamefont {Kumar}, \citenamefont {Jacobson}, \citenamefont {Spielman},\
  and\ \citenamefont {Campbell}}]{PhysRevX.8.021021}%
  \BibitemOpen
  \bibfield  {author} {\bibinfo {author} {\bibfnamefont {S.}~\bibnamefont
  {Eckel}}, \bibinfo {author} {\bibfnamefont {A.}~\bibnamefont {Kumar}},
  \bibinfo {author} {\bibfnamefont {T.}~\bibnamefont {Jacobson}}, \bibinfo
  {author} {\bibfnamefont {I.~B.}\ \bibnamefont {Spielman}},\ and\ \bibinfo
  {author} {\bibfnamefont {G.~K.}\ \bibnamefont {Campbell}},\ }\href
  {https://doi.org/10.1103/PhysRevX.8.021021} {\bibfield  {journal} {\bibinfo
  {journal} {Phys. Rev. X}\ }\textbf {\bibinfo {volume} {8}},\ \bibinfo {pages}
  {021021} (\bibinfo {year} {2018})}\BibitemShut {NoStop}%
\bibitem [{\citenamefont {Ryu}\ \emph {et~al.}(2007)\citenamefont {Ryu},
  \citenamefont {Andersen}, \citenamefont {Clad\'e}, \citenamefont {Natarajan},
  \citenamefont {Helmerson},\ and\ \citenamefont
  {Phillips}}]{PhysRevLett.99.260401}%
  \BibitemOpen
  \bibfield  {author} {\bibinfo {author} {\bibfnamefont {C.}~\bibnamefont
  {Ryu}}, \bibinfo {author} {\bibfnamefont {M.~F.}\ \bibnamefont {Andersen}},
  \bibinfo {author} {\bibfnamefont {P.}~\bibnamefont {Clad\'e}}, \bibinfo
  {author} {\bibfnamefont {V.}~\bibnamefont {Natarajan}}, \bibinfo {author}
  {\bibfnamefont {K.}~\bibnamefont {Helmerson}},\ and\ \bibinfo {author}
  {\bibfnamefont {W.~D.}\ \bibnamefont {Phillips}},\ }\href
  {https://doi.org/10.1103/PhysRevLett.99.260401} {\bibfield  {journal}
  {\bibinfo  {journal} {Phys. Rev. Lett.}\ }\textbf {\bibinfo {volume} {99}},\
  \bibinfo {pages} {260401} (\bibinfo {year} {2007})}\BibitemShut {NoStop}%
\bibitem [{\citenamefont {{Pezz{\`e}}}\ \emph {et~al.}(2024)\citenamefont
  {{Pezz{\`e}}}, \citenamefont {{Xhani}}, \citenamefont {{Daix}}, \citenamefont
  {{Grani}}, \citenamefont {{Donelli}}, \citenamefont {{Scazza}}, \citenamefont
  {{Hernandez-Rajkov}}, \citenamefont {{Kwon}}, \citenamefont {{Del Pace}},\
  and\ \citenamefont {{Roati}}}]{2024NatCo..15.4831P}%
  \BibitemOpen
  \bibfield  {author} {\bibinfo {author} {\bibfnamefont {L.}~\bibnamefont
  {{Pezz{\`e}}}}, \bibinfo {author} {\bibfnamefont {K.}~\bibnamefont
  {{Xhani}}}, \bibinfo {author} {\bibfnamefont {C.}~\bibnamefont {{Daix}}},
  \bibinfo {author} {\bibfnamefont {N.}~\bibnamefont {{Grani}}}, \bibinfo
  {author} {\bibfnamefont {B.}~\bibnamefont {{Donelli}}}, \bibinfo {author}
  {\bibfnamefont {F.}~\bibnamefont {{Scazza}}}, \bibinfo {author}
  {\bibfnamefont {D.}~\bibnamefont {{Hernandez-Rajkov}}}, \bibinfo {author}
  {\bibfnamefont {W.~J.}\ \bibnamefont {{Kwon}}}, \bibinfo {author}
  {\bibfnamefont {G.}~\bibnamefont {{Del Pace}}},\ and\ \bibinfo {author}
  {\bibfnamefont {G.}~\bibnamefont {{Roati}}},\ }\href
  {https://doi.org/10.1038/s41467-024-47759-7} {\bibfield  {journal} {\bibinfo
  {journal} {Nature Communications}\ }\textbf {\bibinfo {volume} {15}},\
  \bibinfo {eid} {4831} (\bibinfo {year} {2024})},\ \Eprint
  {https://arxiv.org/abs/2311.05523} {arXiv:2311.05523 [cond-mat.quant-gas]}
  \BibitemShut {NoStop}%
\bibitem [{\citenamefont {Busch}\ and\ \citenamefont
  {Anglin}(2000)}]{PhysRevLett.84.2298}%
  \BibitemOpen
  \bibfield  {author} {\bibinfo {author} {\bibfnamefont {T.}~\bibnamefont
  {Busch}}\ and\ \bibinfo {author} {\bibfnamefont {J.~R.}\ \bibnamefont
  {Anglin}},\ }\href {https://doi.org/10.1103/PhysRevLett.84.2298} {\bibfield
  {journal} {\bibinfo  {journal} {Phys. Rev. Lett.}\ }\textbf {\bibinfo
  {volume} {84}},\ \bibinfo {pages} {2298} (\bibinfo {year}
  {2000})}\BibitemShut {NoStop}%
\bibitem [{\citenamefont {Wu}\ and\ \citenamefont
  {Niu}(2001)}]{PhysRevA.64.061603}%
  \BibitemOpen
  \bibfield  {author} {\bibinfo {author} {\bibfnamefont {B.}~\bibnamefont
  {Wu}}\ and\ \bibinfo {author} {\bibfnamefont {Q.}~\bibnamefont {Niu}},\
  }\href {https://doi.org/10.1103/PhysRevA.64.061603} {\bibfield  {journal}
  {\bibinfo  {journal} {Phys. Rev. A}\ }\textbf {\bibinfo {volume} {64}},\
  \bibinfo {pages} {061603} (\bibinfo {year} {2001})}\BibitemShut {NoStop}%
\bibitem [{\citenamefont {Hui}\ \emph {et~al.}(2011)\citenamefont {Hui},
  \citenamefont {Barnett}, \citenamefont {Sensarma},\ and\ \citenamefont
  {Das~Sarma}}]{PhysRevA.84.043615}%
  \BibitemOpen
  \bibfield  {author} {\bibinfo {author} {\bibfnamefont {H.-Y.}\ \bibnamefont
  {Hui}}, \bibinfo {author} {\bibfnamefont {R.}~\bibnamefont {Barnett}},
  \bibinfo {author} {\bibfnamefont {R.}~\bibnamefont {Sensarma}},\ and\
  \bibinfo {author} {\bibfnamefont {S.}~\bibnamefont {Das~Sarma}},\ }\href
  {https://doi.org/10.1103/PhysRevA.84.043615} {\bibfield  {journal} {\bibinfo
  {journal} {Phys. Rev. A}\ }\textbf {\bibinfo {volume} {84}},\ \bibinfo
  {pages} {043615} (\bibinfo {year} {2011})}\BibitemShut {NoStop}%
\bibitem [{\citenamefont {Ozawa}\ \emph {et~al.}(2013)\citenamefont {Ozawa},
  \citenamefont {Pitaevskii},\ and\ \citenamefont
  {Stringari}}]{PhysRevA.87.063610}%
  \BibitemOpen
  \bibfield  {author} {\bibinfo {author} {\bibfnamefont {T.}~\bibnamefont
  {Ozawa}}, \bibinfo {author} {\bibfnamefont {L.~P.}\ \bibnamefont
  {Pitaevskii}},\ and\ \bibinfo {author} {\bibfnamefont {S.}~\bibnamefont
  {Stringari}},\ }\href {https://doi.org/10.1103/PhysRevA.87.063610} {\bibfield
   {journal} {\bibinfo  {journal} {Phys. Rev. A}\ }\textbf {\bibinfo {volume}
  {87}},\ \bibinfo {pages} {063610} (\bibinfo {year} {2013})}\BibitemShut
  {NoStop}%
\bibitem [{\citenamefont {Lyu}\ \emph {et~al.}(2024)\citenamefont {Lyu},
  \citenamefont {Chen}, \citenamefont {Zhu},\ and\ \citenamefont
  {Zhang}}]{PhysRevResearch.6.023048}%
  \BibitemOpen
  \bibfield  {author} {\bibinfo {author} {\bibfnamefont {H.}~\bibnamefont
  {Lyu}}, \bibinfo {author} {\bibfnamefont {Y.}~\bibnamefont {Chen}}, \bibinfo
  {author} {\bibfnamefont {Q.}~\bibnamefont {Zhu}},\ and\ \bibinfo {author}
  {\bibfnamefont {Y.}~\bibnamefont {Zhang}},\ }\href
  {https://doi.org/10.1103/PhysRevResearch.6.023048} {\bibfield  {journal}
  {\bibinfo  {journal} {Phys. Rev. Res.}\ }\textbf {\bibinfo {volume} {6}},\
  \bibinfo {pages} {023048} (\bibinfo {year} {2024})}\BibitemShut {NoStop}%
\bibitem [{\citenamefont {Wang}\ and\ \citenamefont
  {Zhang}(2024)}]{PhysRevA.110.033307}%
  \BibitemOpen
  \bibfield  {author} {\bibinfo {author} {\bibfnamefont {C.}~\bibnamefont
  {Wang}}\ and\ \bibinfo {author} {\bibfnamefont {Y.}~\bibnamefont {Zhang}},\
  }\href {https://doi.org/10.1103/PhysRevA.110.033307} {\bibfield  {journal}
  {\bibinfo  {journal} {Phys. Rev. A}\ }\textbf {\bibinfo {volume} {110}},\
  \bibinfo {pages} {033307} (\bibinfo {year} {2024})}\BibitemShut {NoStop}%
\bibitem [{\citenamefont {Mueller}(2002)}]{PhysRevA.66.063603}%
  \BibitemOpen
  \bibfield  {author} {\bibinfo {author} {\bibfnamefont {E.~J.}\ \bibnamefont
  {Mueller}},\ }\href {https://doi.org/10.1103/PhysRevA.66.063603} {\bibfield
  {journal} {\bibinfo  {journal} {Phys. Rev. A}\ }\textbf {\bibinfo {volume}
  {66}},\ \bibinfo {pages} {063603} (\bibinfo {year} {2002})}\BibitemShut
  {NoStop}%
\bibitem [{\citenamefont {{Zhao}}\ \emph {et~al.}(2024)\citenamefont {{Zhao}},
  \citenamefont {{Nie}}, \citenamefont {{Zhao}}, \citenamefont {{Zeng}},
  \citenamefont {{Tian}},\ and\ \citenamefont
  {{Baggioli}}}]{2024JHEP...02..184Z}%
  \BibitemOpen
  \bibfield  {author} {\bibinfo {author} {\bibfnamefont {X.}~\bibnamefont
  {{Zhao}}}, \bibinfo {author} {\bibfnamefont {Z.-Y.}\ \bibnamefont {{Nie}}},
  \bibinfo {author} {\bibfnamefont {Z.-Q.}\ \bibnamefont {{Zhao}}}, \bibinfo
  {author} {\bibfnamefont {H.-B.}\ \bibnamefont {{Zeng}}}, \bibinfo {author}
  {\bibfnamefont {Y.}~\bibnamefont {{Tian}}},\ and\ \bibinfo {author}
  {\bibfnamefont {M.}~\bibnamefont {{Baggioli}}},\ }\href
  {https://doi.org/10.1007/JHEP02(2024)184} {\bibfield  {journal} {\bibinfo
  {journal} {Journal of High Energy Physics}\ }\textbf {\bibinfo {volume}
  {2024}},\ \bibinfo {eid} {184} (\bibinfo {year} {2024})},\ \Eprint
  {https://arxiv.org/abs/2311.08277} {arXiv:2311.08277 [hep-th]} \BibitemShut
  {NoStop}%
\bibitem [{\citenamefont {Kumar}\ \emph {et~al.}(2017)\citenamefont {Kumar},
  \citenamefont {Eckel}, \citenamefont {Jendrzejewski},\ and\ \citenamefont
  {Campbell}}]{PhysRevA.95.021602}%
  \BibitemOpen
  \bibfield  {author} {\bibinfo {author} {\bibfnamefont {A.}~\bibnamefont
  {Kumar}}, \bibinfo {author} {\bibfnamefont {S.}~\bibnamefont {Eckel}},
  \bibinfo {author} {\bibfnamefont {F.}~\bibnamefont {Jendrzejewski}},\ and\
  \bibinfo {author} {\bibfnamefont {G.~K.}\ \bibnamefont {Campbell}},\ }\href
  {https://doi.org/10.1103/PhysRevA.95.021602} {\bibfield  {journal} {\bibinfo
  {journal} {Phys. Rev. A}\ }\textbf {\bibinfo {volume} {95}},\ \bibinfo
  {pages} {021602} (\bibinfo {year} {2017})}\BibitemShut {NoStop}%
\bibitem [{\citenamefont {{Herr}}\ \emph {et~al.}(2014)\citenamefont {{Herr}},
  \citenamefont {{Brasch}}, \citenamefont {{Jost}}, \citenamefont {{Wang}},
  \citenamefont {{Kondratiev}}, \citenamefont {{Gorodetsky}},\ and\
  \citenamefont {{Kippenberg}}}]{2014NaPho...8..145H}%
  \BibitemOpen
  \bibfield  {author} {\bibinfo {author} {\bibfnamefont {T.}~\bibnamefont
  {{Herr}}}, \bibinfo {author} {\bibfnamefont {V.}~\bibnamefont {{Brasch}}},
  \bibinfo {author} {\bibfnamefont {J.~D.}\ \bibnamefont {{Jost}}}, \bibinfo
  {author} {\bibfnamefont {C.~Y.}\ \bibnamefont {{Wang}}}, \bibinfo {author}
  {\bibfnamefont {N.~M.}\ \bibnamefont {{Kondratiev}}}, \bibinfo {author}
  {\bibfnamefont {M.~L.}\ \bibnamefont {{Gorodetsky}}},\ and\ \bibinfo {author}
  {\bibfnamefont {T.~J.}\ \bibnamefont {{Kippenberg}}},\ }\href
  {https://doi.org/10.1038/nphoton.2013.343} {\bibfield  {journal} {\bibinfo
  {journal} {Nature Photonics}\ }\textbf {\bibinfo {volume} {8}},\ \bibinfo
  {pages} {145} (\bibinfo {year} {2014})},\ \Eprint
  {https://arxiv.org/abs/1211.0733} {arXiv:1211.0733 [physics.optics]}
  \BibitemShut {NoStop}%
\bibitem [{\citenamefont {Xu}\ \emph {et~al.}(2020{\natexlab{b}})\citenamefont
  {Xu}, \citenamefont {Chabchoub}, \citenamefont {Pelinovsky},\ and\
  \citenamefont {Kibler}}]{PhysRevResearch.2.033528}%
  \BibitemOpen
  \bibfield  {author} {\bibinfo {author} {\bibfnamefont {G.}~\bibnamefont
  {Xu}}, \bibinfo {author} {\bibfnamefont {A.}~\bibnamefont {Chabchoub}},
  \bibinfo {author} {\bibfnamefont {D.~E.}\ \bibnamefont {Pelinovsky}},\ and\
  \bibinfo {author} {\bibfnamefont {B.}~\bibnamefont {Kibler}},\ }\href
  {https://doi.org/10.1103/PhysRevResearch.2.033528} {\bibfield  {journal}
  {\bibinfo  {journal} {Phys. Rev. Res.}\ }\textbf {\bibinfo {volume} {2}},\
  \bibinfo {pages} {033528} (\bibinfo {year} {2020}{\natexlab{b}})}\BibitemShut
  {NoStop}%
\bibitem [{\citenamefont {{Trypogeorgos}}\ \emph {et~al.}(2025)\citenamefont
  {{Trypogeorgos}}, \citenamefont {{Gianfrate}}, \citenamefont {{Landini}},
  \citenamefont {{Nigro}}, \citenamefont {{Gerace}}, \citenamefont
  {{Carusotto}}, \citenamefont {{Riminucci}}, \citenamefont {{Baldwin}},
  \citenamefont {{Pfeiffer}}, \citenamefont {{Martone}}, \citenamefont {{De
  Giorgi}}, \citenamefont {{Ballarini}},\ and\ \citenamefont
  {{Sanvitto}}}]{trypogeorgos2025emerging}%
  \BibitemOpen
  \bibfield  {author} {\bibinfo {author} {\bibfnamefont {D.}~\bibnamefont
  {{Trypogeorgos}}}, \bibinfo {author} {\bibfnamefont {A.}~\bibnamefont
  {{Gianfrate}}}, \bibinfo {author} {\bibfnamefont {M.}~\bibnamefont
  {{Landini}}}, \bibinfo {author} {\bibfnamefont {D.}~\bibnamefont {{Nigro}}},
  \bibinfo {author} {\bibfnamefont {D.}~\bibnamefont {{Gerace}}}, \bibinfo
  {author} {\bibfnamefont {I.}~\bibnamefont {{Carusotto}}}, \bibinfo {author}
  {\bibfnamefont {F.}~\bibnamefont {{Riminucci}}}, \bibinfo {author}
  {\bibfnamefont {K.~W.}\ \bibnamefont {{Baldwin}}}, \bibinfo {author}
  {\bibfnamefont {L.~N.}\ \bibnamefont {{Pfeiffer}}}, \bibinfo {author}
  {\bibfnamefont {G.~I.}\ \bibnamefont {{Martone}}}, \bibinfo {author}
  {\bibfnamefont {M.}~\bibnamefont {{De Giorgi}}}, \bibinfo {author}
  {\bibfnamefont {D.}~\bibnamefont {{Ballarini}}},\ and\ \bibinfo {author}
  {\bibfnamefont {D.}~\bibnamefont {{Sanvitto}}},\ }\href
  {https://doi.org/10.1038/s41586-025-08616-9} {\bibfield  {journal} {\bibinfo
  {journal} {\nat}\ }\textbf {\bibinfo {volume} {639}},\ \bibinfo {pages} {337}
  (\bibinfo {year} {2025})},\ \Eprint {https://arxiv.org/abs/2407.02373}
  {arXiv:2407.02373 [cond-mat.mes-hall]} \BibitemShut {NoStop}%
\bibitem [{\citenamefont {Carlip}(2014)}]{Carlip:2014pma}%
  \BibitemOpen
  \bibfield  {author} {\bibinfo {author} {\bibfnamefont {S.}~\bibnamefont
  {Carlip}},\ }\href {https://doi.org/10.1142/S0218271814300237} {\bibfield
  {journal} {\bibinfo  {journal} {Int. J. Mod. Phys. D}\ }\textbf {\bibinfo
  {volume} {23}},\ \bibinfo {pages} {1430023} (\bibinfo {year} {2014})},\
  \Eprint {https://arxiv.org/abs/1410.1486} {arXiv:1410.1486 [gr-qc]}
  \BibitemShut {NoStop}%
\bibitem [{\citenamefont {Selinger}(2024)}]{selingerintroduction}%
  \BibitemOpen
  \bibfield  {author} {\bibinfo {author} {\bibfnamefont {J.~V.}\ \bibnamefont
  {Selinger}},\ }\href
  {https://link.springer.com/book/10.1007/978-3-031-70200-6} {\emph {\bibinfo
  {title} {Introduction to topological defects and solitons: in liquid
  crystals, magnets, and related materials}}},\ Vol.\ \bibinfo {volume} {1032}\
  (\bibinfo  {publisher} {Springer Nature},\ \bibinfo {year}
  {2024})\BibitemShut {NoStop}%
\end{thebibliography}%

\end{document}